\documentclass[12pt]{article}\usepackage{etex}
\usepackage[textheight=9in, textwidth=6.5in, letterpaper]{geometry}

\usepackage[utf8]{inputenc}
\usepackage{amsmath}
\usepackage{amssymb}
\usepackage{tikz}
\usepackage{float}
\usepackage{hyperref}
\usepackage{subfig}
\usepackage{physics}
\usepackage{indentfirst}
\usepackage{graphics}
\usepackage{soul}
\usepackage{mathtools}
\usepackage{wrapfig}
\usepackage{cite}

\newcommand{\centered}[1]{\begin{tabular}{l} #1 \end{tabular}}

\newcommand{\rt}{ \rm ret}
\newcommand{\meas}{du \, d^2 z \sqrt{\gamma} \, r^2}
\newcommand{\measnor}{du \, d^2 z \sqrt{\gamma} }

\newcommand{\bk}{\vec{k}}
\newcommand{\bx}{\vec{x}}
\newcommand{\eps}{\epsilon}
\newcommand{\vep}{\varepsilon}
\newcommand{\sg}{\sqrt{-g}}
\newcommand{\bcap}[2]{$|\vec{\beta}_i|$={#1}, $|\vec{\beta}_f|$= {#2}}
\newcommand{\ccap}[2]{$\theta_i$={#1}, $\theta_f$= {#2}}

\def\rcurs{{\mbox{$\resizebox{.09in}{.08in}{\includegraphics[trim= 1em 0 14em 0,clip]{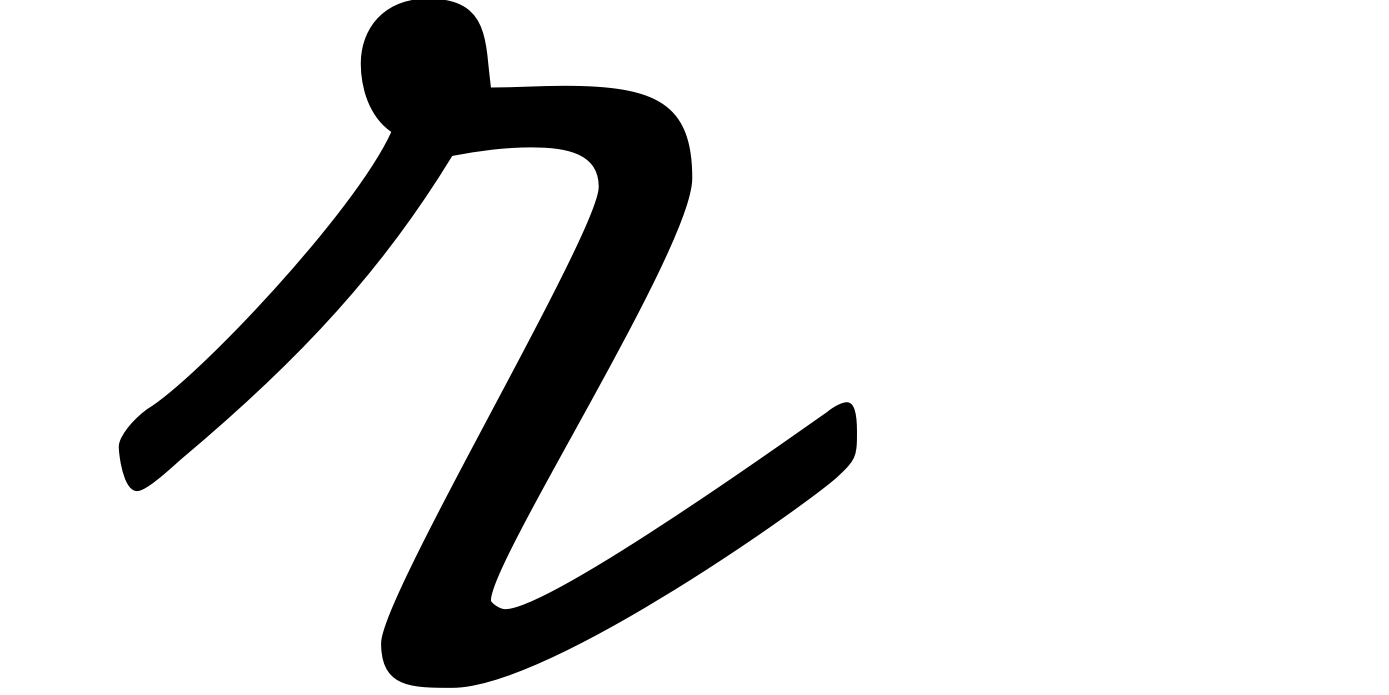}}$}}}

\title{}
\author{}
\date{}

\numberwithin{equation}{section}

\begin{document}

\begin{center}
\LARGE{From Noether's Theorem to Bremsstrahlung:}\\ 
\Large{A pedagogical introduction to}\\
\Large{Large gauge transformations and Classical soft theorems} \vspace{0.2 cm}  
\end{center}


\begin{center}

Noah Miller

\vspace{0.5 cm}

{\it  Center for the Fundamental Laws of Nature, Harvard University,\\
Cambridge, MA 02138, USA}

\vspace{0.5 cm}

\end{center}

\begin{abstract}
    Electromagnetism contains an infinite dimensional symmetry group of large gauge transformations. This gives rise to an infinite number of conserved quantities called ``soft charges'' via Noether's theorem. When charged particles scatter, the conservation of soft charge constrains the overall amount of radiation emitted per angle. Here we describe the physical consequences of soft charge conservation and give fresh accounts of the roles of spacelike and timelike infinity in these conservation laws. We conclude by exploring the possibility of creating a dual boundary theory of electromagnetism.
\end{abstract}

\tableofcontents

\begin{figure}[H]
\begin{center}

\tikzset{every picture/.style={line width=0.75pt}} 

\begin{tikzpicture}[x=0.75pt,y=0.75pt,yscale=-1,xscale=1]

\draw [line width=2.25]    (236,199) -- (321.26,199) ;
\draw [shift={(326.26,199)}, rotate = 180] [fill={rgb, 255:red, 0; green, 0; blue, 0 }  ][line width=0.08]  [draw opacity=0] (16.07,-7.72) -- (0,0) -- (16.07,7.72) -- (10.67,0) -- cycle    ;
\draw [line width=2.25]    (236,199) -- (162.26,199) ;
\draw [shift={(157.26,199)}, rotate = 360] [fill={rgb, 255:red, 0; green, 0; blue, 0 }  ][line width=0.08]  [draw opacity=0] (16.07,-7.72) -- (0,0) -- (16.07,7.72) -- (10.67,0) -- cycle    ;
\draw [line width=2.25]    (312.26,102.3) -- (356.2,159.05) ;
\draw [shift={(359.26,163)}, rotate = 232.25] [fill={rgb, 255:red, 0; green, 0; blue, 0 }  ][line width=0.08]  [draw opacity=0] (16.07,-7.72) -- (0,0) -- (16.07,7.72) -- (10.67,0) -- cycle    ;
\draw [line width=2.25]    (312.26,102.3) -- (268.32,45.56) ;
\draw [shift={(265.26,41.6)}, rotate = 412.25] [fill={rgb, 255:red, 0; green, 0; blue, 0 }  ][line width=0.08]  [draw opacity=0] (16.07,-7.72) -- (0,0) -- (16.07,7.72) -- (10.67,0) -- cycle    ;
\draw [line width=2.25]    (171.86,102.5) -- (215.8,45.76) ;
\draw [shift={(218.86,41.8)}, rotate = 487.75] [fill={rgb, 255:red, 0; green, 0; blue, 0 }  ][line width=0.08]  [draw opacity=0] (16.07,-7.72) -- (0,0) -- (16.07,7.72) -- (10.67,0) -- cycle    ;
\draw [line width=2.25]    (171.86,102.5) -- (127.92,159.25) ;
\draw [shift={(124.86,163.2)}, rotate = 307.75] [fill={rgb, 255:red, 0; green, 0; blue, 0 }  ][line width=0.08]  [draw opacity=0] (16.07,-7.72) -- (0,0) -- (16.07,7.72) -- (10.67,0) -- cycle    ;

\draw (240.29,20.6) node   [align=left] {\begin{minipage}[lt]{39.56pt}\setlength\topsep{0pt}
\begin{center}
Memory\\Effect
\end{center}

\end{minipage}};
\draw (85.69,204.5) node   [align=left] {\begin{minipage}[lt]{100.61pt}\setlength\topsep{0pt}
\begin{center}
Soft Charge\\Conservation\\{\footnotesize (classical soft theorem)}
\end{center}

\end{minipage}};
\draw (390.55,197) node  [font=\normalsize] [align=left] {\begin{minipage}[lt]{100.08pt}\setlength\topsep{0pt}
\begin{center}
Large Gauge\\Symmetry
\end{center}

\end{minipage}};
\draw (243.75,223.4) node   [align=left] {\begin{minipage}[lt]{72.5pt}\setlength\topsep{0pt}
\begin{center}
Noether's 1st\\Theorem
\end{center}

\end{minipage}};
\draw (163.16,91.23) node  [rotate=-307.16] [align=left] {\begin{minipage}[lt]{85.43pt}\setlength\topsep{0pt}
\begin{center}
Integral along $\displaystyle \mathcal{I}${\scriptsize  }
\end{center}

\end{minipage}};
\draw (321.5,91.75) node  [rotate=-52.1] [align=left] {Vacuum Transition};

\end{tikzpicture}

\end{center}
\caption{\label{fig:soft_triangle} The `infrared triangle' of classical gauge theory.}
\end{figure}
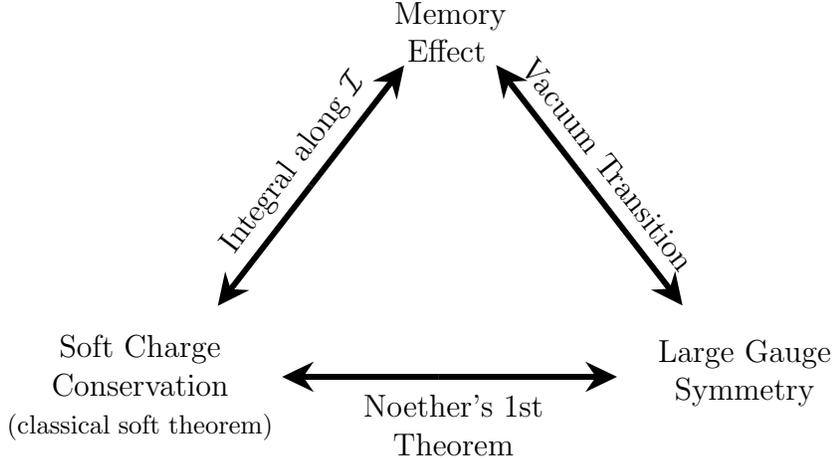

\section{Introduction and Overview}

Every gauge theory transforms under an infinite dimensional symmetry group of gauge transformations. Interestingly, the physical significance of the corresponding conservation laws only became appreciated in recent years when Strominger and collaborators connected them to soft theorems and memory effects. \cite{strominger2014bms,strominger2014asymptotic,he2015bms,he2014new,Kapec:2015ena,pasterski2016new,strominger2016gravitational}. The three main goals of this note are
\begin{enumerate}
    \item Explain how these conservation laws can be understood using Noether's theorem and analyze their physical consequences
    \item Clarify the unique roles that timelike and spacelike infinity play in these conservation laws
    \item Provide a simplified self-contained introduction to the subject suitable for non-experts.
\end{enumerate}
In order to make the exposition as friendly as possible, we will mainly restrict ourselves to the simplest possible setting of classical electromagnetism (EM), with the Liénard–Wiechert potential of a moving point charge playing a central role in our story. It should be noted, however, that the ideas discussed herein can readily be applied to non-abelian Yang Mills theory and gravity as well \cite{strominger2014bms,strominger2014asymptotic,he2015bms,pasterski2016new,strominger2016gravitational, Pate:2017vwa}.

To begin, we will review the standard textbook treatment of radiation, also known as ``bremsstrahlung,'' and discuss how accelerating charges produce an infinite number of low energy (i.e. ``soft'') photons. We will then approach the phenomenon of radiation from the completely different angle of Noether's theorem. It is well known that Maxwell EM has an infinite dimensional gauge symmetry $A_\mu \mapsto A_\mu + \partial_\mu \lambda$. By turning the crank of Noether's theorem on these gauge transformations, we will derive an infinite number of conserved charges $Q_\lambda$ which we call ``soft charges''. We will see that the existence of bremsstrahlung is actually required by the conservation of soft charge! In order to show this, however, we must impose a mysterious ``antipodal matching condition'' on $\lambda$. This antipodal matching condition is required so that no net soft charge ``leaks out'' at spatial infinity when we deform the Cauchy slice on which we evaluate $Q_\lambda$ from $\mathcal{I}^+$ to $\mathcal{I}^-$, as discussed in \cite{Prabhu:2018gzs,campiglia2017asymptotic,Campiglia:2015qka, Herdegen:2016bio}. It will then be shown that the necessity of the antipodal matching condition follows as a straightforward consequence of the finite energy constraint and Lorentz symmetry. The relevance of the finite energy constraint is that it requires for all charged particles to travel with constant velocities at $t = \pm \infty$, because if they were eternally ``wiggling'' or accelerating then the spacetime would contain an infinite amount of radiation energy. By causality, we can then make statements about the electromagnetic fields at $i^0$ using purely Coulombic fields.

Interestingly, in applying Noether's theorem to gauge symmetries, we'll find that the conserved current $J^\mu_\lambda$ is actually the divergence of a surface charge $K^{\mu \nu}_\lambda$. In fact, it can be shown that the Noether currents of gauge symmetries are always divergences of surface charges. This is a fact which has appeared endlessly in the literature (three examples: \cite{barnich2002covariant,compere2018gravitational,avery2016noether}) but desperately needs a name---we will call it ``Noether's 1.5th theorem,'' as it is a sort of combination of Noether's 1st and 2nd theorems. A proof of it is provided in the appendix.

Next, we will take a step back and carefully define the boundary conditions that $A_\mu$ and $F_{\mu \nu}$ ought to satisfy at null infinity by analyzing the Liénard–Wiechert potential. These boundary conditions will include both large $r$ fall-offs as well as antipodal matching conditions. This will allow us to define the ``aymptotic symmetry group'' of large gauge transformations $\mathcal{G}$
\begin{equation}
\begin{split}
    \mathcal{G} = \frac{\text{Gauge transformations preserving the boundary conditions}}{\text{Small gauge transformations}}.
\end{split}
\end{equation}
Here, ``small'' gauge transformations are given by $\lambda$'s that have compact support. Small gauge transformations should be thought of as ``trivial'' transformations which do not change the physical state of $A_\mu$. Large gauge transformations, however, which are given by $\lambda$'s which don't approach 0 at infinity, should rightfully be considered physical symmetries of $A_\mu$. We will find that $\mathcal{G}$ is comprised of all $\lambda$'s which are $u$ ($v$) independent when restricted to $\mathcal{I}^+$ $(\mathcal{I}^-)$.

At this stage, it's usually advantageous to perform a partial gauge fixing which rids us of trivial gauge transformations but faithfully preserves the space of physical gauge transformations $\mathcal{G}$. A very convenient partial gauge fixing is is Lorenz gauge $\partial_\mu A^\mu = 0$. The residual gauge transformations are given by $\lambda$'s which satisfy the wave equation $\partial^2 \lambda = 0$. Lorenz gauge will be an indispensable tool for calculating how the portion of a Cauchy slice which extends over future timlike infinity $i^+$ or past timelike infinity $i^-$ contributes to the total soft charge integral.

At this point, one might ask why these large gauge transformations are \textit{really} physical symmetries of electromagnetism. To answer this question definitively, we'll need to get acquainted with the symplectic form on our phase space of gauge fields. It turns out that small gauge transformations correspond to degenerate directions of our symplectic form while large gauge transformations do not, establishing the triviality of the former and the physicality of the latter.

We will then show that the soft charge $Q_\lambda$ generates the large gauge transformations $\delta A_\mu = \partial_\mu \lambda$ using the Poisson bracket, as one would expect. This will be accompanied by a discussion of the edge modes present at null infinity and their algebra of Poisson brackets. One of these edge modes will be the ``soft photon mode'' $N_A$, which is essentially a measure of how much soft radiation hits null infinity.

This will lead into a discussion of electromagnetic memory. At $u = \pm \infty$ of $\mathcal{I}^+$, the components of $A_\mu$ tangent to the large sphere, denoted $A_A$, can be expressed as total derivatives of functions which only depend on the 2-sphere coordinates (which we denote $z^A$). We can therefore write $A_A(u = \pm \infty) = \partial_A \phi_\pm$ where $\phi_\pm = \phi_\pm(z^A)$. When $A_A$ can be expressed in this form, we say it is in its ``vacuum state'' because it is pure gauge and its field strength will vanish. One would expect that the functions $\phi_\pm$ are not in-and-of-themselves meaningful because they can be changed by a gauge transformation. However, it is important to remember that the only allowed large gauge transformations in $\mathcal{G}$ are $u$-independent, so while we \textit{can} change $\phi_+$ and $\phi_-$ together as
\begin{align}
    \phi_+(z^A) &\mapsto \phi_+(z^A) + \lambda(z^A) \\
    \phi_-(z^A) &\mapsto \phi_-(z^A) + \lambda(z^A)
\end{align}
we \textit{cannot} change difference $\phi_+ - \phi_-$. While it is true that the sum $\phi_+ + \phi_-$ is in some sense arbitrary, the difference $\phi_+ - \phi_-$ will encode information on any scattering that took place in the bulk spacetime. In particular, this difference will only depend on the incoming and outgoing velocities of charged particles. This will imply that the emission of soft radiation during any scattering process induces a `vacuum transition' in the sense of spontaneous symmetry breaking, where the broken symmetry is that of large gauge transformations. The $u \to - \infty$ part of $\mathcal{I}^+$ will be in one vacuum, the $u \to + \infty$ will be in another vacuum, and the two vacuums will not be the same. We will review a thought experiment, considered in \cite{pasterski2017asymptotic,bieri2013electromagnetic}, which could be used to physically detect this vacuum transition.

We will conclude with some speculative thoughts about whether these soft charges can be used to construct a holographic theory of electromagnetism.

The story of large gauge transformations and soft theorems has been excellently reviewed in \cite{strominger2018lectures}. See also the reviews \cite{Heissenberg:2019fbn, aneesh2021celestial,pasterski2021lectures,raclariu2021lectures}. In addition to these reviews, this work was particularly inspired by the discussion of the Noether theorems by Avery and Schwab in \cite{avery2016noether}, the discussion of electromagnetic memory by Pasterski in \cite{pasterski2017asymptotic}, and the discussion of classical soft theorems in Fourier space by Laddha and Sen in 
\cite{Laddha:2019yaj,Laddha:2018rle}. See also \cite{Sahoo:2020ryf,Mao:2017wvx,Delisle:2020uui,Sahoo:2021ctw}.

\subsection{Conventions}
\noindent We use the  ($+$$-$$-$$-$) metric convention and set $\hbar = c = \epsilon_0 = 1$ unless otherwise noted. We write four-vectors as $x^\mu = (x^0, \vec{x}) = (t, \vec{r})$, $k^\mu = (k^0, \vec{k})$ and define $x^2 = x^\mu x_\mu$. We also use $x \cdot y$ = $x^0 y^0 - \vec{x} \cdot \vec{y}$, or even just $k x = k \cdot x$. We denote 2-sphere coordinates using $z^A$ where $A = 1,2$. We define ``$-z^A$'' to be the point antipodal to $z^A$. $\hat n = \hat n(z^A)$ will denote a unit vector with $\vec{r} = r \hat n$.  $\gamma_{AB}$ is the unit 2-sphere metric. We define $D^A$ to be the sphere covariant derivative, raised using $\gamma^{AB}$ instead of the full $g^{AB}$.
\begin{align}
    D_A &\equiv \nabla_A \\
    D^A &\equiv \gamma^{AB} D_B = - r^2 \nabla^A
\end{align}
Finally, we define $A^{(0)}_A(u,z)$ to be the leading piece of $A_A(r, u, z)$ in the large $r$ expansion around $\mathcal{I}^+$.

\newpage

\section{Bremsstrahlung}

\subsection{What is radiation?}

If an electron moves at a constant velocity, it'll produce a Coloumb field. If an electron wiggles or accelerates, a disturbance called ``radiation'' or ``light'' will propagate away from it travelling at $c$. While the Coulomb field strength dies off as $F_{\mu \nu} \sim 1/r^2$ (with energy density $\sim 1/r^4$), the radiation field dies off much slower, with $F_{\mu \nu} \sim 1/r$ (and energy density $\sim 1/r^2$). Therefore, at distances far away from the electron, it is only these propagating ``radiative modes'' of the electromagnetic field which can be felt.

\begin{figure}[H]
\input{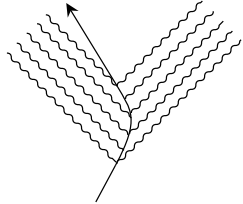}
\caption{\label{fig:electronaccel} When a charged particle accelerates, it releases radiation.}
\end{figure}

In QFT, the electromagnetic field is made up of photons. Therefore, whenever charged particles scatter and undergo some acceleration, photons will inevitably be produced. This suggests that the best basis in which to write an $\mathcal{S}$-matrix is not comprised of pure electron states, but rather ``dressed'' states where the final electron state is dressed with these photons.

The reader must be warned that the word ``dressed'' evokes a misleading mental image. 
One should not envisage that there is somehow a ``cloud'' of photons hovering around the electron. These electron-dressing photons are the bremsstrahlung which is produced while the electron is accelerating. This bremsstrahlung will fly off in all directions as shown in Figure \ref{fig:electronaccel}. The Coulomb field, which \textit{does} follow the electron around, is already automatically accounted for in unembellished introductory QFT. This is evidenced by the fact that the standard Feynman rules reproduce the Rutherford scattering amplitude which is indicative of the characteristic $\sim 1/r^2$ electric field. (In fact, one could reasonably consider the Coulomb field to be part of the electron itself, because it is the whole \{electron + Coulomb field\} object which has a combined mass of 511 keV. This can be seen by using $E=mc^2$ and observing that electrons with their Coulomb fields are not infinitely massive objects.)

\subsection{Semiclassical infrared divergence}

We will now review the semiclassical origin of the infrared divergence in the production of bremsstrahlung, essentially following section 1.3.2 of the classic textbook by Itzykson and Zuber \cite{itzykson1980quantum}. The classical EM Lagrangian in flat space is
\begin{equation}
    \mathcal{L} = - \frac{1}{4} F^{\mu \nu} F_{\mu \nu} - j^\mu A_\mu.
\end{equation}
Here, $j^\mu$ is an external charge current. In principle, $j^\mu$ could be the charge current of a dynamical matter field, but for our purposes this is unimportant so we treat it as a background source. It must satisfy
\begin{equation}
    \partial_\mu j^\mu = 0.
\end{equation}
We can Fourier decompose $j^\mu$ as
\begin{equation}\label{jfourier}
    j^\mu(x) = \frac{1}{(2 \pi)^4}\int d^4 k \; e^{-i k x} \tilde{j}^\mu(k) \hspace{1.5 cm} \tilde{j}^\mu(k) = \int d^4 x \; e^{i k x} j^\mu(x).
\end{equation}

The formula for the exact amount of energy radiated by the EM field Fourier mode with wavevector $k^\mu = (k^0, \vec{k})$, where $k^2 = 0$ and $k^0 > 0$, is
\begin{equation}\label{phaseE}
    d \mathcal{E} = \frac{-1}{(2 \pi)^3} \frac{d^3 k}{2 k^0} k^0 \tilde{j}^\mu \tilde{j}^*_\mu.
\end{equation}
This equation is proven in appendix \ref{app_EM} as equation \eqref{dEsummed}. Recall that in quantum field theory, each photon has an energy of $\hbar k^0$. While \eqref{phaseE} is derived classically, we can reasonably expect for the number of quantum photons emitted with momentum $k^\mu$ to roughly be (with $\hbar=1$)
\begin{equation}
\begin{split}
d N = \frac{d \mathcal{E}}{k^0} = \frac{-1}{(2 \pi)^3} \frac{d^3 k}{2 k^0} \tilde{j}^\mu \tilde{j}^*_\mu.
\end{split}
\end{equation}

We shall now apply this formula to a simple physical situation. Imagine a charged particle which has an initial momentum $p_i$. It then experiences a ``kick'' at $t = 0$ which changes its momentum to $p_f$. The path of the particle is given by
\begin{equation}\label{path}
    x^\mu(\tau) = \begin{cases} \frac{p^\mu_i}{m} \tau & \tau < 0 \\
    \frac{p^\mu_f}{m} \tau & \tau > 0 \\ \end{cases}.
\end{equation}
The charge current is given by
\begin{equation}\label{jkick}
    j^\mu(x) = e \int d \tau \frac{d x^\mu}{d \tau} \delta^4[ x - x(\tau)]
\end{equation}
where $e$ is the particle's electric charge. We want to compute the Fourier transform $\tilde{j}^\mu(k)$ of this function. 
Plugging \eqref{jkick} into \eqref{jfourier}, we find
\begin{equation}\label{jfourierpole}
    \tilde{j}^\mu(k) = i e \left( \frac{p_f^\mu}{k \cdot p_f} - \frac{p_i^\mu}{k \cdot p_i} \right).
\end{equation}
Plugging this into \eqref{phaseE}, we get that emitted energy per $k^\mu$ is
\begin{align}
    d \mathcal{E} &= \frac{e^2}{2 (2 \pi)^3} d^3 k \left( 2 \frac{p_i\cdot p_f}{(k \cdot p_i)(k \cdot p_f)} - \frac{m^2}{(k \cdot p_f)^2} - \frac{m^2}{(k \cdot p_i)^2} \right).
\end{align}
We can express this in terms of energy radiated in the $\hat n$ direction by substituting
\begin{align}
    k^\mu = |\vec{k}|(1, \hat n) \hspace{1cm}
    p_i^\mu = m \gamma_i (1, \vec{\beta}_i) \hspace{1cm}
    p_f^\mu = m \gamma_f (1, \vec{\beta}_f)
\end{align}
which gives us
\begin{align}
    d \mathcal{E} = \frac{e^2}{2(2 \pi)^3} \frac{d^3 k}{|\vec{k}|^2} \left( \frac{2(1 - \vec{\beta}_i \cdot \vec{\beta}_f )}{(1 - \hat n \cdot \vec{\beta}_i)(1 - \hat n \cdot \vec{\beta}_f)} - \frac{1 - \beta_f^2}{(1 - \hat n \cdot \vec{\beta}_f)^2} - \frac{1 - \beta_i^2}{(1 - \hat n \cdot \vec{\beta}_i)^2} \right)
\end{align}
Note that the quantity in parenthesis, which we shall now refer to as $f(\hat n)$, doesn't depend on the magnitude of $\vec{k}$ at all. We can therefore write
\begin{equation}
    d \mathcal{E} = \frac{e^2}{2 (2 \pi)^3} \frac{d^3 k}{|\vec{k}|^2} f(\hat n).
\end{equation}
If we wish to compute $\mathcal{E} = \int d \mathcal{E}$, the integral over the magnitude of $\vec{k}$ will simply be $\int_0^\infty k^2 d k \tfrac{1}{k^2}$. This has no divergence at small $k$, meaning the total radiated energy is finite. You may be concerned that the integral does diverge at large $k$. However, this ultraviolet divergence is due to the fact that our particle's path \eqref{path} undergoes an infinite acceleration at $t = 0$. If this `kink' were to be smoothed out to a path with some finite acceleration $a$, then the energy radiated per $k$ would essentially vanish when $k \gg a$.

Having said this, the \textit{number} of photons, which is
\begin{equation}
    d N = \frac{e^2}{2 (2 \pi)^3} \frac{d^3 k}{|\vec{k}|^3} f(\hat n).
\end{equation}
\textit{will} diverge at low momenta, as the integral will take the form $\int_0^\infty k^2 d k \tfrac{1}{k^3}$. We have now encountered the famous fact that an infinite number of low energy particles will be radiated any time charged objects scatter. Funnily enough, we can also see that this isn't a problem in spacetime dimension higher than 4. For instance, if the the dimension is $D$, then the integral is $\int_0^\infty k^{D-2} d k \tfrac{1}{k^3}$ which has no infrared divergence when $D>4$ due to the phase space factor in the integral.

\subsection{Universality of soft radiation}

\begin{figure}[H]
\centering
\includegraphics[width=0.8\textwidth]{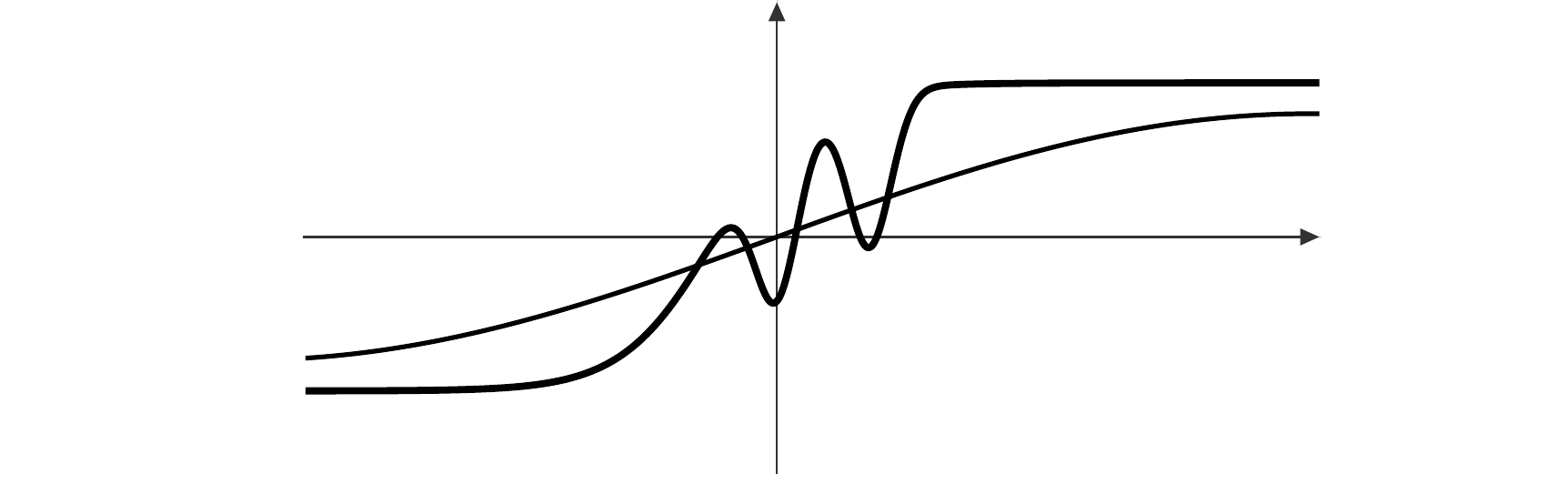}
\caption{\label{fig:fourier} If a function (thick line) asymptotes to constant values at $\pm \infty$ then the long wave length Fourier components (thin line) will only depend on those asymptotic values. More specifically, its Fourier transform will exhibit poles characteristic of the Fourier transform of the Heaviside step function.}
\end{figure}

From \eqref{jfourier}, we see that the radiation emitted per $k^\mu$ only depends on the Fourier mode $\tilde{j}^\mu(k)$ of the current density. Therefore, the emission of soft radiation only depends on the long time behavior of the charges. This implies that in scattering processes, where particles come in from infinity with some initial velocities and go out to infinity with some final velocities, the soft radiation will only depend on those initial and final velocities. It will be completely blind to the specific details of the path the particle took at intermediate times, as pointed out in \cite{Laddha:2019yaj,Laddha:2018rle}. This is why we say that soft radiation is ``universal,'' as it only depends on the ingoing and outgoing data of charged particles. In QFT, the universality of soft radiation can be seen from the fact that soft photons only couple to the external legs of Feynman diagrams.

\subsection{Angular emission profile}

\begin{figure}[H]
\begin{tabular}{ccc}
{\centered{\includegraphics[width = 0.25\textwidth]{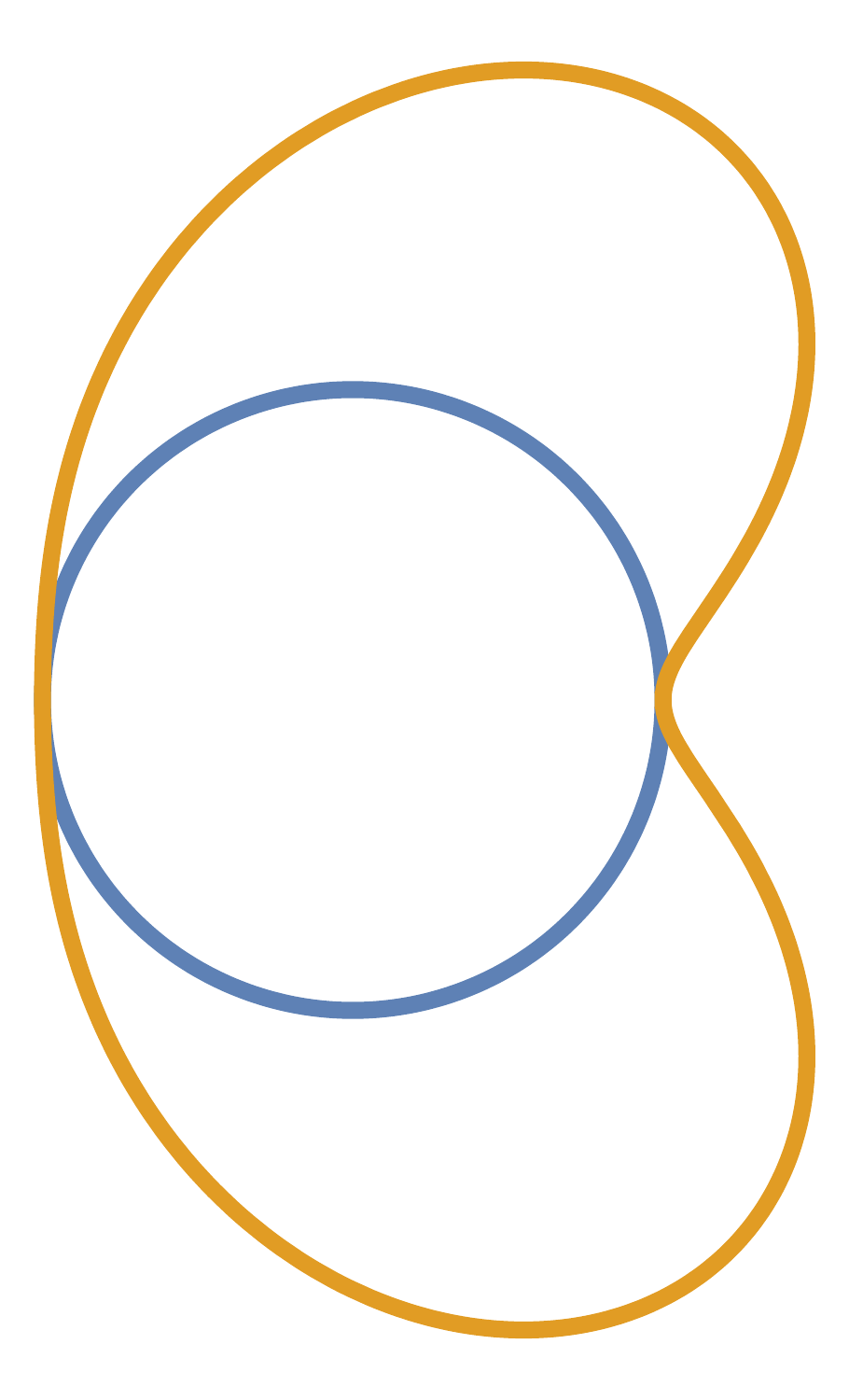}}} &
{\centered{\includegraphics[width = 0.25\textwidth]{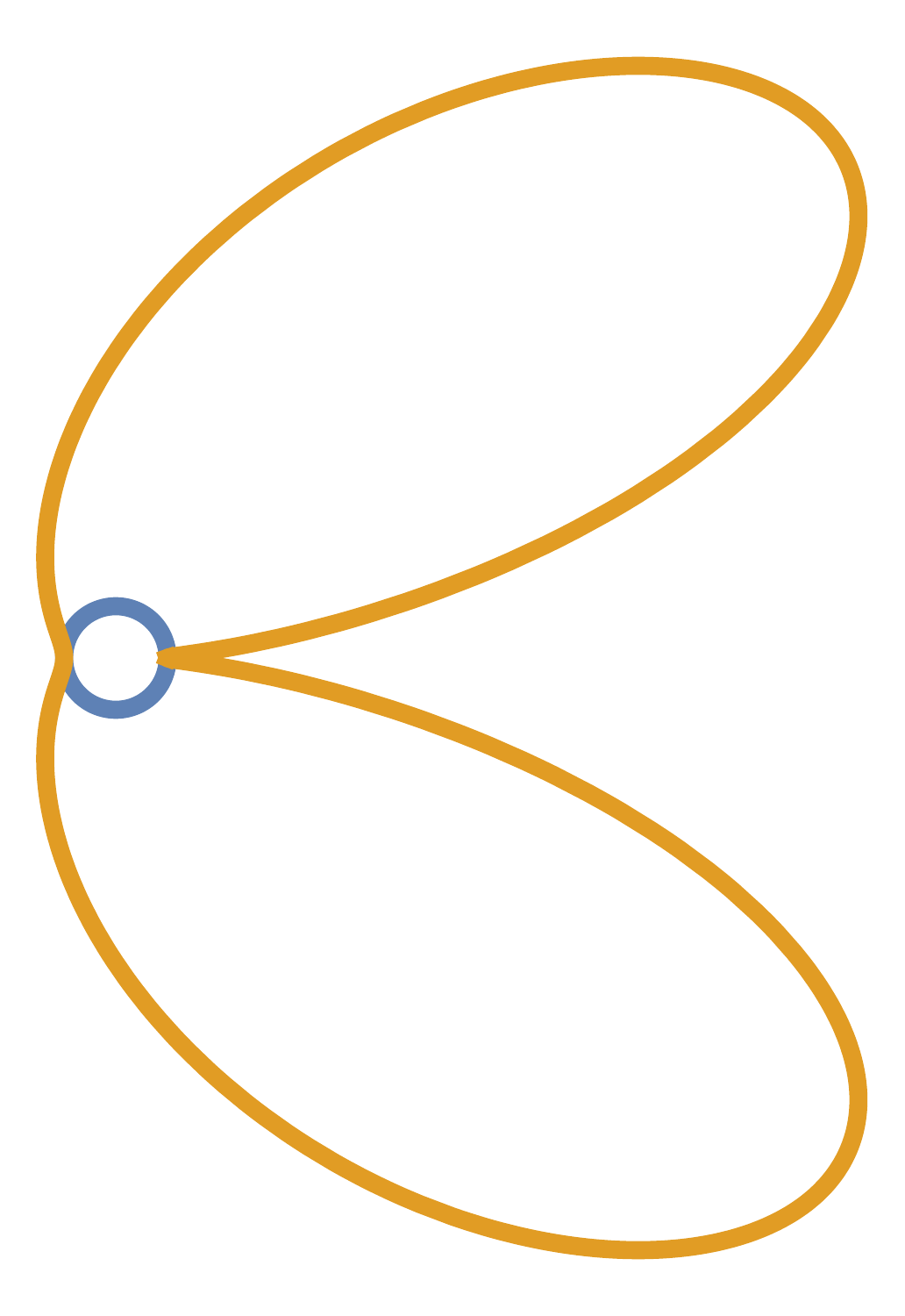}}} &
{\centered{\includegraphics[width = 0.25\textwidth]{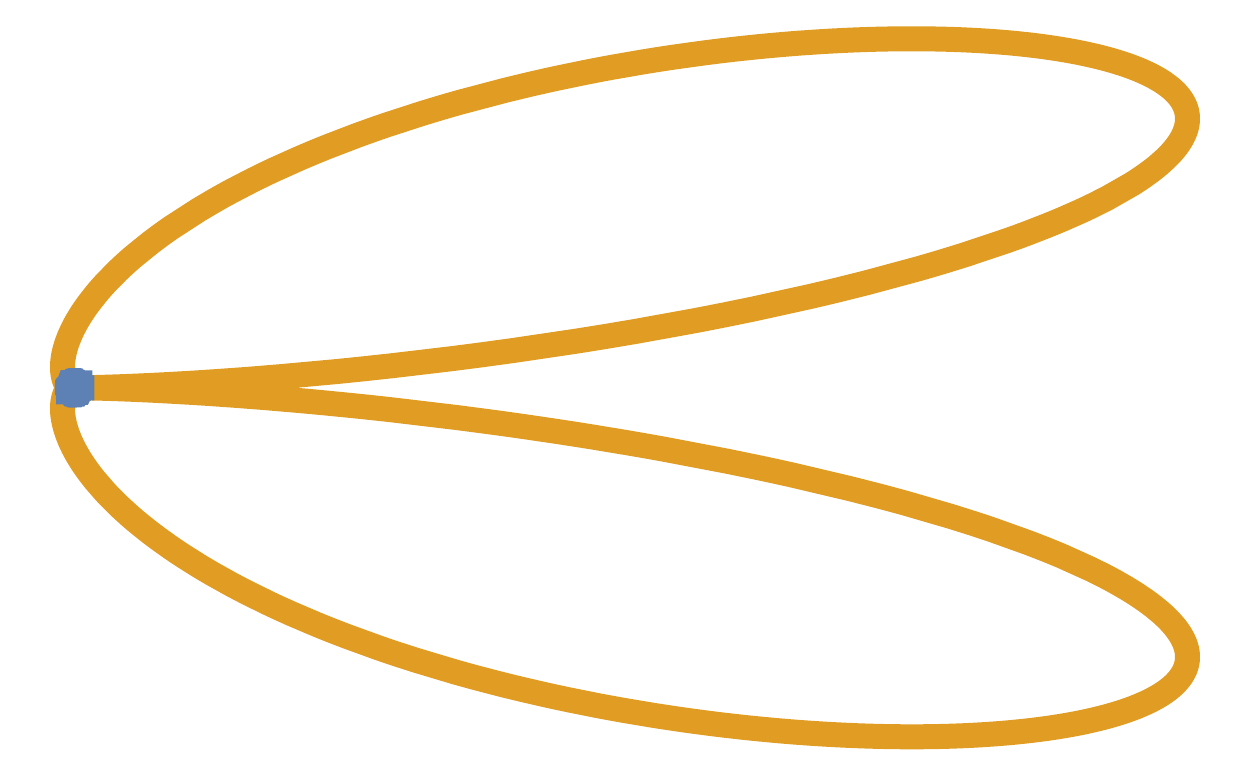}}} \\

\bcap{0.1}{0.4} & \bcap{0.1}{0.8} & \bcap{0.1}{0.97}

\\
{\centered{\includegraphics[width = 0.25\textwidth]{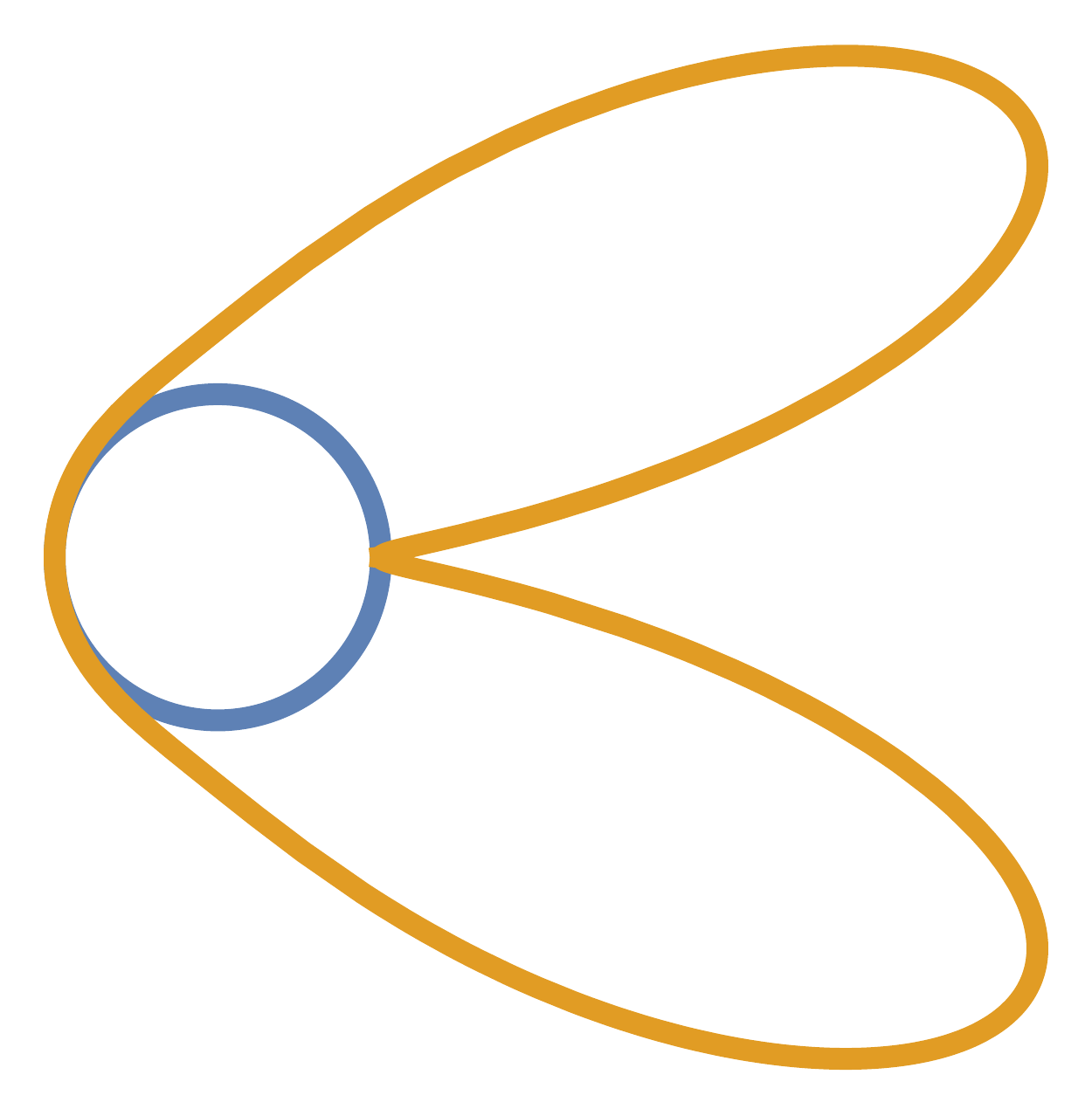}}} &
{\centered{\includegraphics[width = 0.25\textwidth]{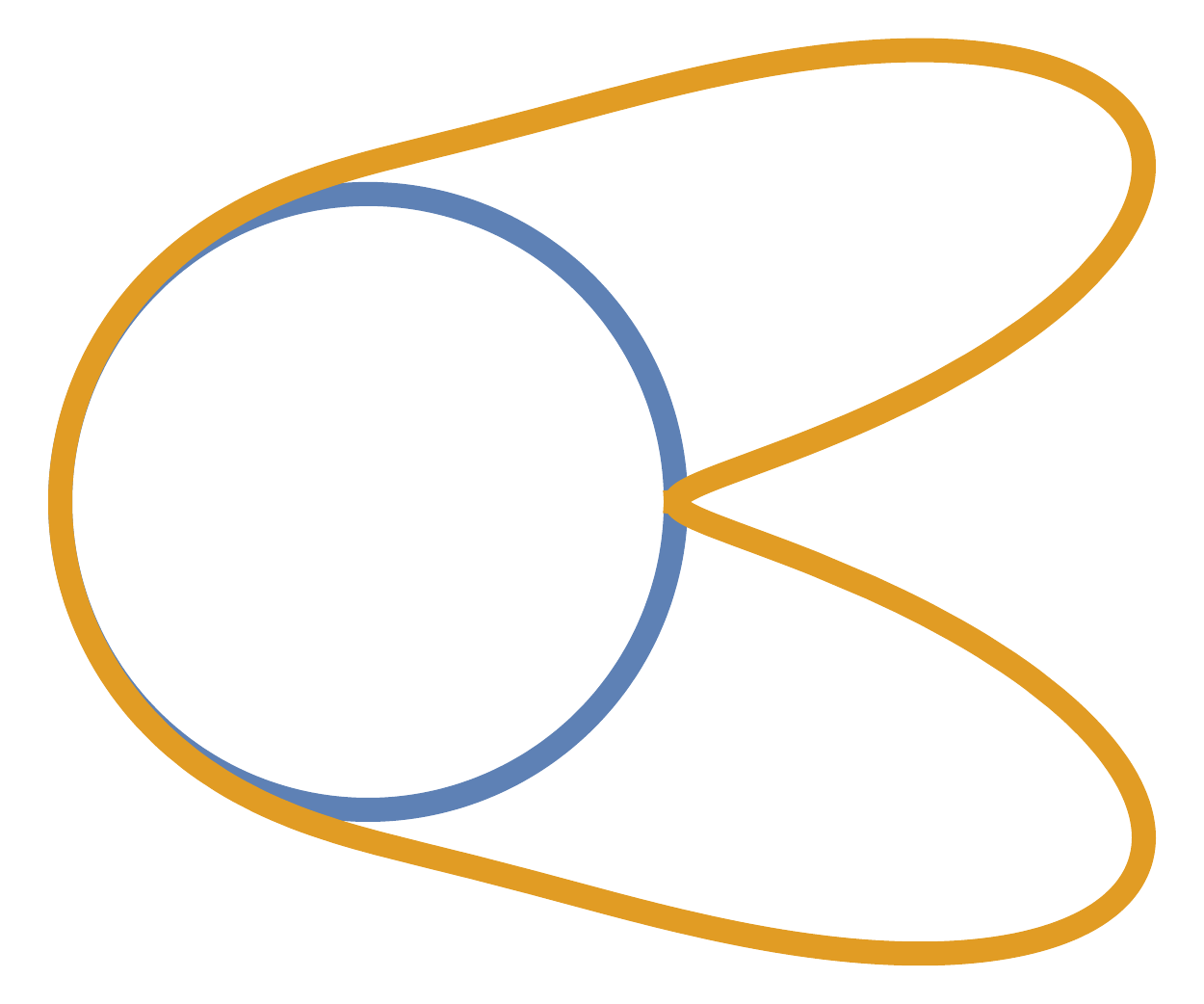}}} &
{\centered{\includegraphics[width = 0.25\textwidth]{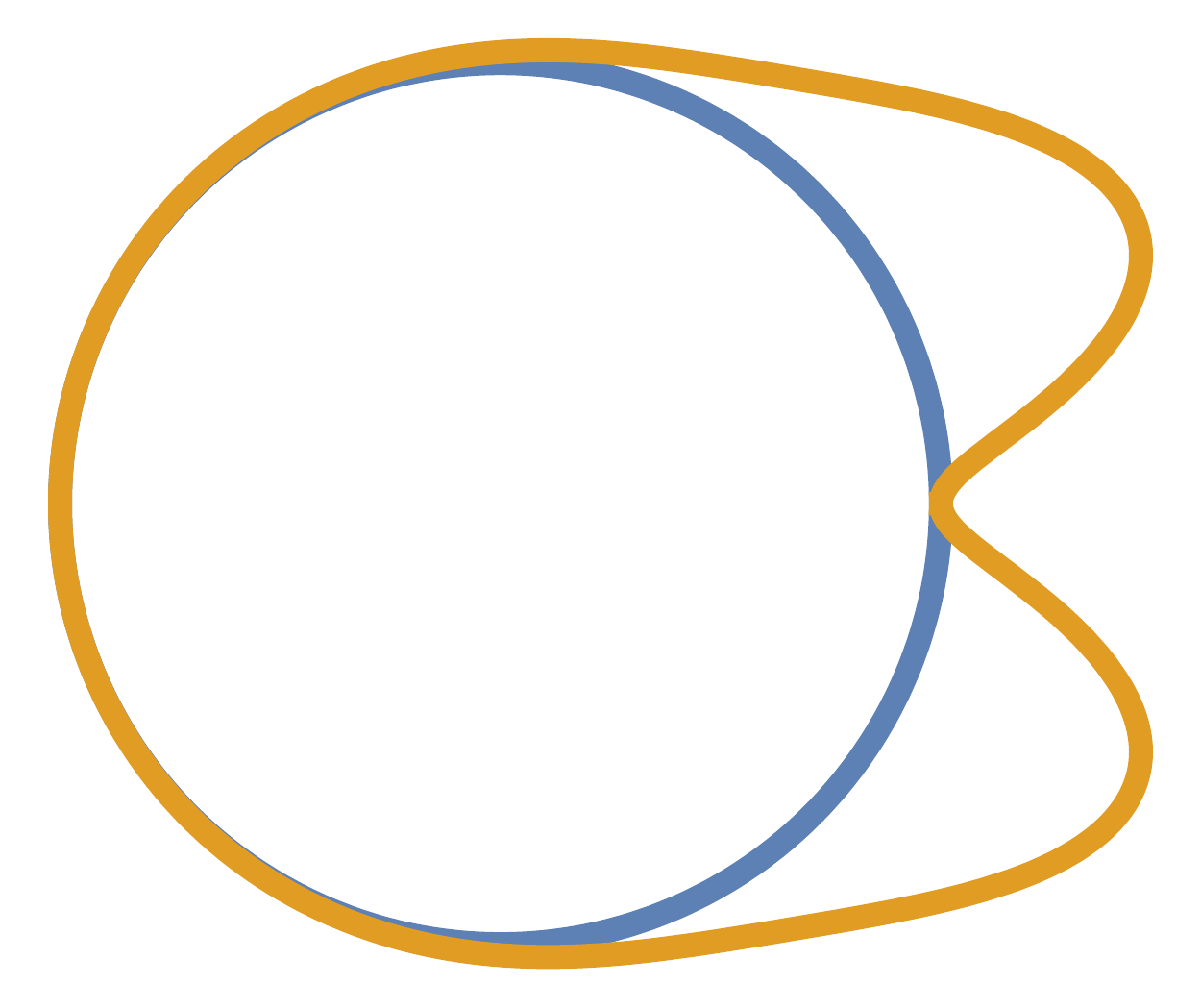}}} \\
\bcap{0.6}{0.8} & \bcap{0.7}{0.8} & \bcap{0.75}{0.8}
\end{tabular}
\caption{\label{angular1} Bremsstrahlung emission per angle when $\vec{\beta}_i$ points in the same direction as $\vec{\beta}_f$. $f(\theta)$ from \eqref{fcollinear} is graphed for different values of $|\vec{\beta}_i|$ and $|\vec{\beta}_f|$ where both velocity vectors point towards the right. The blue circle represents $f = 0$. (Each graph is scaled to be the same size, and the relative scales can be inferred from the blue circle.)}
\end{figure}

We have seen that the energy and number of soft photons radiated per angle $\hat n$ is proportional to
\begin{equation}
    f(\hat n) = \frac{2(1 - \vec{\beta}_i \cdot \vec{\beta}_f )}{(1 - \hat n \cdot \vec{\beta}_i)(1 - \hat n \cdot \vec{\beta}_f)} - \frac{1 - \beta_f^2}{(1 - \hat n \cdot \vec{\beta}_f)^2} - \frac{1 - \beta_i^2}{(1 - \hat n \cdot \vec{\beta}_i)^2}
\end{equation}
where $\vec{\beta}_i$ ($\vec{\beta}_f$) is the initial (final) velocity of the charged particle. Let's build some intuition for how this function behaves. First, if $\vec{\beta_i}$ points in the same direction as $\vec{\beta}_f$, and if $\hat n$ makes an angle of $\theta$ with both of them, then
\begin{equation}\label{fcollinear}
    f(\theta)_{\vec{\beta}_i || \vec{\beta}_f} = \frac{(|\vec{\beta}_f| - |\vec{\beta}_i|)^2 \sin^2\theta}{(1 - |\vec{\beta}_i| \cos \theta)^2(1 - |\vec{\beta}_f| \cos \theta)^2}.
\end{equation}
If $\theta = 0$, the expression is $0$. This means that if a particle accelerates along its velocity vector, no light will be emitted in the direction of its acceleration. Furthermore, the value $\theta_{\rm max}$ for which $f(\theta)$ is maximized will shrink as $|\vec{\beta_i}|$ and $|\vec{\beta}_f|$ approach 1. This is displayed in Figure \ref{angular1} and is called the ``relativistic beaming'' effect. Therefore, when an object accelerates along its velocity vector, it will emit a ``ring'' of light which has a smaller angular size the faster the particle is moving.



\newpage

\begin{figure}[H]
\begin{tabular}{ccc}
{\centered{\includegraphics[width = 0.25\textwidth]{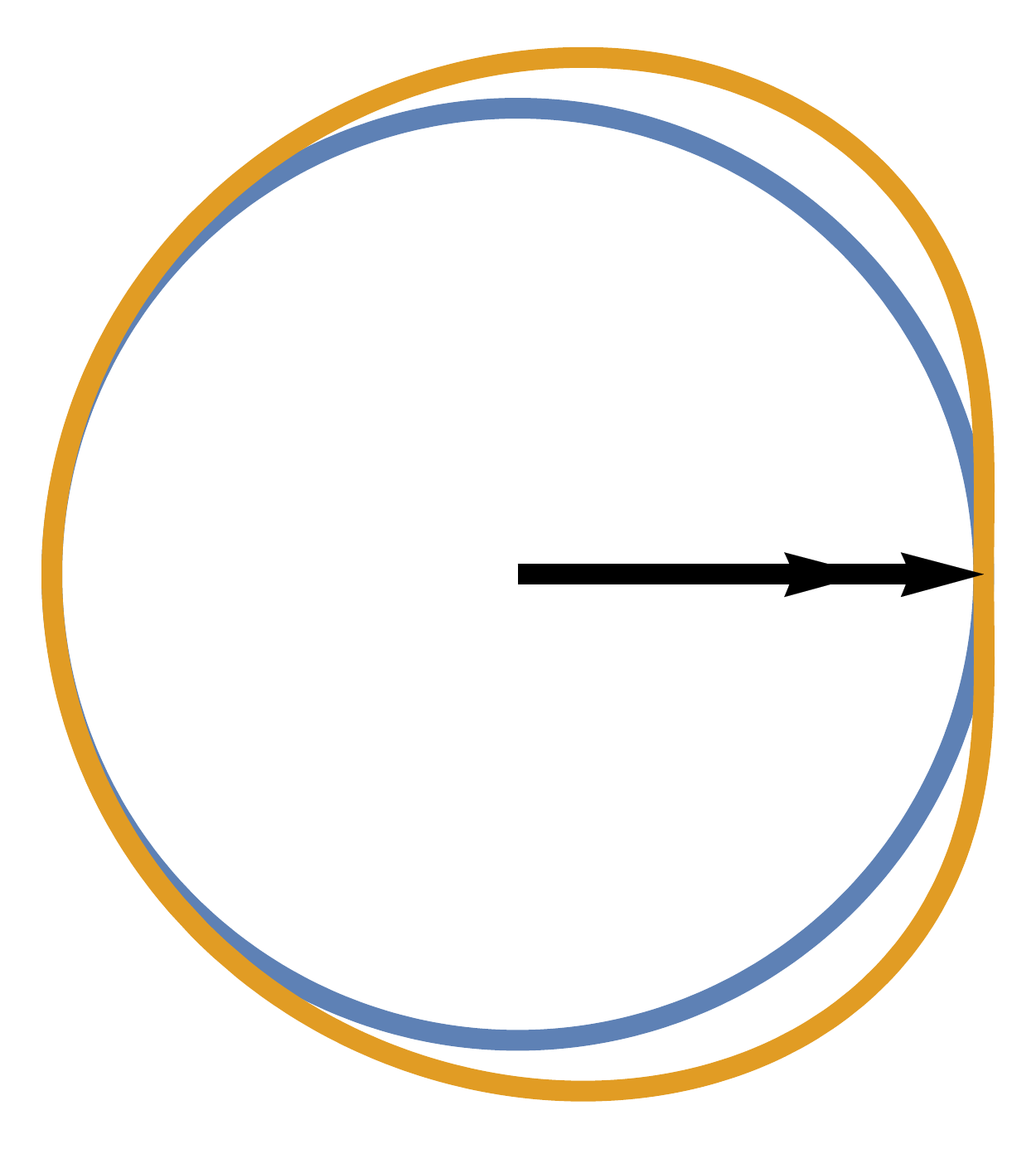}}} &
{\centered{\includegraphics[width = 0.25\textwidth]{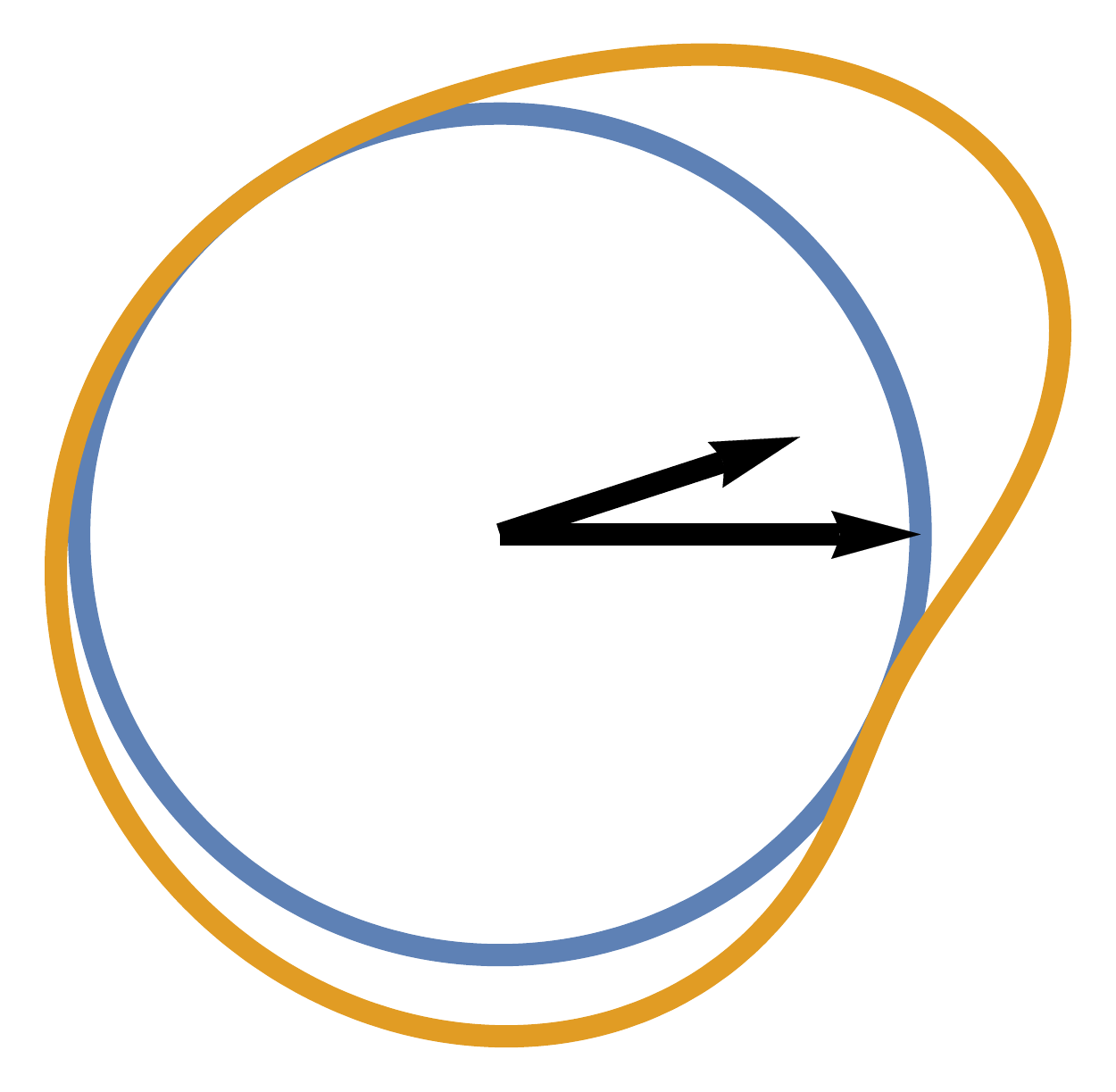}}} &
{\centered{\includegraphics[width = 0.25\textwidth]{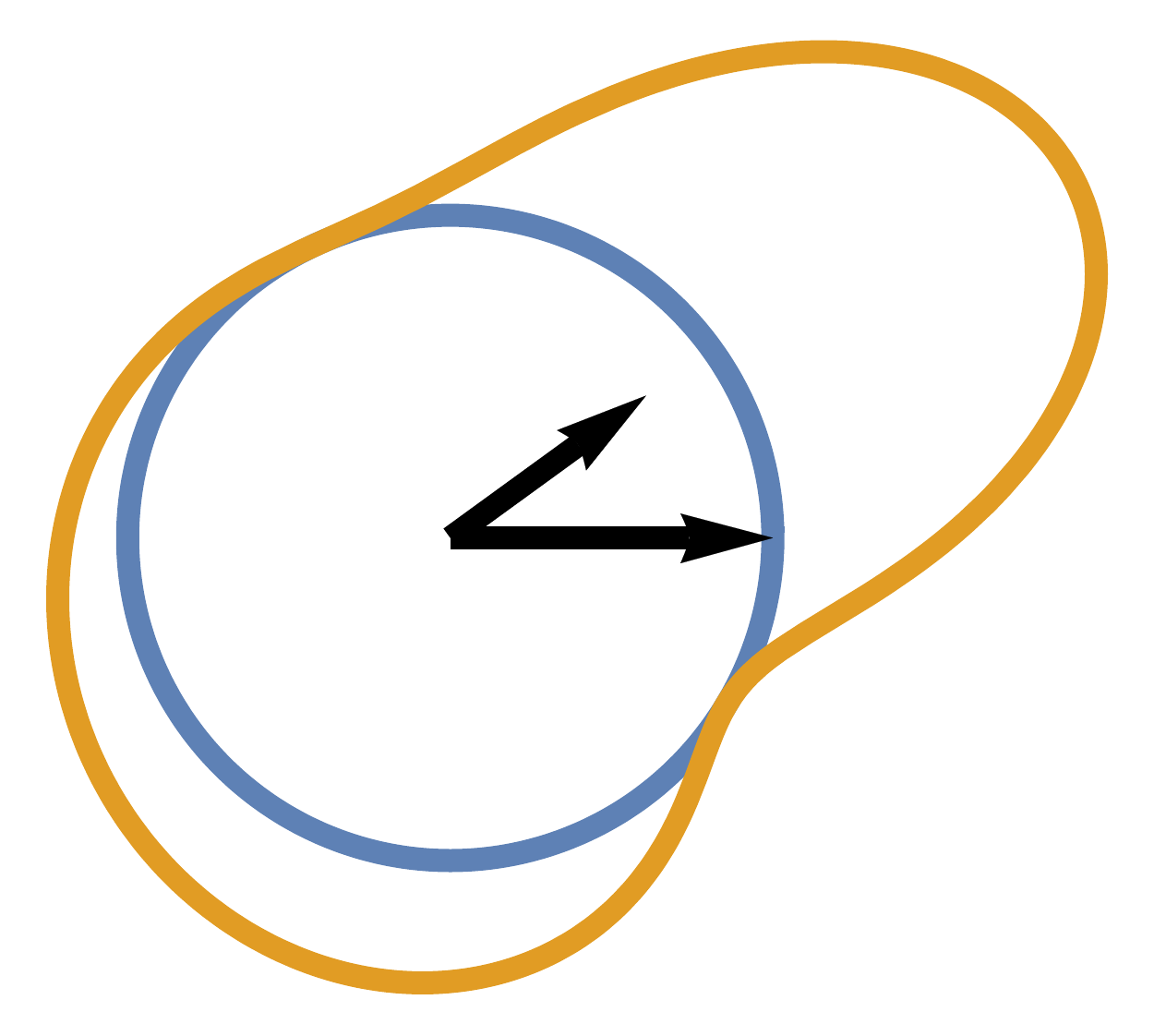}}} \\

\ccap{0}{0} & \ccap{0}{$\tfrac{1}{10} \pi$} & \ccap{0}{$\tfrac{1}{5} \pi$}

\\
{\centered{\includegraphics[width = 0.25\textwidth]{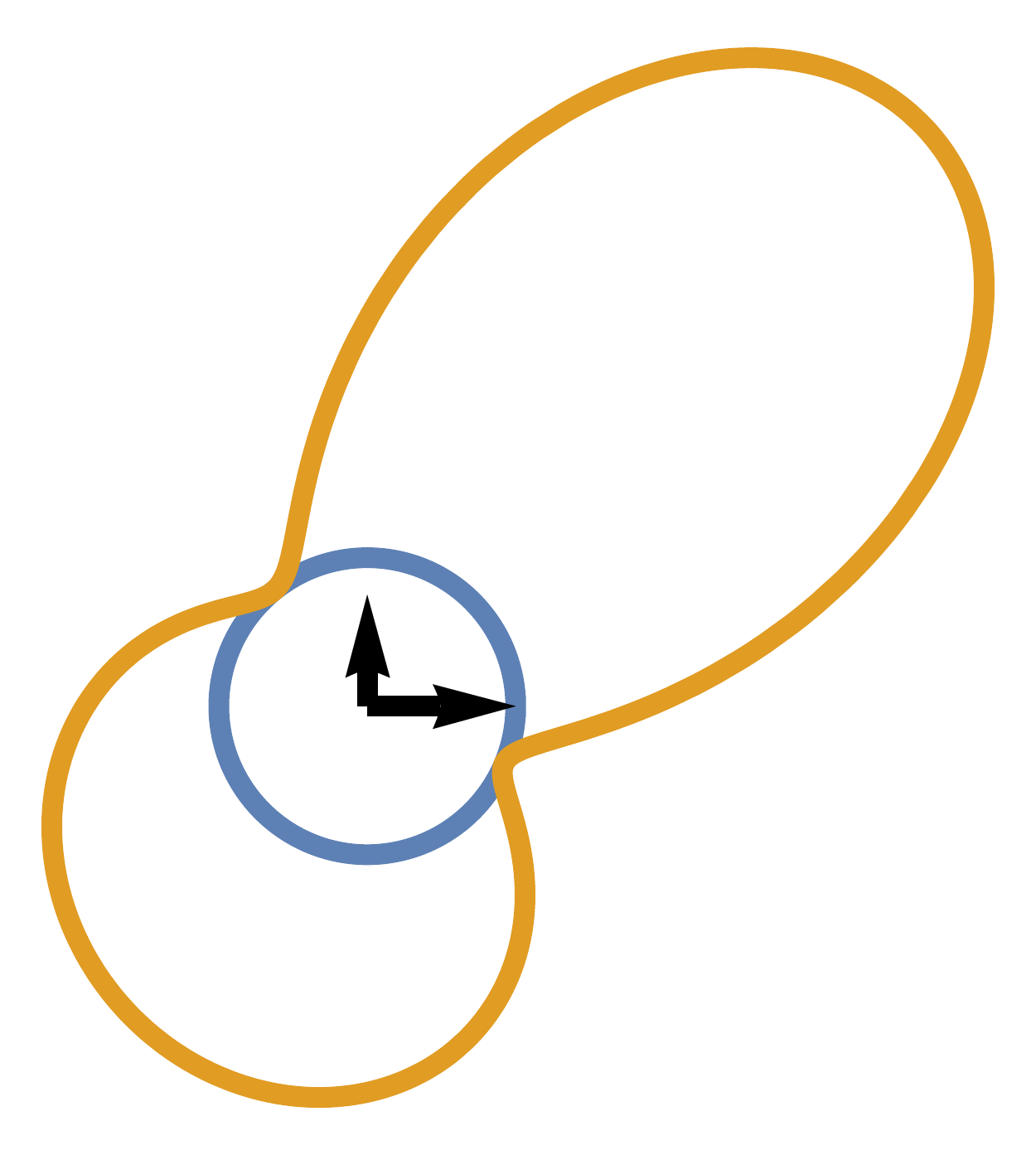}}} &
{\centered{\includegraphics[width = 0.25\textwidth]{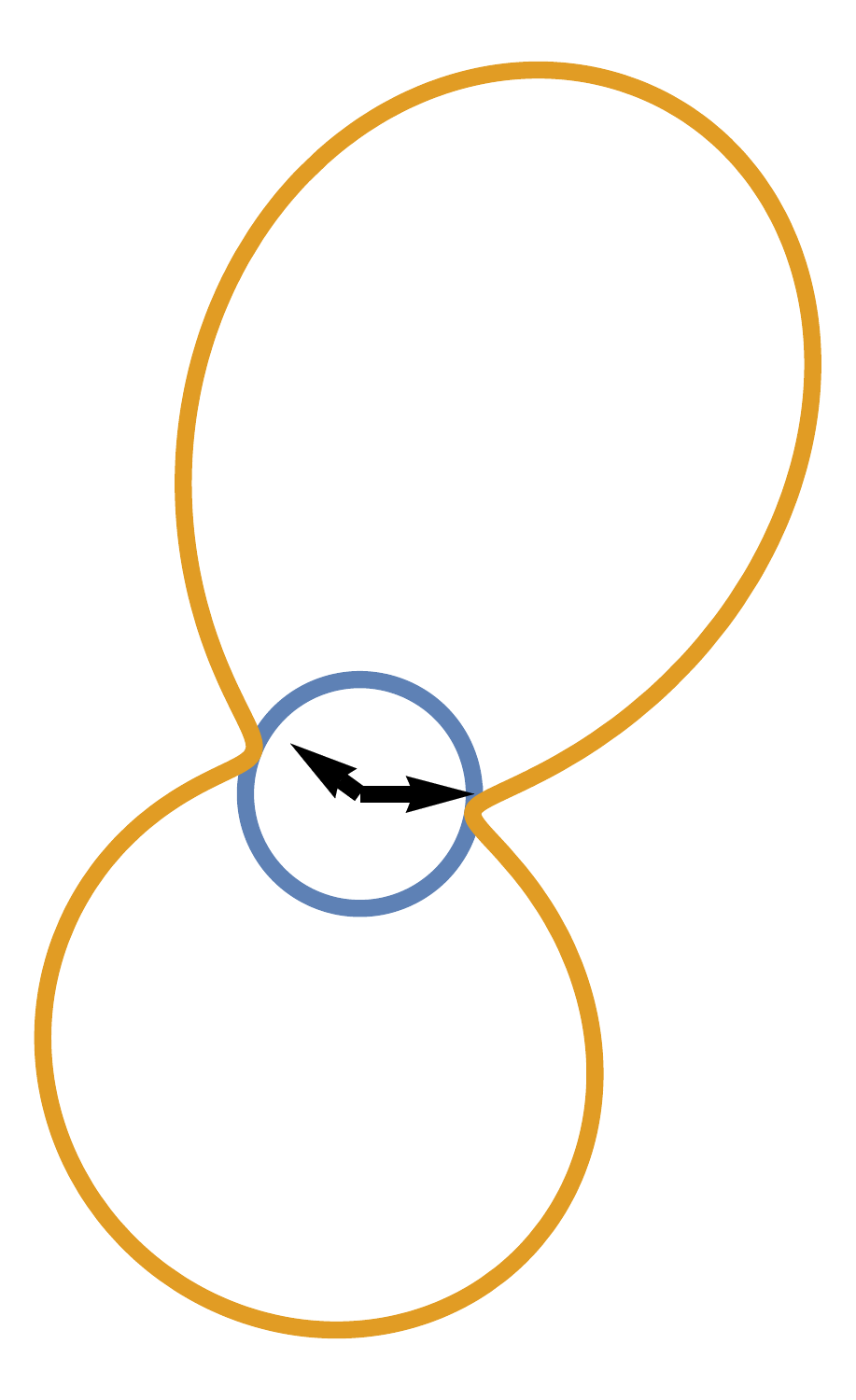}}} &
{\centered{\includegraphics[width = 0.25\textwidth]{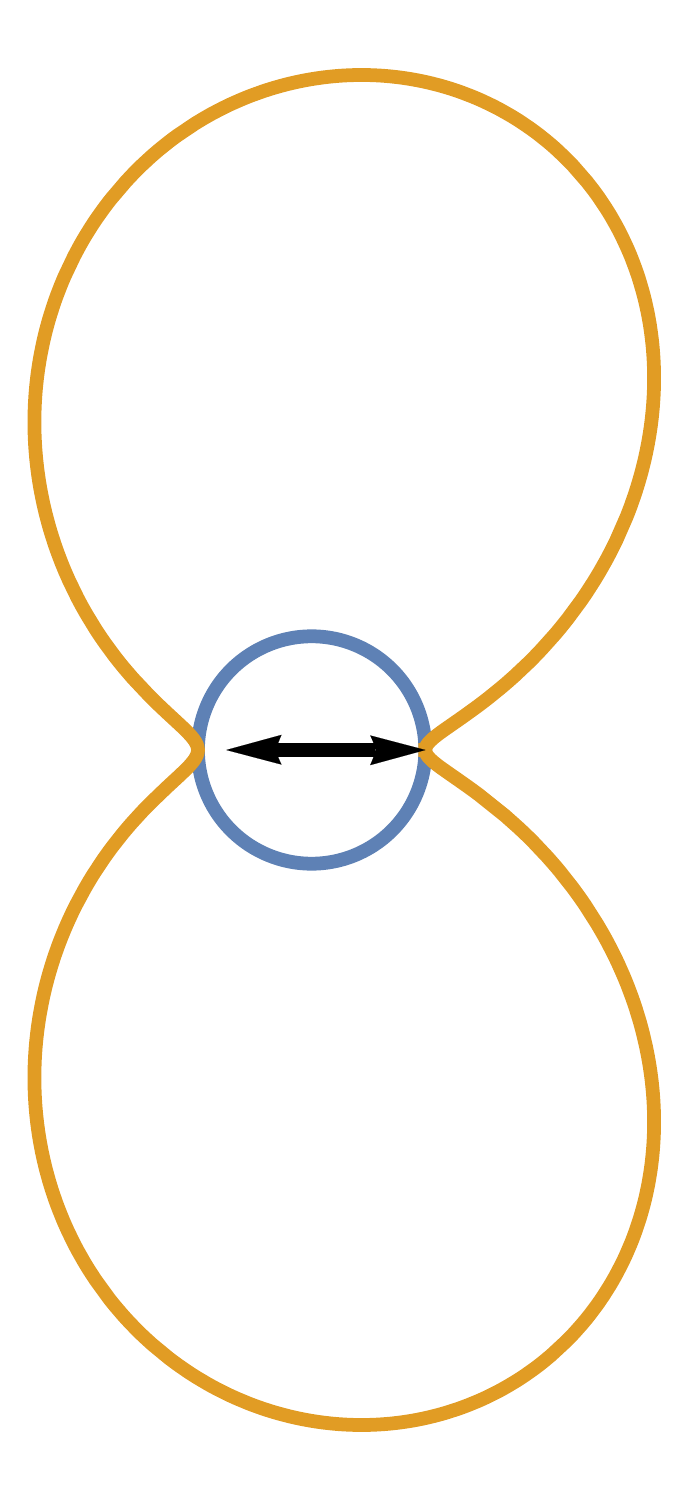}}} \\
\ccap{0}{$\tfrac{1}{2} \pi$} & \ccap{0}{$\tfrac{4}{5} \pi$} & \ccap{0}{$\pi$}
\end{tabular}
\caption{\label{angular2} Bremsstrahlung emission per angle when the direction of $\vec{\beta}_f$ is varied. Here we are graphing $f(\hat n)$ when $\hat n$ lies in the plane of $\vec{\beta}_i$ and $\vec{\beta}_f$. In each case, $|\vec{\beta}_i| = 0.4$ and $|\vec{\beta}_f| = 0.3$, with the arrows showing the relative magnitudes and directions of these vectors. The angles the vectors make with the horizontal are written below the plots, and the blue circle represents $f = 0$.}
\end{figure}

What if $\vec{\beta}_i$ is not collinear to $\vec{\beta}_f$? Then $f(\hat n)$ will not have an azimuthal rotational symmetry and in general we would need a 2-dimensional graph to visualize it. However, for simplicity, we'll restrict ourselves to graphing $f(\hat n)$ in the plane spanned by $\vec{\beta}_i$ and $\vec{\beta}_f$. The result is shown in Figure \ref{angular2}.

\newpage
\section{Noether's Theorem $\implies$ Bremsstrahlung}

Symmetries and conservation laws constrain the way in which physical systems can evolve. For example, a very complicated scattering process must always have its total initial momentum equal its total final momentum due to spatial translational symmetry. Amazingly, gauge symmetry in electromagnetism can be used to derive an infinite number of conservation laws which place very strict constraints on the way in which the electromagnetic field can evolve. In this section, we will see that these conservation laws explicitly require that accelerating charges emit radiation!

\subsection{Gauge symmetry and Noether's 1.5th theorem}

Once again, the classical EM action is
\begin{equation}
    S = \int d^4 x \left( - \frac{1}{4} F^{\mu \nu} F_{\mu \nu} - j^\mu A_\mu\right)
\end{equation}
where
\begin{equation}
    F_{\mu \nu} = \partial_\mu A_\nu - \partial_\nu A_\mu
\end{equation}
and $j^\mu$ is an external current source satisfying $\partial_\mu j^\mu = 0$. A gauge transformation acts on the the field by
\begin{equation}
    \delta A_\mu = \partial_\mu \lambda.
\end{equation}
The action is invariant under this transformation up to a boundary term, meaning it is a symmetry.

In order to find the conserved Noether current $J^\mu_\lambda$ corresponding to this symmetry, we perform the Noether procedure. First, we multiply the variation by an infinitesimal spacetime dependent function $\eps = \eps(x)$.
\begin{equation}\label{varepslambda}
\begin{split}
    \delta A_\mu = \eps \, \partial_\mu \lambda
\end{split}
\end{equation}
The variation in the action is
\begin{align}
    \delta S &= \int d^4 x \left( - F^{\mu \nu} \partial_\mu \delta A_\nu - j^\mu \delta A_\mu \right) \\
    &= \int d^4 x \left( - \partial_\mu \eps \, F^{\mu \nu} \partial_\nu \lambda - \eps j^\mu \partial_\mu \lambda  \right) \\
    &= \int d^4 x \big( \partial_\mu (- \, \eps F^{\mu \nu} \partial_\nu \lambda) + \eps \, \partial_\mu(F^{\mu \nu} \partial_\nu \lambda - \lambda j^\mu)\big).
\end{align}
On a solution to the equations of motion, $\delta S = 0$ for any variation, including our variation \eqref{varepslambda}. Therefore, if we choose $\eps$ to have compact support, we can see that on solutions to the equations of motion
\begin{align}
    \partial_\mu J^\mu_\lambda = 0
\end{align}
where
\begin{equation}\label{Jlambda_def}
    J^\mu_\lambda \equiv -F^{\mu \nu} \partial_\nu \lambda + \lambda j^\mu.
\end{equation}
It is sometimes claimed that the Noether currents derived from gauge symmetries vanish. This is patently false---we just calculated the Noether current of a gauge transformation and found that it's clearly not $0$! However, something does seem strange, as this is almost too good to be true. Gauge transformations comprise an infinite dimensional space of symmetries, and we just used them to derive an infinite number of conserved currents $J_\lambda^\mu$. What's the catch?

Well, there is no catch, these really are honest-to-goodness conserved currents. Having said that, there is something a bit peculiar about them. If we use the equations of motion
\begin{equation}
    \partial_\mu F^{\mu \nu} = j^\nu \hspace{0.75 cm} \text{(e.o.m.)}
\end{equation}
then we can rewrite the conserved current as
\begin{equation}
    J^\mu_\lambda = \partial_\nu K^{\mu \nu}_\lambda
\end{equation}
where
\begin{equation}
    K^{\mu \nu}_\lambda \equiv -\lambda F^{\mu \nu}.
\end{equation}
$K^{\mu \nu}$ is manifestly anti-symmetric, which means that the conservation of $J^\mu_\lambda$ follows directly from the trivial equation $\partial_\mu \partial_\nu K^{\mu \nu}_\lambda = 0$, which holds both on- and off-shell. So while the current $J^\mu_\lambda$ itself isn't trivial, the fact that it's conserved is trivial because it's the divergence of an anti-symmetric tensor.

In fact, Noether currents derived from gauge symmetry are always divergences of anti-symmetric tensors. This is established by a result, which has appeared many times in the literature \cite{barnich2002covariant,avery2016noether}, that we will call ``Noether's 1.5th theorem.''\footnote{All of the Noether theorems are reviewed and proven in Appendix \ref{app_noether}.}

\vspace{0.6cm}

\begin{minipage}{0.9\textwidth}
\begin{center}\textbf{\underline{Noether's 1.5th theorem}:} \end{center} If a Lagrangian has a gauge symmetry parameterized infinitesimally by a spacetime function $f$, then (up to the equations of motion) the Noether current $J^\mu_f$ will be equal to $\partial_\nu K^{\mu \nu}_f$ for some anti-symmetric tensor $K^{\mu \nu}_f$ which is locally constructed out of the fields and the gauge parameter $f$.
\end{minipage}

\vspace{1cm}

\noindent The fact that this anti-symmetric tensor is ``locally constructed'' (meaning $K^{\mu \nu}_f(x)$ only depends on the values of the fields and $f$ at the point $x$) is the non-trivial part of the theorem. This is because if any current $J^\mu$ satisfies $\partial_\mu J^\mu = 0$, then a $K^{\mu \nu}$ satisfying $\partial_\nu K^{\mu \nu} = J^\mu$ can always be found by integrating $J^\mu$ as long as the spacetime has trivial de Rham cohomology. However, because the $K^{\mu \nu}$ produced by such a process involves an integral (meaning that its value at one spacetime point will depend on the the field and gauge parameter at other spacetime points) it will generically not be locally constructed.

One might wonder what happens when $\lambda = 1$. Then, $K^{\mu \nu}_1 = -F^{\mu \nu}$ and $J^\mu_1 = j^\mu$ which satisfies $J^\mu_1 = 0$ away from charges. This vanishing of $J^\mu_1$ is due to the fact that, when $\lambda$ is a constant $\delta A_\mu = \partial_\mu 1 = 0$ and the symmetry transformation is empty, meaning the Noether current cannot depend on $A_\mu$.\footnote{One might be confused why $J^\mu_1 = j^\mu$ instead of $J^\mu_1 = 0$ even though $\delta A_\mu = \partial_\mu 1 = 0$ identically. Mechanistically this is because if we first start with generic $\lambda$, the Lagrangian changes by the total derivative $\partial_\mu( -\lambda j^\mu)$, and it is this term $\lambda j^\mu$ in the total derivative which gets included in $J^\mu_1$. More enlighteningly, note that if $j^\mu$ were not an external current but rather, say, the Noether current of the global $U(1)$ symmetry of a matter field, then then equation $J^\mu_1 = j^\mu$ would be completely expected. The variation in the gauge field would vanish while the variation in the matter field wouldn't, meaning only the latter will contribute to $J^\mu_1$.} By Stoke's theorem, if we integrate the flux of $K^{\mu \nu}_1$ through a closed 2-dimensional surface, the integral will be unchanged as we deform the surface as long as it does not pass through any charges. This is just Gauss's theorem, which we can see is given by a special case of Noether's 1.5th theorem.\footnote{Actually, very similar logic can be used in general relativity. For each infinitesimal diffeomorphism $\xi$, there is an associated current $J^\mu_\xi = \partial_\nu K^{\mu \nu}_\xi$. If one chooses a Killing symmetry $\xi$ of a particular metric, which generates the empty transformation $\delta g_{\mu \nu} = \mathcal{L}_\xi g_{\mu \nu} = 0$ on that metric, then $\partial_\mu K_\xi^{\mu \nu} = J^\mu_\xi = 0$ away from matter sources. One can then use, say, $\xi = \partial_t$ in Schwarzschild, and deform the 2D surface of integration of $K_\xi^{\mu \nu}$ from the black hole horizon (more specifically the bifurcation 2-sphere) to spatial infinity and derive its first law of thermodynamics \cite{wald1993black, compere2006introduction}.} The fact that empty gauge transformations give vanishing currents and divergence-free surface charges is called the ``Generalized Noether theorem'' in the literature \cite{compere2019advanced, Barnich:2000zw}.

\subsection{Soft charge conservation and radiation}\label{cons_rad}

Now that we have our current $J^\mu_\lambda$, we can integrate it over a Cauchy slice $\Sigma$ in order to get a conserved charge
\begin{equation}
\begin{split}
    Q_\lambda \equiv \int_\Sigma d \Sigma_\mu J^\mu_\lambda.
\end{split}
\end{equation}
In order to most easily study radiation, we will be pushing $\Sigma$ to either future or past null infinity $\mathcal{I}^+$ and $\mathcal{I}^-$. Due to charge conservation, if we define
\begin{align}
    Q_\lambda^+ &= \int_{\mathcal{I}^+} d \Sigma_\mu J^\mu_\lambda \\
    Q_\lambda^- &= \int_{\mathcal{I}^-} d \Sigma_\mu J^\mu_\lambda
\end{align}
then
\begin{equation}
\begin{split}
    Q_\lambda^+ = Q_\lambda^-.
\end{split}
\end{equation}
Let's use a set of coordinates well-adapted for studying null infinity. If $r$ is the radial coordinate, then we define the retarded and advanced times $u$ and $v$ to be
\begin{align}
    u &= t - r \\
    v &= t + r.
\end{align}
We can then write the Minkowski metric
\begin{equation}
    ds^2 = dt^2 - dr^2 - r^2 \gamma_{A B} \, dz^A dz^B
\end{equation}
in retarded coordinates or advanced coordinates as
\begin{align}
    ds^2 &= d u^2 + 2 du dr - r^2 \gamma_{A B} \, dz^A dz^B \\
    &= d v^2 - 2 dv dr - r^2 \gamma_{A B}\, dz^A dz^B.
\end{align}
Here, $z^A$ with $A = 1,2$ are coordinates which parameterize the 2-sphere and $\gamma_{AB}$ is the unit sphere metric.

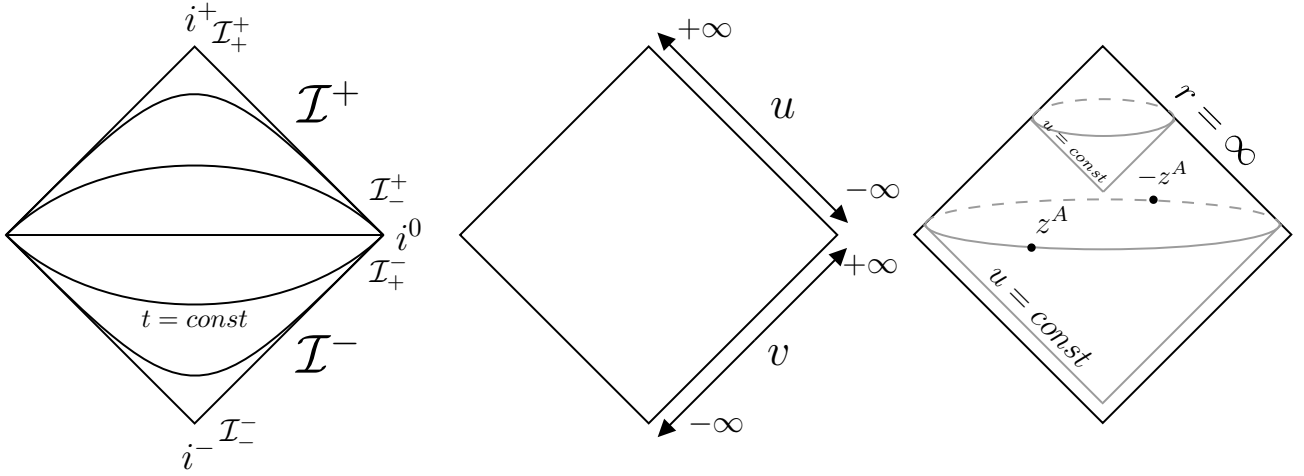
\begin{figure}[H]
\begin{center}

\tikzset{every picture/.style={line width=0.75pt}} 

\begin{tikzpicture}[x=0.75pt,y=0.75pt,yscale=-1,xscale=1]

\draw   (97.25,34.14) -- (192.36,129.25) -- (97.25,224.36) -- (2.14,129.25) -- cycle ;
\draw [color={rgb, 255:red, 0; green, 0; blue, 0 }  ,draw opacity=1 ]   (2.14,129.25) .. controls (96.6,34.67) and (97,34.27) .. (192.36,129.25) ;
\draw [color={rgb, 255:red, 0; green, 0; blue, 0 }  ,draw opacity=1 ]   (2.14,129.25) .. controls (97,223.87) and (97.8,223.87) .. (192.36,129.25) ;
\draw [color={rgb, 255:red, 0; green, 0; blue, 0 }  ,draw opacity=1 ]   (2.14,129.25) -- (192.36,129.25) ;
\draw [color={rgb, 255:red, 0; green, 0; blue, 0 }  ,draw opacity=1 ]   (2.14,129.25) .. controls (49,82.27) and (145.8,82.67) .. (192.36,129.25) ;
\draw [color={rgb, 255:red, 0; green, 0; blue, 0 }  ,draw opacity=1 ]   (2.14,129.25) .. controls (49,176.23) and (145.8,175.83) .. (192.36,129.25) ;
\draw   (326.25,34.08) -- (421.36,129.19) -- (326.25,224.3) -- (231.14,129.19) -- cycle ;
\draw    (380.2,76.6) -- (424.68,121.08) ;
\draw [shift={(426.8,123.2)}, rotate = 225] [fill={rgb, 255:red, 0; green, 0; blue, 0 }  ][line width=0.08]  [draw opacity=0] (8.93,-4.29) -- (0,0) -- (8.93,4.29) -- cycle    ;
\draw    (380.2,76.6) -- (332.72,29.12) ;
\draw [shift={(330.6,27)}, rotate = 405] [fill={rgb, 255:red, 0; green, 0; blue, 0 }  ][line width=0.08]  [draw opacity=0] (8.93,-4.29) -- (0,0) -- (8.93,4.29) -- cycle    ;
\draw    (379.7,182.36) -- (424.18,137.88) ;
\draw [shift={(426.3,135.76)}, rotate = 495] [fill={rgb, 255:red, 0; green, 0; blue, 0 }  ][line width=0.08]  [draw opacity=0] (8.93,-4.29) -- (0,0) -- (8.93,4.29) -- cycle    ;
\draw    (379.7,182.36) -- (332.89,229.17) ;
\draw [shift={(330.77,231.3)}, rotate = 315] [fill={rgb, 255:red, 0; green, 0; blue, 0 }  ][line width=0.08]  [draw opacity=0] (8.93,-4.29) -- (0,0) -- (8.93,4.29) -- cycle    ;
\draw   (555.25,33.92) -- (650.36,129.02) -- (555.25,224.13) -- (460.14,129.02) -- cycle ;
\draw  [draw opacity=0] (591.07,69.97) .. controls (591.07,69.97) and (591.07,69.97) .. (591.07,69.97) .. controls (591.07,75.1) and (575.16,79.27) .. (555.53,79.27) .. controls (535.91,79.27) and (520,75.1) .. (520,69.97) -- (555.53,69.97) -- cycle ; \draw  [color={rgb, 255:red, 155; green, 155; blue, 155 }  ,draw opacity=1 ] (591.07,69.97) .. controls (591.07,69.97) and (591.07,69.97) .. (591.07,69.97) .. controls (591.07,75.1) and (575.16,79.27) .. (555.53,79.27) .. controls (535.91,79.27) and (520,75.1) .. (520,69.97) ;
\draw  [draw opacity=0] (644.47,123.87) .. controls (644.47,123.87) and (644.47,123.87) .. (644.47,123.87) .. controls (644.47,130.83) and (604.49,136.47) .. (555.17,136.47) .. controls (505.85,136.47) and (465.87,130.83) .. (465.87,123.87) .. controls (465.87,123.87) and (465.87,123.87) .. (465.87,123.87) -- (555.17,123.87) -- cycle ; \draw  [color={rgb, 255:red, 155; green, 155; blue, 155 }  ,draw opacity=1 ] (644.47,123.87) .. controls (644.47,123.87) and (644.47,123.87) .. (644.47,123.87) .. controls (644.47,130.83) and (604.49,136.47) .. (555.17,136.47) .. controls (505.85,136.47) and (465.87,130.83) .. (465.87,123.87) .. controls (465.87,123.87) and (465.87,123.87) .. (465.87,123.87) ;
\draw [color={rgb, 255:red, 155; green, 155; blue, 155 }  ,draw opacity=1 ]   (519,70.97) -- (555.5,107.47) ;
\draw [color={rgb, 255:red, 155; green, 155; blue, 155 }  ,draw opacity=1 ]   (555.5,107.47) -- (591.67,71.3) ;
\draw  [draw opacity=0][dash pattern={on 4.5pt off 4.5pt}] (520,69.97) .. controls (520,69.97) and (520,69.97) .. (520,69.97) .. controls (520,64.83) and (535.91,60.67) .. (555.53,60.67) .. controls (575.16,60.67) and (591.07,64.83) .. (591.07,69.97) -- (555.53,69.97) -- cycle ; \draw  [color={rgb, 255:red, 155; green, 155; blue, 155 }  ,draw opacity=1 ][dash pattern={on 4.5pt off 4.5pt}] (520,69.97) .. controls (520,69.97) and (520,69.97) .. (520,69.97) .. controls (520,64.83) and (535.91,60.67) .. (555.53,60.67) .. controls (575.16,60.67) and (591.07,64.83) .. (591.07,69.97) ;
\draw  [draw opacity=0][dash pattern={on 4.5pt off 4.5pt}] (465.87,123.87) .. controls (465.87,116.91) and (505.85,111.27) .. (555.17,111.27) .. controls (604.49,111.27) and (644.47,116.91) .. (644.47,123.87) -- (555.17,123.87) -- cycle ; \draw  [color={rgb, 255:red, 155; green, 155; blue, 155 }  ,draw opacity=1 ][dash pattern={on 4.5pt off 4.5pt}] (465.87,123.87) .. controls (465.87,116.91) and (505.85,111.27) .. (555.17,111.27) .. controls (604.49,111.27) and (644.47,116.91) .. (644.47,123.87) ;
\draw [color={rgb, 255:red, 155; green, 155; blue, 155 }  ,draw opacity=1 ]   (464.87,123.87) -- (554.87,213.87) ;
\draw [color={rgb, 255:red, 155; green, 155; blue, 155 }  ,draw opacity=1 ]   (645.47,123.87) -- (554.87,214.47) ;
\draw  [fill={rgb, 255:red, 0; green, 0; blue, 0 }  ,fill opacity=1 ] (517.72,135.61) .. controls (517.72,134.72) and (518.44,134) .. (519.33,134) .. controls (520.21,134) and (520.93,134.72) .. (520.93,135.61) .. controls (520.93,136.5) and (520.21,137.22) .. (519.33,137.22) .. controls (518.44,137.22) and (517.72,136.5) .. (517.72,135.61) -- cycle ;
\draw  [fill={rgb, 255:red, 0; green, 0; blue, 0 }  ,fill opacity=1 ] (579.23,111.44) .. controls (579.23,110.56) and (579.95,109.84) .. (580.83,109.84) .. controls (581.72,109.84) and (582.44,110.56) .. (582.44,111.44) .. controls (582.44,112.33) and (581.72,113.05) .. (580.83,113.05) .. controls (579.95,113.05) and (579.23,112.33) .. (579.23,111.44) -- cycle ;

\draw (165.2,61.9) node  [font=\LARGE]  {$\mathcal{I}^{+}$};
\draw (164.86,188.1) node  [font=\LARGE]  {$\mathcal{I}^{-}$};
\draw (205.59,129.06) node  [font=\large]  {$i^{0}$};
\draw (100.8,19.06) node  [font=\large]  {$i^{+}$};
\draw (99.35,240.06) node  [font=\large]  {$i^{-}$};
\draw (195.6,107.5) node  [font=\normalsize]  {$\mathcal{I}_{-}^{+}$};
\draw (116.2,28.3) node  [font=\normalsize]  {$\mathcal{I}_{+}^{+}$};
\draw (195.26,148.5) node  [font=\normalsize]  {$\mathcal{I}_{+}^{-}$};
\draw (119.86,229.1) node  [font=\normalsize]  {$\mathcal{I}_{-}^{-}$};
\draw (393.88,65.35) node  [font=\Large]  {$u$};
\draw (391.81,190.05) node  [font=\Large]  {$v$};
\draw (339.08,18.57) node [anchor=north west][inner sep=0.75pt]    {$+\infty $};
\draw (423.58,99.57) node [anchor=north west][inner sep=0.75pt]    {$-\infty $};
\draw (344.58,218.07) node [anchor=north west][inner sep=0.75pt]    {$-\infty $};
\draw (422.58,137.57) node [anchor=north west][inner sep=0.75pt]    {$+\infty $};
\draw (97.33,171.83) node  [font=\footnotesize,color={rgb, 255:red, 0; green, 0; blue, 0 }  ,opacity=1 ]  {$t=const$};
\draw (523.88,170.18) node  [font=\normalsize,rotate=-45]  {$u=const$};
\draw (541.39,86.79) node  [font=\tiny,rotate=-45]  {$u=const$};
\draw (519.71,129.68) node [anchor=south west] [inner sep=0.75pt]  [font=\normalsize]  {$z^{A}$};
\draw (612.4,74.4) node  [font=\large,rotate=-45]  {$r=\infty $};
\draw (571.04,105.95) node [anchor=south west] [inner sep=0.75pt]  [font=\footnotesize]  {$-z^{A}$};

\end{tikzpicture}

\end{center}
\caption{\label{fig:penrose} The Penrose diagram for flat space. $\mathcal{I}^+$ ($\mathcal{I}^-$) is reached by holding $u$ ($v$) fixed and taking $r$ to $\infty$. $\mathcal{I}^+_\pm$ ($\mathcal{I}^-_\pm$) is then reached when $u$ ($v$) is taken to $\pm\infty$.}
\end{figure}

$\mathcal{I}^+$ ($\mathcal{I}^-$) is reached by holding $u$ ($v$) fixed and sending $r \to \infty$. The charge integral is equal to\footnote{In appendix \ref{app_null_integrate} we review how to integrate over null infinity.}
\begin{align}\label{Qplus1}
    Q^+_\lambda &= \int_{\mathcal{I}^+} d \Sigma_\mu J^\mu_\lambda = \int_{\mathcal{I}^+} \meas (J_\lambda)_u \\
    &= \int_{\mathcal{I}^+} \meas \left( -F_{u \nu} \partial^\nu \lambda + \lambda j_u\right). \label{Qplus2}
\end{align}
We now define $\lambda^+$ ($\lambda^-$) to be the value of $\lambda$ when restricted to $\mathcal{I}^+$ ($\mathcal{I}^-$).
\begin{align}
    \lambda^+(u, z^A) &\equiv \lim_{r \to \infty}\lambda(u, r, z^A) \\
    \lambda^-(v, z^A) &\equiv \lim_{r \to \infty}\lambda(v, r, z^A) 
\end{align}
Up until now, we have allowed for $\lambda$ to be completely arbitrary. It is time to impose two important restrictions on it. First, we require that $\lambda^\pm$ be functions of the sphere coordinates $z^A$ alone, not depending on $u$ or $v$.
\begin{equation}
    \lambda^\pm = \lambda^\pm(z^A)
\end{equation}
Second, we require that they are \textit{antipodally matched}
\begin{equation}
    \lambda^+(z^A) = \lambda^-(-z^A)
\end{equation}
where $-z^A$ is our notation for the point antipodal to $z^A$ on the 2-sphere. This antipodal matching condition seems to come out of the blue, but we'll return to it shortly and see why it is required for $Q^+_\lambda = Q^-_\lambda$ to hold assuming finite energy.

We now rewrite \eqref{Qplus2} as
\begin{align}
    \underbrace{Q^+_\lambda}_{\text{soft charge}} &= \int_{\mathcal{I}^+} du \, d^2 z \sqrt{\gamma} \, \big( \underbrace{ F_{uA} D^A \lambda }_{\text{soft part}}+ \underbrace{r^2 \lambda j_u}_{\text{hard part}}\big)
\end{align}
As a piece of terminology, we call $Q_\lambda^+$ the ``soft charge.'' The first piece, which depends on the EM field and is only non-zero when $\lambda$ isn't constant, is called the ``soft part of the soft charge.'' Much later we will see that it measures how much soft radiation has hit null infinity. The second piece, the ``hard part of the soft charge,'' is the electric charge flux integral weighted by $\lambda$. It is finite because $j_u \sim 1/r^2$. We denote the decomposition into the hard and soft pieces as
\begin{equation}
\begin{split}
    Q^\pm_\lambda = Q^{\pm,S}_\lambda + Q^{\pm,H}_\lambda.
\end{split}
\end{equation}

The conservation of soft charge $Q_\lambda$ gives
\begin{equation}
\begin{split}
    Q^{+,S}_\lambda + Q^{+,H}_\lambda = Q^{-,S}_\lambda + Q^{-,H}_\lambda.
\end{split}
\end{equation}
This can be rewritten as
\begin{equation}\label{SH_1}
\begin{split}
    Q^{+,H}_\lambda - Q^{-,H}_\lambda = - Q^{+,S}_\lambda + Q^{-,S}_\lambda.
\end{split}
\end{equation}

\noindent Let us now make one last simplifying assumption and say our charged particles are massless, hitting null infinity instead of timelike infinity. We do this to avoid having to integrate $J^\mu_\lambda$ over timelike infinity, as doing so requires a bit of care. It should be emphasized, however, that this is not a necessary assumption to make and once we are acquainted with Lorenz gauge in section \ref{sec_lorenz} we will learn how to properly handle massive particles.

We now leverage the antipodal matching condition. Consider a collection of charges labelled by $k = 1 \ldots N$ which have some initial/final velocities $\vec{\beta}_{k,i}$ and $\vec{\beta}_{k,f}$. If $\vec{\beta}_{k,i}=\vec{\beta}_{k,f}$ for all $k$, then the particles will hit $\mathcal{I}^+$ at the exact antipodal angles they entered $\mathcal{I}^-$. This would imply that $Q^{+,H}_\lambda - Q^{-,H}_\lambda = 0$. However, if some of the particles leave with different velocities than they entered, then the difference in hard charge will be non-zero. Therefore,

\begin{equation}\label{SH_2}
\begin{split}
    Q^{+,H}_\lambda - Q^{-,H}_\lambda = \begin{cases} 0 & \text{ if } \vec{\beta}_{k,i} = \vec{\beta}_{k,f} \text{ for all } k \\ \mathcal{O}(1) &  \text{ if } \vec{\beta}_{k,i} \neq \vec{\beta}_{k,f} \text{ for some } k\end{cases}
\end{split}
\end{equation}
\vskip.15cm
\noindent where $\mathcal{O}(1)$ denotes some finite number in the $r \to \infty$ limit. (Incidentally, the above equation will still hold when the particles are massive.)

We now turn to the soft part of the soft charge, which depends on $F_{uA}$. In order to come to any conclusions about whether $Q^{\pm, S}_\lambda$ is finite or 0, we need to know what the large $r$ fall-offs of $F_{uA}$ are. To do this, we will use the stress energy tensor of the EM field
\begin{equation}
\begin{split}
    T_{\mu \nu} = -F_\mu^{\; \alpha} F_{\nu \alpha} + \tfrac{1}{4} g_{\mu \nu} F^{\alpha \beta} F_{\alpha \beta}.
\end{split}
\end{equation}
We will use that the energy density $T_{tt} \sim \mathcal{O}(1/r^4)$ for a Coulomb field and $T_{tt} \sim \mathcal{O}(1/r^2)$ for a radiation field. Doing a coordinate change, we find
\begin{equation}
\begin{split}
    T_{tt} = T_{uu} = \frac{1}{r^2} \gamma^{AB} F_{u A} F_{u B} + \ldots
\end{split}
\end{equation}
Therefore, $F_{uA} \sim \mathcal{O}(1/r)$ for a Coulomb field and $F_{uA} \sim \mathcal{O}(1)$ for a radiation field. We now see that $Q^{\pm,S}_\lambda$ is 0 when no radiation reaches $\mathcal{I}^\pm$ and is $\mathcal{O}(1)$ when radiation does reach $\mathcal{I}^\pm$.

\vskip.25cm
\begin{equation}\label{SH3}
    Q^{\pm,S}_\lambda = \begin{cases} 0 & \text{ if no radiation passes through } \mathcal{I}^\pm \\
    \mathcal{O}(1) & \text{ if radiation passes through } \mathcal{I}^\pm \end{cases}.
\end{equation}
\vskip.35cm
Let's assume that no radiation enters through $\mathcal{I}^-$, which is a physically natural assumption. This would imply that $Q^{-,S}_\lambda = 0$. Combining \eqref{SH_1}, \eqref{SH_2}, \eqref{SH3}, we conclude that radiation must hit $\mathcal{I}^+$ if any charged particle's final velocity doesn't equal its initial velocity, meaning it experienced net acceleration. Therefore, Noether's theorem implies the existence of bremsstrahlung. QED!

\subsection{Finite energy and scattering} \label{sec_finite_energy}

In physics, it is often useful to assume that the system that you are studying has a finite amount of energy. If we impose this condition, we can deduce something interesting.\footnote{We will only be concerned with finite energy at large $r$. The divergence of the energy density of a Coulomb field near a point charge is an inevitable feature of non-renormalized classical EM.} Because wiggling/accelerating charges radiate energy-carrying light waves, if there is a finite amount of energy in our electromagnetic field then these particles cannot have been wiggling/accelerating infinitely far in the past. At $t = - \infty$, they must start by travelling at a constant velocity and at $t = + \infty$ they must end by travelling at a constant velocity. This is the exact sort of scenario that one studies in scattering problems.

From here on out, we will therefore only be interested in studying situations where the particles start and end by travelling at constant velocities. This will turn out to be a crucial assumption which, for instance, lies at the heart of the mysterious antipodal matching condition.

\subsection{The reason for antipodal matching}\label{antipodal_reason}

When we deformed the Cauchy surface from $\mathcal{I}^+$ to $\mathcal{I}^-$ and proclaimed that $Q^+_\lambda = Q^-_\lambda$, we weren't being very careful. This is because even though the Penrose diagram visually makes it look like $\mathcal{I}^+_-$ is the same as $\mathcal{I}^-_+$, this is simply not the case. The Penrose diagram violently distorts spatial infinity, and if we want to push a surface from $\mathcal{I}^+$ to $\mathcal{I}^-$ we must be mindful of the behavior there.

\begin{figure}
\begin{center}

\tikzset{every picture/.style={line width=0.75pt}} 

\begin{tikzpicture}[x=0.75pt,y=0.75pt,yscale=-1,xscale=1]

\draw    (26.33,214.67) -- (204.59,214.67) ;
\draw [shift={(207.59,214.67)}, rotate = 180] [fill={rgb, 255:red, 0; green, 0; blue, 0 }  ][line width=0.08]  [draw opacity=0] (8.93,-4.29) -- (0,0) -- (8.93,4.29) -- cycle    ;
\draw    (26.33,206.67) -- (204.59,206.67) ;
\draw [shift={(207.59,206.67)}, rotate = 180] [fill={rgb, 255:red, 0; green, 0; blue, 0 }  ][line width=0.08]  [draw opacity=0] (8.93,-4.29) -- (0,0) -- (8.93,4.29) -- cycle    ;
\draw    (26.33,198.67) -- (204.59,198.67) ;
\draw [shift={(207.59,198.67)}, rotate = 180] [fill={rgb, 255:red, 0; green, 0; blue, 0 }  ][line width=0.08]  [draw opacity=0] (8.93,-4.29) -- (0,0) -- (8.93,4.29) -- cycle    ;
\draw    (18,174.67) -- (140.3,52.36) ;
\draw [shift={(142.42,50.24)}, rotate = 495] [fill={rgb, 255:red, 0; green, 0; blue, 0 }  ][line width=0.08]  [draw opacity=0] (8.93,-4.29) -- (0,0) -- (8.93,4.29) -- cycle    ;
\draw    (12,168.67) -- (134.3,46.36) ;
\draw [shift={(136.42,44.24)}, rotate = 495] [fill={rgb, 255:red, 0; green, 0; blue, 0 }  ][line width=0.08]  [draw opacity=0] (8.93,-4.29) -- (0,0) -- (8.93,4.29) -- cycle    ;
\draw    (6,162) -- (128.3,39.7) ;
\draw [shift={(130.42,37.58)}, rotate = 495] [fill={rgb, 255:red, 0; green, 0; blue, 0 }  ][line width=0.08]  [draw opacity=0] (8.93,-4.29) -- (0,0) -- (8.93,4.29) -- cycle    ;
\draw    (26.67,194.33) -- (196.44,131.94) ;
\draw [shift={(199.25,130.91)}, rotate = 519.8199999999999] [fill={rgb, 255:red, 0; green, 0; blue, 0 }  ][line width=0.08]  [draw opacity=0] (8.93,-4.29) -- (0,0) -- (8.93,4.29) -- cycle    ;
\draw    (24,185.67) -- (193.77,123.28) ;
\draw [shift={(196.59,122.24)}, rotate = 519.8199999999999] [fill={rgb, 255:red, 0; green, 0; blue, 0 }  ][line width=0.08]  [draw opacity=0] (8.93,-4.29) -- (0,0) -- (8.93,4.29) -- cycle    ;
\draw    (20.67,177) -- (190.44,114.61) ;
\draw [shift={(193.25,113.58)}, rotate = 519.8199999999999] [fill={rgb, 255:red, 0; green, 0; blue, 0 }  ][line width=0.08]  [draw opacity=0] (8.93,-4.29) -- (0,0) -- (8.93,4.29) -- cycle    ;

\draw (218.33,186.33) node [anchor=north west][inner sep=0.75pt]   [align=left] {to spatial infinity\\$\displaystyle t$ fixed, $\displaystyle r\ \rightarrow \infty $};
\draw (149.33,14) node [anchor=north west][inner sep=0.75pt]   [align=left] {to null infinity\\$\displaystyle t-r$ fixed, $\displaystyle r\ \rightarrow \infty $};
\draw (208.67,97.33) node [anchor=north west][inner sep=0.75pt]   [align=left] {to spatial infinity\\$\displaystyle t-\alpha r$ fixed, $\displaystyle r\ \rightarrow \infty $};

\end{tikzpicture}

\end{center}
\caption{\label{fig:to_infinity}How to get to null and spatial infinity: hold $t - \alpha r$ fixed while taking $r \to \infty$. If $\alpha = 1$, you will reach $\mathcal{I}^+$. If $-1 < \alpha < 1$, you will reach $i^0$. Note that $t/r \to \alpha$ in the $r \to \infty$ limit.}
\end{figure}
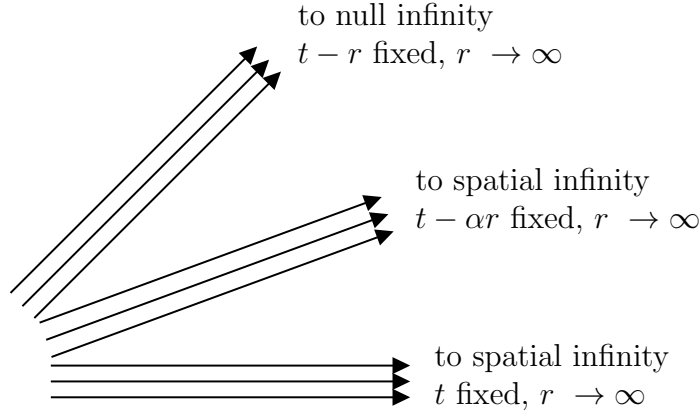

In particular, the current flux of $J^\mu_\lambda$ passing through $i^0$ will account for the difference between $Q^+_\lambda$ and $Q^-_\lambda$.
\begin{align}
    Q^+_\lambda - Q^-_\lambda &= \int_{i^0} d \Sigma_\mu J^\mu_\lambda \\
    &= - \int_{i^0} d \Sigma_\mu F^{\mu \nu} \nabla_\nu \lambda. \label{i0_int}
\end{align}
A priori, there is no reason for the term on the RHS to be 0. However, remember that we are only interested in scattering configurations where the particles enter with some constant velocity. If we use the retarded propagator, then by causality the EM field at $i^0$, $\mathcal{I}^+_-$, and $\mathcal{I}^-_+$ will only depend on the behavior of the particles in the distant past. Therefore, the fields there will be exactly equal to the Liénard–Wiechert field of a particle moving with a constant velocity $\vec{\beta}$. If we assume the particle passes through the origin for simplicity (because the leading $1/r$ behavior will be unaffected by any constant position shift) the field is given by
\begin{equation}
\begin{split}
    A_\mu(t, \vec{r}) = \frac{e}{4 \pi} \frac{(1, -\vec{\beta}) }{\sqrt{ (\vec{r} - \vec{\beta} t)^2 + (\vec{r} \cdot \vec{\beta} )^2 - \beta^2 r^2 }}.
\end{split}
\end{equation}
This formula is derived in appendix \ref{app_LW} as equation \eqref{LW_tr}.

\begin{figure}
\input{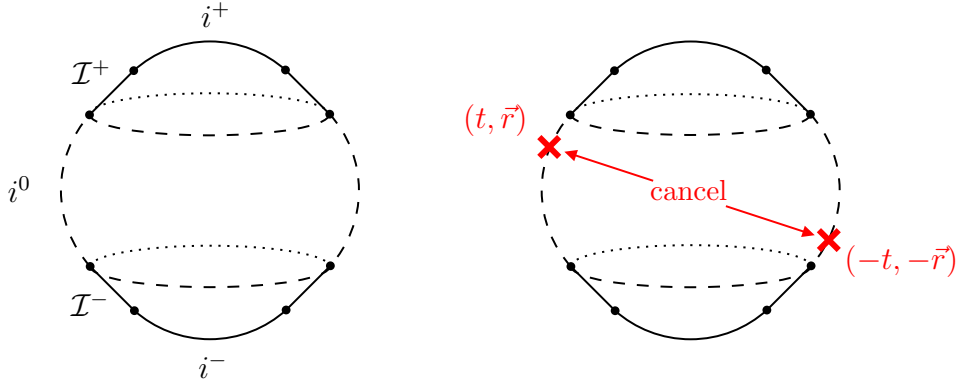}
\caption{\label{fig:antipodal_cancel} When we deform the Cauchy slice from $\mathcal{I}^+ \cup i^+$ to $\mathcal{I}^- \cup i^-$, we need  $\int_{i^0} d \Sigma_\mu J^\mu_\lambda$ to vanish if we want $Q^+_\lambda = Q^-_\lambda$. If $\lambda$ is an even function at spatial infinity, then the contribution to the flux integral at $(t, \vec{r})$ will be cancelled by the contribution at $(-t, - \vec{r})$. This is because when $(t, \vec{r}) \mapsto (-t, - \vec{r})$, we'll have $F_{\mu \nu} \mapsto - F_{\mu \nu}$, $\partial_\mu \lambda \mapsto -\partial_\mu\lambda $, $d\Sigma_\mu \mapsto - d \Sigma_\mu$.}
\end{figure}

We can see that the field has a very interesting property. Namely, satisfies
\begin{equation}
\begin{split}
    A_\mu(-t, -\vec{r}) = A_\mu(t, \vec{r}) \hspace{1 cm} \text{(at }i^0)
\end{split}
\end{equation}
meaning it is an \textit{even} function when the spacetime point is negated. This means that $F_{\mu \nu}$, which is made of derivatives of $A_\mu$, will be an \textit{odd} function when the spacetime point is negated.
\begin{equation}
\begin{split}
    F_{\mu, \nu}(-t, - \vec{r}) = -F_{\mu \nu}(t, \vec{r}) \hspace{1 cm} \text{(at }i^0)
\end{split}
\end{equation}
This means that the the integral in \eqref{i0_int} will vanish identically \textit{if} we specify that $\lambda$ must be an even function as well!
\begin{equation}\label{lambdaeven}
\begin{split}
    \lambda(-t, - \vec{r}) = \lambda(t, \vec{r}) \hspace{1 cm} \text{(at }i^0)
\end{split}
\end{equation}
By continuity, \eqref{lambdaeven} implies that $\lambda^+(z^A) = \lambda^-(-z^A)$ which is exactly the antipodal matching condition! Therefore, the antipodal matching condition is necessary to ensure that no net ``soft flux'' $J^\mu_\lambda$ leaves through spatial infinity.

\begin{figure}[H]
\begin{center}

\tikzset{every picture/.style={line width=0.75pt}} 

\begin{tikzpicture}[x=0.75pt,y=0.75pt,yscale=-1,xscale=1]

\draw  [dash pattern={on 4.5pt off 4.5pt}]  (10.2,92.8) -- (205.7,92.8) ;
\draw  [fill={rgb, 255:red, 0; green, 0; blue, 0 }  ,fill opacity=1 ] (105.15,92.8) .. controls (105.15,91.25) and (106.4,90) .. (107.95,90) .. controls (109.5,90) and (110.75,91.25) .. (110.75,92.8) .. controls (110.75,94.35) and (109.5,95.6) .. (107.95,95.6) .. controls (106.4,95.6) and (105.15,94.35) .. (105.15,92.8) -- cycle ;
\draw [line width=2.25]    (141.2,93) -- (174.6,93) ;
\draw [shift={(179.6,93)}, rotate = 180] [fill={rgb, 255:red, 0; green, 0; blue, 0 }  ][line width=0.08]  [draw opacity=0] (14.29,-6.86) -- (0,0) -- (14.29,6.86) -- cycle    ;
\draw [line width=2.25]    (76.2,93) -- (42.8,93) ;
\draw [shift={(37.8,93)}, rotate = 360] [fill={rgb, 255:red, 0; green, 0; blue, 0 }  ][line width=0.08]  [draw opacity=0] (14.29,-6.86) -- (0,0) -- (14.29,6.86) -- cycle    ;
\draw  [dash pattern={on 4.5pt off 4.5pt}]  (234.89,125.13) -- (419.81,61.67) ;
\draw  [fill={rgb, 255:red, 0; green, 0; blue, 0 }  ,fill opacity=1 ] (324.7,94.31) .. controls (324.2,92.85) and (324.98,91.25) .. (326.44,90.75) .. controls (327.9,90.25) and (329.5,91.03) .. (330,92.49) .. controls (330.5,93.95) and (329.72,95.55) .. (328.26,96.05) .. controls (326.8,96.55) and (325.2,95.77) .. (324.7,94.31) -- cycle ;
\draw [line width=2.25]    (358.86,82.8) -- (390.46,71.96) ;
\draw [shift={(395.19,70.33)}, rotate = 521.06] [fill={rgb, 255:red, 0; green, 0; blue, 0 }  ][line width=0.08]  [draw opacity=0] (14.29,-6.86) -- (0,0) -- (14.29,6.86) -- cycle    ;
\draw [line width=2.25]    (297.38,103.89) -- (265.79,114.74) ;
\draw [shift={(261.06,116.36)}, rotate = 341.06] [fill={rgb, 255:red, 0; green, 0; blue, 0 }  ][line width=0.08]  [draw opacity=0] (14.29,-6.86) -- (0,0) -- (14.29,6.86) -- cycle    ;

\draw (109.64,33.6) node    {$t=0$};
\draw (328.15,32.2) node    {$\gamma ( t-\beta x) =0$};

\end{tikzpicture}

\end{center}
\caption{\label{fig:boosted} The fact that $F(-t,-\vec{r}) = - F(t, \vec{r})$ at $i^0$ is a consequence of boost symmetry.}
\end{figure}
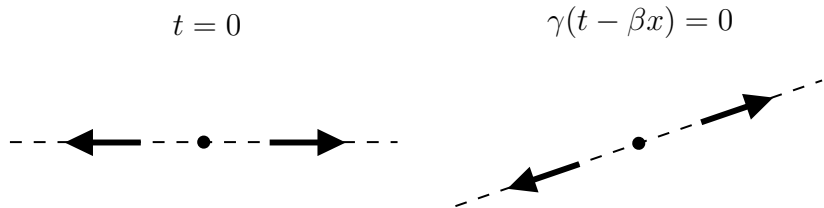

Let us now understand why the odd-ness of $F_{\mu \nu}$ is a consequence of Lorentz invariance. It is well known that the electric and magnetic fields generated by a non-accelerating moving point charge at the origin will negate when $\vec{r} \mapsto - \vec{r}$. If we then do a boost, as in Figure \ref{fig:boosted}, we see that fields must also be negated when $(t, \vec{r}) \mapsto (-t, -\vec{r})$, which is exactly what we wanted to show.\footnote{The connection between finite energy, even/oddness conditions at $i^0$, and soft charge conservation were discussed in \cite{Herdegen:2016bio, Herdegen:1995nf}, although the techniques and conclusions were different.}

\section{The Asymptotic Symmetry Group}

\noindent The general motto is that the asymptotic symmetry group $\mathcal{G}$ of a gauge field is \cite{ruzziconi2019asymptotic}
\begin{equation}\label{gdef}
    \mathcal{G} = \frac{\text{Gauge transformations preserving the boundary conditions}}{\text{Trivial gauge transformations}}.
\end{equation}
In order to make sense of the numerator, we must carefully impose boundary conditions on $A_\mu$ and $F_{\mu \nu}$ at infinity, which will be done in the next section. The denominator, ``trivial gauge transformations,'' refers to gauge transformations $\lambda$ that have compact support. One should think of $\mathcal{G}$, which is the group of large gauge transformations where $\lambda$ is not $0$ at infinity, as consisting of physical transformations of $A_\mu$. Later, after we study the symplectic form on our phase space of gauge fields, we will return to this concept and elaborate on this subtle point.

\subsection{Boundary conditions}

\subsubsection{Coulombic vs. radiating boundary conditions}

The boundary conditions we will impose at $\mathcal{I}^\pm$ and $i^0$ will include large $r$ fall-offs and antipodal matching conditions. However, how do we know which boundary conditions to choose a priori? Well, we don't. There is no completely general algorithm for determining what the ``correct'' falls-offs are for a field theory to satisfy. Some artistry is always involved. Having said that, there are a few commonly used criteria, such as demanding that energy, energy flux, canonical charges, and the symplectic form should all be finite. In this note we will take an even more pedestrian approach: we will just take the well-known expression for the electromagnetic field generated by a moving point charge, find its fall-offs, and then adopt those. The EM field generated by a finite number of charges moving along some arbitrary paths will also obey these fall-offs.

Actually, we must be a bit more subtle. As discussed in section \ref{sec_finite_energy}, if we require that our EM field holds a finite amount of energy then we must demand that at $t = \pm \infty$, the charged particles in our spacetime must have constant velocities. Therefore, we will actually need to specify two different sets of boundary conditions: one for early/late retarded times $u$, when the charged particles are travelling at constant velocities and generate Coulombic fields, and another for generic $u$ when they can accelerate and generate radiative fields.

\subsubsection{Field strength fall-offs}

We begin by finding the large $r$ field strength fall-offs. To do this, we will use the exact expression for the electric and magnetic fields generated by a moving point charge which is given in most EM textbooks with varying notations. We will use the expression as written in \cite{Griffiths:1492149}, but equivalent expressions can for instance be found in \cite{Jackson:100964, Feynman:1494701}.

The electric and magnetic fields at a point $(t,\vec{r})$ will only depend on the position, velocity, and acceleration of the charged particle on the past lightcone of $(t, \vec{r})$.  Following \cite{Griffiths:1492149}, let us define $\vec{\rcurs}$ as the distance vector from $\vec{r}$ to the retarded position $\vec{r}_{\rm ret}$ of the particle on the past lightcone of $(t, \vec{r})$.
\begin{equation}
    \vec{\rcurs} \equiv \vec{r} - \vec{r}_{\rm ret}
\end{equation}
Furthermore, denote $\rcurs = |\vec{\rcurs}|$ and $\hat{\rcurs} = \vec{\rcurs}/\rcurs$. For convenience, let us also define
\begin{equation}
    \vec{s} \equiv \hat{\rcurs} - \vec{\beta}
\end{equation}
where $\vec{\beta}$ is the velocity of the charged particle at the retarded position.

The electric and magnetic fields created by the point charge are
\begin{align}\label{EB}
    \vec{E}(t, \vec{r}) &= \frac{e}{4 \pi} \frac{\rcurs}{(\vec{\rcurs} \cdot \vec{s})^3 } \left( (1 - \beta^2) \vec{s} + \vec{\rcurs} \times (\vec{s} \times \vec{a}) \right) \\
    \vec{B}(t, \vec{r}) &= \hat{\rcurs} \times \vec{E}(t, \vec{r}) \label{EB2}
\end{align}
where $\vec{a}$ is the acceleration of the particle (\textit{not} the proper acceleration) at the retarded position.

If we parameterize $(t, \vec{r})$ with $(u, r, z^A)$ as
\begin{equation}
\begin{split}
    x^\mu = ( t , \vec{r} ) = ( u + r , r\, \hat n( z^A) )
\end{split}
\end{equation}
we'll have in the fixed $u$ large $r$ limit
\begin{equation}\label{rcurs_r}
    \vec{\rcurs} = r \hat n + \mathcal{O}(1)
\end{equation}
where $\mathcal{O}(1)$ is a constant vector for large $r$. (It is a small exercise to show \eqref{rcurs_r} holds explicitly, for instance, on the path of a particle travelling with constant velocity). One can use \eqref{rcurs_r} to extract the fall-offs of the electric and magnetic fields in both the radiative $\vec{a} \neq 0$ and Coulombic $\vec{a} = 0$ cases.

In particular, we are interested in finding the fall-offs for the components of $\vec{E}$ and $\vec{B}$ which point either radially outward from or tangent to the large sphere at infinity. We denote $E_r = \hat n \cdot \vec{E}$ and $E_T = \hat m \cdot \vec{E}$ where $\hat m$ is some unit vector which is orthogonal to $\hat n$. A small bit of work reveals the fall-offs when $\vec{a} \neq 0$ are
\begin{equation}\label{EB_coulomb}
\begin{split}
    E_r \sim \mathcal{O}(1/r^2)
    \hspace{0.75 cm}
    E_T \sim \mathcal{O}(1/r)
    \hspace{0.75 cm}
    B_r \sim \mathcal{O}(1/r^2)
    \hspace{0.75 cm}
    B_T \sim \mathcal{O}(1/r)
\end{split}
\end{equation}
and the fall-offs when $\vec{a} = 0$ are
\begin{equation}\label{EB_rad}
\begin{split}
    E_r \sim \mathcal{O}(1/r^2) \hspace{0.75 cm}
    E_T \sim \mathcal{O}(1/r^2)
    \hspace{0.75 cm}
    B_r \sim \mathcal{O}(1/r^3)
    \hspace{0.75 cm}
    B_T \sim \mathcal{O}(1/r^2).
\end{split}
\end{equation}

Let's now find the fall-offs for $F_{\mu \nu}$ with $(u,r,z^A)$ indices. Under a change of coordinates $x^\mu \mapsto y^\nu(x)$, a covariant vector $V_\mu$ with lowered indices transforms as

\begin{equation}
    V_\nu = \frac{\partial x^{\mu}}{\partial y^{\nu}} V_{\mu}.
\end{equation}

\noindent If we denote the spatial coordinates of $\vec{r}$ with $i = 1,2,3$, with $V_\mu = (V_0, V_i)$ in rectilinear coordinates, then the components in $(u,r,z^A)$ coordinates are
\begin{align} \label{covariant_1}
    V_u &= V_0 \\
    V_r &= V_0 + \hat n^i V_i \\
    V_A &= (\partial_A \hat n^i) V_i.  \label{covariant_3}
\end{align}
Note that $\partial_A \hat n$ orthogonal to $\hat n$, which can be seen by differentiating $\hat n \cdot \hat n = 1$. We can therefore take $\hat m \propto \partial_A \hat n$ and see that

\begin{equation}
    \begin{matrix*}[l]
        F_{ur} &=& \hat{n}^i F_{0i} &\propto E_r \\
        F_{uA} &=& r (\partial_A \hat{n}^i ) F_{0i} &\propto r E_T \\
        F_{Ar} &=& r (\partial_A \hat{n}^i)( F_{i0} + \hat n^j F_{ij} ) &\propto r (E_T + B_T) \label{FAr}\\
        F_{AB} &=& r^2 (\partial_A \hat{n}^i) (\partial_B \hat{n}^j ) F_{ij} & \propto r^2 B_r.
    \end{matrix*}
\end{equation}
We are almost ready to use \eqref{EB_coulomb} and \eqref{EB_rad} to get the $F_{\mu \nu}$ fall-offs. However, a small subtlety arises with $F_{Ar}$ as cancellations will occur between $E_T$ and $B_T$ as written. From $\nabla^\mu F_{\mu r} = 0$ we have $D^A F_{Ar} = -\partial_r(r^2 F_{ur})$ which implies $F_{Ar} \sim \mathcal{O}(1/r^2)$.


Using this, along with \eqref{EB_coulomb}, we have that the radiative (generic $u$) fall-offs are
\begin{equation}
\begin{split}
    F_{ur} \sim \mathcal{O}(1/r^2)  \hspace{0.75 cm} F_{uA} \sim \mathcal{O}(1) \hspace{0.75 cm} F_{Ar} \sim \mathcal{O}(1/r^2) \hspace{0.75 cm} F_{AB} \sim \mathcal{O}(1).
\end{split}
\end{equation}
Using \eqref{EB_rad}, the Coulombic ($u = \pm \infty$) fall-offs are
\begin{equation}
\begin{split}
    \hspace{1 cm} F_{ur} \sim \mathcal{O}(1/r^2)  \hspace{0.75 cm} F_{uA} \sim \mathcal{O}(1/r) \hspace{0.75 cm} F_{Ar} \sim \mathcal{O}(1/r^2) \hspace{0.75 cm} F_{AB} \sim \mathcal{O}(1/r).
\end{split}
\end{equation}
One can verify that $T_{uu} \sim \mathcal{O}(1/r^2)$ for the radiative fall-offs and $T_{uu} \sim \mathcal{O}(1/r^4)$ for the Coulombic fall-offs.


\subsubsection{Liénard–Wiechert fall-offs, $A_A$ is pure gauge at $\mathcal{I}^+_\pm$}

The retarded Liénard–Wiechert gauge field of a moving point charge in Lorenz gauge $\partial_\mu A^\mu = 0$ is
\begin{equation}
    A_\mu(x) = \frac{e}{4 \pi} \frac{(\dot x_{\rm ret})_\mu}{\dot x_{\rm ret} \cdot (x - x_{\rm ret})}
\end{equation}

\noindent where $x_{\rm ret}$ is the unique point on the particle's path in the past light cone of $x$, defined by
\begin{align}
    &x^0 > x_{\rm ret}^0 \\
    &(x - x_{\rm ret})^2 = 0
\end{align}
and $\dot x_{\rm ret}$ is the four-velocity of the particle at position $x_{\rm ret}$. If we denote
\begin{equation}
    \dot x_{\rm ret}^\mu = \gamma ( 1, \vec{\beta} ) \hspace{1cm}(\dot x_{\rm ret})_\mu = \gamma (1 , -\vec{\beta} )
\end{equation}
then
\begin{equation}\label{Aret_1}
    A_\mu(x) = \frac{e}{4 \pi} \frac{(1,- \vec{\beta})}{(1, \vec{\beta}) \cdot  (x - x_{\rm ret})}.
\end{equation}
If we use retarded coordinates $u = t -r$ and set $\vec{x} = r \hat n$ for unit vector $\hat n$,
\begin{equation}
    x^\mu = (r + u, r \hat n )
\end{equation}
then
\begin{align}
    (1, \vec{\beta}) \cdot (x - x_{\rm ret}) &= r+u-(x_{\rm ret})^0 - r \vec{\beta} \cdot \hat n + \vec{\beta} \cdot \vec{x}_{\rm ret} \\
    &= r(1 - \vec{\beta} \cdot \hat n) + \mathcal{O}(1). \label{Aret_2}
\end{align}
Note that $x_{\rm ret}$ and $\vec{\beta}$ secretly depend on $(u,r,z^A)$. If we fix $u, \hat n$ and take $r$ large, then these quantities approach $(u, z^A)$-dependent $\mathcal{O}(1)$ numbers as the value of $x_{\rm ret}$ zeroes-in on a single point on the particle's path. 

Using \eqref{Aret_1}, \eqref{Aret_2}, and \eqref{covariant_1}--\eqref{covariant_3}, we can see that
\begin{align}
    A_u &= \frac{e}{4 \pi r} \frac{1}{1 - \vec{\beta} \cdot \hat n} + \mathcal{O}(1/r^2) \\
    A_r &= \frac{e}{4 \pi r} + \mathcal{O}(1/r^2) \label{Ar} \\
    A_A &= -\frac{e}{4 \pi} \frac{ \partial_A \hat n \cdot \vec{\beta}}{1 - \vec{\beta} \cdot \hat n} + \mathcal{O}(1/r) \label{AA_vel0}
\end{align}
When our particle is travelling at a constant velocity, the $A_A$ component holds a surprise. $\vec{\beta}$ will be a constant, which allows us to write
\begin{equation}\label{AA_orig}
    A_A = \frac{e}{4 \pi} \partial_A \log(1 - \vec{\beta} \cdot \hat n) + \mathcal{O}(1/r) .
\end{equation}
Amazingly, $A_A$ is ``pure gauge,'' i.e. a total derivative! Inspired by this observation, we will adopt the requirement that $A_A$ must be pure gauge at $u = \pm \infty$ into our boundary conditions.

It should be noted our expression for $A_r$ disagrees with \cite{strominger2018lectures} which gives the fall-off $A_r \sim \mathcal{O}(1/r^2)$. \cite{strominger2018lectures} deduced its $A_r$ fall-off from $F_{Ar} \sim \mathcal{O}(1/r^2)$ which implies $\partial_A A_r \sim \mathcal{O}(1/r^2)$. While it was assumed this meant $A_r \sim \mathcal{O}(1/r^2)$, the possibility that the leading $1/r$ piece of $A_r$ was angle-independent, as is the case in \eqref{Ar}, was overlooked.

\subsubsection{Matching at spatial infinity}

In section \ref{antipodal_reason}, we noted that Coulombic field strengths satisfy
\begin{equation}
    F_{\mu \nu}(-t, - \vec{r}) = - F_{\mu \nu}(t, \vec{r}) \hspace{1 cm} \text{(at }i^0).
\end{equation}
We will choose to adopt this relation into our set of boundary conditions.

\subsubsection{Summary of boundary conditions}

The radiative generic $u$ field strength fall-offs are
\begin{align} \label{falloff2}
    F_{ur} \sim O(1/r^2) \hspace{0.5 cm} F_{uA} \sim O(1) \hspace{0.5 cm} F_{Ar} \sim O(1/r^2) \hspace{0.5 cm} F_{AB} \sim O(1).
\end{align}
For $u = \pm \infty$, we have the more restrictive Coulombic fall-offs
\begin{align}
    F_{ur} \sim O(1/r^2) \hspace{0.5 cm} F_{uA} \sim O(1/r) \hspace{0.5 cm} F_{Ar} \sim O(1/r^2) \hspace{0.5 cm} F_{AB} \sim O(1/r).
\end{align}
We also require that $F_{\mu \nu}$ is odd at spatial infinity.
\begin{equation}
    F_{\mu \nu}(-t, - \vec{r}) = - F_{\mu \nu}(t, \vec{r}) \hspace{1 cm} \text{(at }i^0).
\end{equation}
The $A_\mu$ fall-offs are
\begin{equation}\label{bdy_A_rad}
    A_u \sim O(1/r) \hspace{0.5 cm} A_r \sim O(1/r) \hspace{0.5 cm} A_A \sim O(1).
\end{equation}
As a consequence of the finite energy constraint, we also require that $A_A^{(0)}$ is pure gauge at $u = \pm \infty$, so
\begin{equation}
    A_A^{(0)}(u = \pm \infty)= \partial_A \phi_\pm
\end{equation}
for some functions $\phi_\pm = \phi_\pm(z^A)$. All of the equations in this section have exact analogs on $\mathcal{I}^-$ in advanced coordinates.

\subsection{Non-trivial gauge transformations}

$A_\mu$, of course, has the gauge symmetry
\begin{equation}
    \delta A_u = \partial_u \lambda, \hspace{0.75 cm} \delta A_r = \partial_r \lambda, \hspace{0.75 cm} \delta A_A = \partial_A \lambda.
\end{equation}
In order to write down $\mathcal{G}$ via \eqref{gdef}, we must find what conditions $\lambda$ must satisfy to preserve the boundary conditions given in the previous section. If we expand $\lambda$ around $\mathcal{I}^+$ and $\mathcal{I}^-$ as
\begin{align}
    \lambda(u, r, z^A) = \lambda^+(u, z^A) + \frac{\lambda^+_{(1)}(u, z^A)}{r} + \ldots \hspace{0.5 cm} \text{near }\mathcal{I}^+ \\
    \lambda(v,r, z^A) = \lambda^-(u, z^A) + \frac{\lambda^-_{(1)}(v, z^A)}{r} + \ldots \hspace{0.5 cm} \text{near }\mathcal{I}^-
\end{align}
then we can see that in order to preserve \eqref{bdy_A_rad} we crucially must have the that leading piece in $r$ is $u$-independent.
\begin{align}
    \lambda^+ &= \lambda^+(z^A) \\
    \lambda^- &= \lambda^-(z^A)
\end{align}

There is one other condition we might as well impose on $\lambda$. In section \ref{antipodal_reason} we observed that if we demand that
\begin{equation}
    \lambda^+(z^A) = \lambda^-(-z^A).
\end{equation}
then $Q^+_\lambda = Q^-_\lambda$ and soft charge will be conserved.

We have now finished characterizing the group of asymptotic symmetries of electromagnetism:

\begin{equation}\label{G_final}
    \mathcal{G} = \left\{ \begin{matrix} \lambda\text{'s for which }\lambda^+\text{ and }\lambda^-\text{ are  }u\text{ and }v\text{ independent at } \mathcal{I}^+\\ \text{ and }\mathcal{I}^-\text{ and are antipodally matched }\lambda^+(z^A) = \lambda^-(-z^A) \end{matrix} \right\}.
\end{equation}

\newpage
\section{The Utility of Lorenz Gauge}\label{sec_lorenz}

\subsection{Residual gauge transformations}\label{sec_resid}

We physicists like for the data on one Cauchy slice to uniquely determine the data on all other Cauchy slices. In gauge theories, however, the ability to perform gauge transformations spoils the uniqueness of time evolution. Therefore, it is desirable to enforce a partial gauge fixing which 
\begin{enumerate}
    \item Ensures that time evolution of $A_\mu$ is unique
    \item Faithfully preserves $\mathcal{G}$ in allowing for large transformations to remain as residual gauge transformations, but fixes away the ability to perform small gauge transformations.
    \begin{equation*}
        \mathcal{G} = \frac{\text{Allowed  gauge trans.}}{\text{Trivial gauge trans.}} \xrightarrow{\text{partial gauge fixing}} \frac{\text{Allowed residual  gauge trans.}}{\text{\st{Trivial gauge trans.}}} = \mathcal{G}
    \end{equation*}
\end{enumerate}
One nice partial gauge fixing which satisfies these two properties is Lorenz gauge,
\begin{equation}
    \partial_\mu A^\mu = 0.
\end{equation}
This has the residual gauge transformation $A_\mu \mapsto A_\mu + \partial_\mu \lambda$ for $\lambda$'s which solve
\begin{equation}
    \partial^2 \lambda = 0.
\end{equation}
Let's study these residual $\lambda$'s and see how they behave at null infinity. We will happily find that they satisfy the antipodal matching condition all on their own, confirming that Lorenz gauge is a convenient partial gauge fixing for us to use.

First let's look at the most protypical example of a solution to the wave equation: the plane wave.
\begin{equation}
    \lambda = e^{-ikx}
\end{equation}
We are interested in what happens to this plane wave at both $\mathcal{I}^+$ and $\mathcal{I}^-$. Let's therefore set $k^\mu = (|\vec{k}|, \vec{k})$ and write $x^\mu$ in both retarded and advanced coordinates as
\begin{align}
    x^\mu &= (u + r , r \hat n) \\
    &= (v - r, r \hat n ).
\end{align}
We then write $\lambda$ in both sets of coordinates as
\begin{align} \label{lambdau}
    \lambda &= e^{- i |\vec{k}| u } e^{i r(|\vec{k}| -  \hat n \cdot \vec{k})} \\
    &= e^{- i |\vec{k}| v } e^{i r(-|\vec{k}| -  \hat n \cdot \vec{k})}. \label{lambdav}
\end{align}
$\mathcal{I}^+$ is reached by holding $u, \hat n$ fixed and sending $r \to \infty$, while $\mathcal{I}^-$ is reached by holding $v, \hat{n}$ fixed and sending $r \to \infty$.
\begin{figure}[H]
\begin{center}

\tikzset{every picture/.style={line width=0.75pt}} 

\begin{tikzpicture}[x=0.75pt,y=0.75pt,yscale=-1,xscale=1]

\draw    (174.33,224.28) -- (266.25,132.36) ;
\draw    (125.08,224.28) -- (32.5,131.69) ;
\draw  [draw opacity=0] (266.75,131.15) .. controls (266.75,131.18) and (266.75,131.22) .. (266.75,131.25) .. controls (266.75,137.46) and (214.3,142.5) .. (149.61,142.5) .. controls (85.91,142.5) and (34.08,137.62) .. (32.5,131.54) -- (149.61,131.25) -- cycle ; \draw   (266.75,131.15) .. controls (266.75,131.18) and (266.75,131.22) .. (266.75,131.25) .. controls (266.75,137.46) and (214.3,142.5) .. (149.61,142.5) .. controls (85.91,142.5) and (34.08,137.62) .. (32.5,131.54) ;
\draw  [draw opacity=0][dash pattern={on 4.5pt off 4.5pt}] (32.66,131.96) .. controls (32.66,131.93) and (32.65,131.89) .. (32.65,131.86) .. controls (32.65,126.69) and (85.28,122.5) .. (150.2,122.5) .. controls (213.92,122.5) and (265.79,126.54) .. (267.7,131.57) -- (150.2,131.86) -- cycle ; \draw  [dash pattern={on 4.5pt off 4.5pt}] (32.66,131.96) .. controls (32.66,131.93) and (32.65,131.89) .. (32.65,131.86) .. controls (32.65,126.69) and (85.28,122.5) .. (150.2,122.5) .. controls (213.92,122.5) and (265.79,126.54) .. (267.7,131.57) ;
\draw   (124.08,27.92) .. controls (124.08,26.67) and (135.11,25.67) .. (148.71,25.67) .. controls (162.31,25.67) and (173.33,26.67) .. (173.33,27.92) .. controls (173.33,29.16) and (162.31,30.17) .. (148.71,30.17) .. controls (135.11,30.17) and (124.08,29.16) .. (124.08,27.92) -- cycle ;
\draw  [draw opacity=0] (174.56,223.63) .. controls (173.96,226.14) and (163.1,228.15) .. (149.8,228.15) .. controls (136.64,228.15) and (125.88,226.19) .. (125.07,223.71) -- (149.8,223.41) -- cycle ; \draw   (174.56,223.63) .. controls (173.96,226.14) and (163.1,228.15) .. (149.8,228.15) .. controls (136.64,228.15) and (125.88,226.19) .. (125.07,223.71) ;
\draw  [draw opacity=0][dash pattern={on 4.5pt off 4.5pt}] (124.84,224.36) .. controls (125.45,221.84) and (136.3,219.84) .. (149.6,219.84) .. controls (162.76,219.84) and (173.52,221.8) .. (174.33,224.28) -- (149.6,224.58) -- cycle ; \draw  [dash pattern={on 4.5pt off 4.5pt}] (124.84,224.36) .. controls (125.45,221.84) and (136.3,219.84) .. (149.6,219.84) .. controls (162.76,219.84) and (173.52,221.8) .. (174.33,224.28) ;
\draw    (173.33,28.22) -- (265.25,120.14) ;
\draw    (124.08,28.22) -- (31.5,120.81) ;
\draw  [draw opacity=0][dash pattern={on 4.5pt off 4.5pt}] (265.75,121.35) .. controls (265.75,121.32) and (265.75,121.28) .. (265.75,121.25) .. controls (265.75,115.04) and (213.3,110) .. (148.61,110) .. controls (84.91,110) and (33.08,114.88) .. (31.5,120.96) -- (148.61,121.25) -- cycle ; \draw  [dash pattern={on 4.5pt off 4.5pt}] (265.75,121.35) .. controls (265.75,121.32) and (265.75,121.28) .. (265.75,121.25) .. controls (265.75,115.04) and (213.3,110) .. (148.61,110) .. controls (84.91,110) and (33.08,114.88) .. (31.5,120.96) ;
\draw  [draw opacity=0] (31.66,120.54) .. controls (31.66,120.57) and (31.65,120.61) .. (31.65,120.64) .. controls (31.65,125.81) and (84.28,130) .. (149.2,130) .. controls (212.92,130) and (264.79,125.96) .. (266.7,120.93) -- (149.2,120.64) -- cycle ; \draw   (31.66,120.54) .. controls (31.66,120.57) and (31.65,120.61) .. (31.65,120.64) .. controls (31.65,125.81) and (84.28,130) .. (149.2,130) .. controls (212.92,130) and (264.79,125.96) .. (266.7,120.93) ;
\draw    (495.33,224.28) -- (587.25,132.36) ;
\draw    (446.08,224.28) -- (353.5,131.69) ;
\draw  [draw opacity=0] (587.75,131.15) .. controls (587.75,131.18) and (587.75,131.22) .. (587.75,131.25) .. controls (587.75,137.46) and (535.3,142.5) .. (470.61,142.5) .. controls (406.91,142.5) and (355.08,137.62) .. (353.5,131.54) -- (470.61,131.25) -- cycle ; \draw   (587.75,131.15) .. controls (587.75,131.18) and (587.75,131.22) .. (587.75,131.25) .. controls (587.75,137.46) and (535.3,142.5) .. (470.61,142.5) .. controls (406.91,142.5) and (355.08,137.62) .. (353.5,131.54) ;
\draw  [draw opacity=0][dash pattern={on 4.5pt off 4.5pt}] (353.66,131.96) .. controls (353.66,131.93) and (353.65,131.89) .. (353.65,131.86) .. controls (353.65,126.69) and (406.28,122.5) .. (471.2,122.5) .. controls (534.92,122.5) and (586.79,126.54) .. (588.7,131.57) -- (471.2,131.86) -- cycle ; \draw  [dash pattern={on 4.5pt off 4.5pt}] (353.66,131.96) .. controls (353.66,131.93) and (353.65,131.89) .. (353.65,131.86) .. controls (353.65,126.69) and (406.28,122.5) .. (471.2,122.5) .. controls (534.92,122.5) and (586.79,126.54) .. (588.7,131.57) ;
\draw   (445.08,27.92) .. controls (445.08,26.67) and (456.11,25.67) .. (469.71,25.67) .. controls (483.31,25.67) and (494.33,26.67) .. (494.33,27.92) .. controls (494.33,29.16) and (483.31,30.17) .. (469.71,30.17) .. controls (456.11,30.17) and (445.08,29.16) .. (445.08,27.92) -- cycle ;
\draw  [draw opacity=0] (495.56,223.63) .. controls (494.96,226.14) and (484.1,228.15) .. (470.8,228.15) .. controls (457.64,228.15) and (446.88,226.19) .. (446.07,223.71) -- (470.8,223.41) -- cycle ; \draw   (495.56,223.63) .. controls (494.96,226.14) and (484.1,228.15) .. (470.8,228.15) .. controls (457.64,228.15) and (446.88,226.19) .. (446.07,223.71) ;
\draw  [draw opacity=0][dash pattern={on 4.5pt off 4.5pt}] (445.84,224.36) .. controls (446.45,221.84) and (457.3,219.84) .. (470.6,219.84) .. controls (483.76,219.84) and (494.52,221.8) .. (495.33,224.28) -- (470.6,224.58) -- cycle ; \draw  [dash pattern={on 4.5pt off 4.5pt}] (445.84,224.36) .. controls (446.45,221.84) and (457.3,219.84) .. (470.6,219.84) .. controls (483.76,219.84) and (494.52,221.8) .. (495.33,224.28) ;
\draw    (494.33,28.22) -- (586.25,120.14) ;
\draw    (445.08,28.22) -- (352.5,120.81) ;
\draw  [draw opacity=0][dash pattern={on 4.5pt off 4.5pt}] (586.75,121.35) .. controls (586.75,121.32) and (586.75,121.28) .. (586.75,121.25) .. controls (586.75,115.04) and (534.3,110) .. (469.61,110) .. controls (405.91,110) and (354.08,114.88) .. (352.5,120.96) -- (469.61,121.25) -- cycle ; \draw  [dash pattern={on 4.5pt off 4.5pt}] (586.75,121.35) .. controls (586.75,121.32) and (586.75,121.28) .. (586.75,121.25) .. controls (586.75,115.04) and (534.3,110) .. (469.61,110) .. controls (405.91,110) and (354.08,114.88) .. (352.5,120.96) ;
\draw  [draw opacity=0] (352.66,120.54) .. controls (352.66,120.57) and (352.65,120.61) .. (352.65,120.64) .. controls (352.65,125.81) and (405.28,130) .. (470.2,130) .. controls (533.92,130) and (585.79,125.96) .. (587.7,120.93) -- (470.2,120.64) -- cycle ; \draw   (352.66,120.54) .. controls (352.66,120.57) and (352.65,120.61) .. (352.65,120.64) .. controls (352.65,125.81) and (405.28,130) .. (470.2,130) .. controls (533.92,130) and (585.79,125.96) .. (587.7,120.93) ;
\draw    (469.71,222.8) -- (469.71,30.17) ;
\draw [line width=1.5]    (385.53,163.2) -- (506.93,41.8) ;

\draw (221.29,70.78) node [anchor=south west] [inner sep=0.75pt]    {$\lambda ^{+}\left( u,z^{A}\right)$};
\draw (76.79,181.38) node [anchor=north east] [inner sep=0.75pt]    {$\lambda ^{-}\left( v,z^{A}\right)$};
\draw (494.69,68.21) node  [font=\large,rotate=-315]  {$u=v_{0}$};
\draw (452.84,211) node [anchor=north west][inner sep=0.75pt]  [rotate=-270]  {$r=0$};
\draw (416.02,148.21) node  [font=\large,rotate=-315]  {$v=v_{0}$};
\draw (383.53,166.6) node [anchor=north east] [inner sep=0.75pt]  [font=\normalsize]  {$\hat{n} =-\hat{k}$};
\draw (508.93,38.4) node [anchor=south west] [inner sep=0.75pt]  [font=\normalsize]  {$\hat{n} =\hat{k}$};

\end{tikzpicture}

\end{center}
\caption{\label{fig:lambda_bdy} On the left: $\lambda^+$ ($\lambda^-$) is $\lambda$ when restricted to $\mathcal{I}^+$ ($\mathcal{I}^-$). On the right: the line $v = v_0$, $\hat n = - \hat k$ when continued through $r = 0$ becomes the line $u = v_0, \hat n = \hat k$. The plane wave $e^{-ikx}$ will be constant along this line.}
\end{figure}
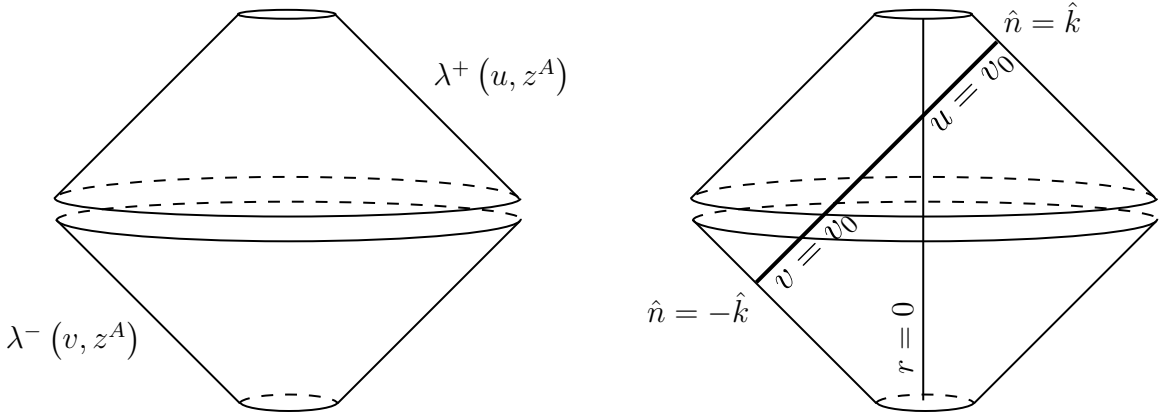
Now, notice that the RHS of \eqref{lambdau} equals $e^{-i|\vec{k}| u}$ if $\hat n = \hat k \equiv \vec{k}/|\vec{k}|$. However, if $\hat n$ points in any other direction, the function will rapidly oscillate as $\hat n$ is varied in the large $r$ limit. This means that, in a distributional sense, the function must be 0 when $\hat n \neq \hat k$. We can express this as

\begin{equation}\label{planeIp}
    \eval{e^{-ikx}}_{\mathcal{I}^+} = \begin{cases} e^{-ik_0 u } & \text{ if } \hat n = \hat k\\
    0 & \text{ for any other angle}.
    \end{cases}
\end{equation}
\vskip .2cm
\noindent We have an analogous equation on $\mathcal{I}^-$, except it is only non-zero when $\hat n = - \hat k$.

\begin{equation}\label{planeIm}
    \eval{e^{-ikx}}_{\mathcal{I}^-} = \begin{cases} e^{-ik_0 v } & \text{ if } \hat n = -\hat k\\
    0 & \text{ for any other angle}.
    \end{cases}
\end{equation}
\vskip .2cm
\noindent Furthermore, it can be argued that the functions \eqref{planeIp} and \eqref{planeIm} are antipodally matched. If one draws a line of $v = v_0$ constant and $\hat n = -\hat k$, the plane wave will be constant along this whole line. At $r = 0$, one must switch to retarded coordinates, where the line becomes $u = v_0$ and $\hat n = \hat k$. This is drawn in Figure \ref{fig:lambda_bdy}. Because the function must be constant along this line, we can see that
\begin{equation}
    \lambda^-(v_0, z^A) = \lambda^+(v_0, - z^A).
\end{equation}
What's more, any $\lambda$ which is a linear combination of plane waves will also satisfy the above equation. It appears the residual gauge transformations in Lorenz gauge ``know'' about the antipodal map. This makes Lorenz gauge a natural partial gauge fixing to use.

However, there is something concerning about these plane waves. At null infinity, they depend on $u$ and $v$. However, the gauge transformations allowed in $\mathcal{G}$ do not, per \eqref{G_final}. We can make them $u$-independent if we take $|\vec{k}| \to 0$, but then all the plane waves just become constant functions! 

Nonetheless, we could still use these $|\vec{k}| \to 0$ plane waves to construct other non-constant $z^A$-dependent but $u$-independent functions at $\mathcal{I}^+$ if we create linear combinations of them, adding waves with $\hat k$ pointing in multiple directions, and taking the low frequency limit while keeping the ratios of the Fourier components of different $\hat k$'s fixed.

Having said that, such functions are strictly speaking not linear combinations of plane waves (even though they are limits of linear combinations of plane waves). Sadly, not every function actually has a Fourier transform. In particular, solutions to the wave equation which do have Fourier transforms cannot be $u$ and $v$ independent at $\mathcal{I}^\pm$. While we must now abandon the plane waves, they have nonetheless provided us with some helpful intuition.

We now explain how to construct acceptable $\lambda$'s by means of a Green's function. In particular, this Green's function will solve the homogeneous wave equation and localize to an delta function on the 2-sphere at null infinity. It is given by
\begin{equation}
    G(x, \hat q) = \frac{1}{4 \pi} \frac{x^2}{(q \cdot x + i \eps)^2}
\end{equation}
where $q^\mu = (1, \hat q)$ is a null vector. It is straightforward to show that it solves the wave equation.
\begin{align}
    \partial^2 G(x, \hat q) = \frac{1}{4 \pi} \left( \frac{8}{(q \cdot x + i \eps)^2} - \frac{4 q \cdot x}{(q \cdot x + i \eps)^3} - \frac{4 q \cdot x}{(q \cdot x + i \eps)^3} + \frac{6 q^2 x^2}{(q \cdot x + i \eps)^4} \right) = 0
\end{align}
We must now show it becomes a delta function on the boundary. More specifically, if we parameterize $x^\mu = (u + r, r \hat n(z^A) )$, then we want to show
\begin{equation}
    \lim_{r \to \infty} G(x, \hat{q}) = \frac{1}{\sqrt{\gamma}} \delta^2( z^A - q^A)
\end{equation}
where we define $q^A$ by $\hat q = \hat n(q^A)$. We can do this by writing $G(x, \hat q)$ explicitly with $x = (u + r, r \hat n)$ as
\begin{equation}
    G(x, \hat q) = \frac{1}{4 \pi} \frac{u (u + 2r)}{(u + r(1 - \hat q \cdot \hat n) +i \eps)^2}.
\end{equation}
\vskip .2cm
\noindent
For large $r$, this vanishes unless $\hat q = \hat n$, at which point it blows up. We can confirm that it has the right normalization constant by integrating over $\hat n$. We do this in polar coordinates where $\sqrt{\gamma} d^2 z = \sin \theta d \theta d \phi$ and $\hat q \cdot \hat n = \cos \theta$:
\begin{align}
    &\lim_{r \to \infty} \int_{S^2}  d^2 z \sqrt{\gamma} \; G(x, \hat q) \\
    &= \lim_{r \to \infty} \int_0^{2 \pi } d \phi \int_{-1}^1 d (\cos \theta) \frac{1}{4 \pi} \frac{u (u + 2r)}{(u + r(1 - \cos \theta) +i \eps)^2} \\
    &= 1.
\end{align}
The $+i \epsilon$ is there to avoid the pole in the integrand. We could have just as well used $-i \eps$ and $G(x,\hat q)$ would not have changed overall.

If one wants to solve for the residual gauge transformation $\lambda(x)$ in the bulk, one can simply convolve $\lambda^+(z^A)$ with the Green's function over $\hat q$ as
\begin{equation}
\begin{split}
    \lambda(x) = \int_{S^2} d^2 z \sqrt{\gamma} \; \lambda^+(z^A) G\big(x, \hat q(z^A) \big) \; .
\end{split}
\end{equation}
Note that our Green's function manifestly satisfies the antipodal matching condition because
\begin{equation}
\begin{split}
    G(-x, \hat q) = G(x, \hat q).
\end{split}
\end{equation}

\subsection{The soft charge of massive particles at $i^\pm$}

Up until now we have pretended that $\mathcal{I}^+$ is a Cauchy slice. However, this supposed Cauchy slice has an infinitely large hole right in the middle of it: timelike infinity.

In section \ref{cons_rad} we showed that conservation of soft charge $Q_\lambda$ requires that radiation be released if charged particles have unequal initial and final velocities. However, because we didn't want to worry about integrating $J^\mu_\lambda$ over timelike infinity, we assumed that our charged particles were massless and hit null infinity instead of timelike infinity. Now that we are acquainted with Lorenz gauge, we have the tools necessary to remedy this oversight \cite{Campiglia:2015qka}. Recalling that we can write the soft charge as the sum of a ``soft'' and ``hard'' piece
\begin{align}
    Q_\lambda &= Q_\lambda^S + Q_\lambda^H \\
    &=\int_\Sigma d \Sigma_\mu (- F^{\mu \nu} \nabla_\nu \lambda) + \int_\Sigma d \Sigma_\mu \lambda j^\mu
\end{align}
we will show that the integral of the soft piece over $i^\pm$ is 0 and the integral of the hard piece over $i^\pm$ will only depend on the initial/final velocities of the particles in a very precise way.

Let's first try to understand how timelike infinity is defined and how it attaches to $\mathcal{I}^+_+$. 
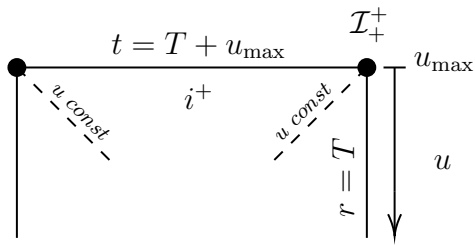
\begin{figure}[h]
\begin{center}

\tikzset{every picture/.style={line width=0.75pt}} 

\begin{tikzpicture}[x=0.75pt,y=0.75pt,yscale=-1,xscale=1]

\draw    (92,63) -- (92,149) ;
\draw    (92,63) -- (268.5,63) ;
\draw  [fill={rgb, 255:red, 0; green, 0; blue, 0 }  ,fill opacity=1 ] (87.75,63) .. controls (87.75,60.65) and (89.65,58.75) .. (92,58.75) .. controls (94.35,58.75) and (96.25,60.65) .. (96.25,63) .. controls (96.25,65.35) and (94.35,67.25) .. (92,67.25) .. controls (89.65,67.25) and (87.75,65.35) .. (87.75,63) -- cycle ;
\draw  [fill={rgb, 255:red, 0; green, 0; blue, 0 }  ,fill opacity=1 ] (264.25,63) .. controls (264.25,60.65) and (266.15,58.75) .. (268.5,58.75) .. controls (270.85,58.75) and (272.75,60.65) .. (272.75,63) .. controls (272.75,65.35) and (270.85,67.25) .. (268.5,67.25) .. controls (266.15,67.25) and (264.25,65.35) .. (264.25,63) -- cycle ;
\draw    (282.5,125.44) -- (282.5,63) ;
\draw [shift={(282.5,63)}, rotate = 450] [color={rgb, 255:red, 0; green, 0; blue, 0 }  ][line width=0.75]    (0,5.59) -- (0,-5.59)   ;
\draw    (282.5,125.44) -- (282.5,146) ;
\draw [shift={(282.5,148)}, rotate = 270] [color={rgb, 255:red, 0; green, 0; blue, 0 }  ][line width=0.75]    (10.93,-3.29) .. controls (6.95,-1.4) and (3.31,-0.3) .. (0,0) .. controls (3.31,0.3) and (6.95,1.4) .. (10.93,3.29)   ;
\draw  [dash pattern={on 4.5pt off 4.5pt}]  (92,63) -- (139.44,110.44) ;
\draw  [dash pattern={on 4.5pt off 4.5pt}]  (268.5,63) -- (221.06,110.44) ;
\draw    (268.5,63) -- (268.5,149) ;

\draw (307.59,60.16) node    {$u_{\mathrm{max}}$};
\draw (306.72,110.94) node    {$u$};
\draw (183.79,51.72) node    {$t=T+u_{\mathrm{max}}$};
\draw (256.67,118.51) node  [rotate=-270]  {$r=T$};
\draw (183.83,77.61) node    {$i^{+}$};
\draw (124.28,85.08) node  [font=\scriptsize,rotate=-45]  {$u\ const$};
\draw (235.95,85.08) node  [font=\scriptsize,rotate=-315]  {$u\ const$};
\draw (269.97,41.61) node    {$\mathcal{I}_{+}^{+}$};
\draw (35,67) node [anchor=north west][inner sep=0.75pt]  [color={rgb, 255:red, 255; green, 255; blue, 255 }  ,opacity=1 ] [align=left] {.};

\end{tikzpicture}

\end{center}
\caption{\label{fig:corners} How $i^+$ attaches to $\mathcal{I}^+_+$ . On the sides we have the surface $r = T$. The upper bound of $u$ is $u_{\rm max}$. Timelike infinity is $t = T + u_{\rm max}$ where $|\vec{r}| \leq T$. Because we take the $T$ large limit before the $u_{\rm max}$ large limit, $u_{\rm max} \ll T$.}
\end{figure}

\noindent To get to $\mathcal{I}^+_+$, one must hold $u$ fixed, take $r \to \infty$, and then take the $u \to \infty$ limit. Therefore, while both $r$ and $u$ are infinite at $\mathcal{I}^+_+$, $r \gg u$. Because $i^+$ can be spanned by a $t = const$ surface, we can glue $i^+$ to $\mathcal{I}^+_+$ in the following way. First, consider the surface of $r = T$ where $T$ is a large constant. The surface is parameterized by $\hat n$ and $u < u_{\rm max}$. $u_{\rm max}$ is taken to be large but must satisfy $u_{\rm max} \ll T$. From Figure \ref{fig:corners}, we can see that $i^+$ can be taken to be\footnote{This is a nonstandard definition of $i^+$. The common method involves hyperbolic slicings.}
\begin{equation}
\begin{split}
    i^+ = \lim_{u_{\rm max} \to \infty} \lim_{T \to \infty} \{ (t, \vec{r}) \; |\; t = T + u_{\rm max}, |\vec{r}| \leq T \}. 
\end{split}
\end{equation}
Actually, because $u_{\rm max}$ is small compared to $T$, we are free to ignore it unless we are interested in $\mathcal{O}(1/T)$ corrections. Henceforth we regard $i^+$ to simply be the $t = T$ slice with $|\vec{r}| \leq T$.

\begin{figure}
\begin{center}

\tikzset{every picture/.style={line width=0.75pt}} 

\begin{tikzpicture}[x=0.75pt,y=0.75pt,yscale=-1,xscale=1]

\draw   (113,84.83) .. controls (113,50.73) and (189.67,23.09) .. (284.26,23.09) .. controls (378.84,23.09) and (455.52,50.73) .. (455.52,84.83) .. controls (455.52,118.93) and (378.84,146.57) .. (284.26,146.57) .. controls (189.67,146.57) and (113,118.93) .. (113,84.83) -- cycle ;
\draw  [fill={rgb, 255:red, 0; green, 0; blue, 0 }  ,fill opacity=1 ] (392.23,79.03) .. controls (392.23,77.63) and (393.36,76.5) .. (394.76,76.5) .. controls (396.16,76.5) and (397.29,77.63) .. (397.29,79.03) .. controls (397.29,80.42) and (396.16,81.55) .. (394.76,81.55) .. controls (393.36,81.55) and (392.23,80.42) .. (392.23,79.03) -- cycle ;
\draw  [fill={rgb, 255:red, 0; green, 0; blue, 0 }  ,fill opacity=1 ] (266.97,58.89) .. controls (266.97,57.49) and (268.1,56.36) .. (269.5,56.36) .. controls (270.89,56.36) and (272.03,57.49) .. (272.03,58.89) .. controls (272.03,60.28) and (270.89,61.42) .. (269.5,61.42) .. controls (268.1,61.42) and (266.97,60.28) .. (266.97,58.89) -- cycle ;
\draw  [fill={rgb, 255:red, 0; green, 0; blue, 0 }  ,fill opacity=1 ] (204.99,107.17) .. controls (204.99,105.77) and (206.12,104.64) .. (207.52,104.64) .. controls (208.91,104.64) and (210.05,105.77) .. (210.05,107.17) .. controls (210.05,108.56) and (208.91,109.69) .. (207.52,109.69) .. controls (206.12,109.69) and (204.99,108.56) .. (204.99,107.17) -- cycle ;
\draw    (207.52,107.17) -- (187.35,117.02) ;
\draw [shift={(184.66,118.34)}, rotate = 333.96000000000004] [fill={rgb, 255:red, 0; green, 0; blue, 0 }  ][line width=0.08]  [draw opacity=0] (10.72,-5.15) -- (0,0) -- (10.72,5.15) -- (7.12,0) -- cycle    ;
\draw    (269.5,58.89) -- (264.83,51.38) ;
\draw [shift={(263.25,48.84)}, rotate = 418.14] [fill={rgb, 255:red, 0; green, 0; blue, 0 }  ][line width=0.08]  [draw opacity=0] (7.14,-3.43) -- (0,0) -- (7.14,3.43) -- (4.74,0) -- cycle    ;
\draw    (394.76,79.03) -- (434.14,79.03) ;
\draw [shift={(437.14,79.03)}, rotate = 180] [fill={rgb, 255:red, 0; green, 0; blue, 0 }  ][line width=0.08]  [draw opacity=0] (10.72,-5.15) -- (0,0) -- (10.72,5.15) -- (7.12,0) -- cycle    ;

\draw (295.77,181.12) node  [font=\Large]  {$t=T\ \text{slice}$};
\draw (180.62,29.27) node [anchor=south east] [inner sep=0.75pt]    {$| \vec{r} |=T$};
\draw (209.63,97.92) node [anchor=south] [inner sep=0.75pt]  [font=\small]  {$\vec{\beta }_{1,f} T$};
\draw (278.07,54.86) node [anchor=south west] [inner sep=0.75pt]  [font=\small]  {$\vec{\beta }_{2,f} T$};
\draw (395.9,71.13) node [anchor=south] [inner sep=0.75pt]  [font=\small]  {$\vec{\beta }_{3,f} T$};
\draw (8,77.7) node [anchor=west] [inner sep=0.75pt]  [font=\fontsize{3.53em}{4.24em}\selectfont]  {$i^{+} :$};
\draw (549,82.7) node [anchor=west] [inner sep=0.75pt]  [font=\fontsize{3.53em}{4.24em}\selectfont,color={rgb, 255:red, 255; green, 255; blue, 255 }  ,opacity=1 ]  {$.$};

\end{tikzpicture}

\end{center}
\caption{\label{fig:timelike} The view at timelike infinity. $i^+$ is a 3-dimensional ball with a radius of $T$. While the ball expands as $T$ is increased, the relative positions of the particles in the ball are fixed in place. The particles become conformally frozen at timelike infinity.}
\end{figure}
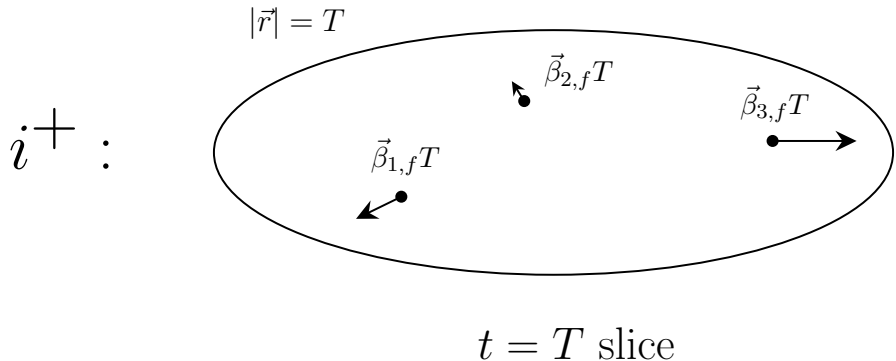

At $i^+$, all of our particles will be travelling with constant velocities, meaning the position of the $k^{\rm th}$ particle is given by
\begin{equation}
\begin{split}
    \vec{r}_{k} = \vec{\beta}_{k,f} T + \vec{r}_{k}^{\;0}
\end{split}
\end{equation}
for some constant shift $\vec{r}_{k}^{\;0}$. As $T$ gets large, these constant shifts become irrelevant as they are $\mathcal{O}(1/T)$ relative corrections to the particle's position. This means that at $i^+$, the particle's position is determined only by its velocity. This is shown in Figure \ref{fig:timelike}.

Let's now assume we are working in Lorenz gauge. This implies that $\lambda$ will solve the wave equation $\partial^2 \lambda = 0$. Furthermore, $\lambda(x)$ will be some convolution of the propagator $G(\hat q, x)$ integrated over the unit 2-sphere, which implies that $\lambda(x)$ will inherit some algebraic properties of $G(\hat q, x)$. Importantly, $G$ has a spacetime scaling symmetry
\begin{equation}
\begin{split}
    G(s x, \hat q) = G(x, \hat q) \hspace{0.5 cm} \text{for all } s \in \mathbb{R}.
\end{split}
\end{equation}
This means that in Lorenz gauge,
\begin{equation}
\begin{split}
    \lambda(s x) = \lambda(x) \hspace{0.5 cm} \text{for all } s \in \mathbb{R}.
\end{split}
\end{equation}
In particular, this implies that $\lambda$ is really only a function of $\vec{r}/t$.
\begin{equation}
\begin{split}
    \lambda(t, \vec{r}) = \lambda(1,\vec{r}/t)
\end{split}
\end{equation}

This has two important implications for large $T$. The first is that the hard charge of massive particles will only depend on the final velocities of those particles as $\vec{r}_k/T $ approaches $\vec{\beta}_{k,f}$ for $T$ large. In particular,
\begin{equation}
    Q^{+,H}_\lambda = \sum_{k} e_{k} \lambda(1, \vec{\beta}_{k,f})
\end{equation}
where $e_{k}$ is the charge of the $k^{\rm th}$ particle.

The second implication, noted in \cite{campiglia2017asymptotic,Campiglia:2015qka}, is that $\lambda$ is actually a harmonic function at $i^+$, meaning it solves the equation
\begin{equation}
\begin{split}
    \Delta \lambda = 0 \hspace{1 cm} ( \text{at } i^\pm).
\end{split}
\end{equation}
This can be seen by expanding the Lorenz gauge condition at large $T$, where the $\partial_t^2$ derivative is subleading in $1/T$ compared to $\Delta \lambda$.
\begin{align}
    0 &= \eval{ (\partial_t^2 - \Delta ) \lambda(1, \vec{r}/t)}_{t = T} \\
    &= \mathcal{O}\left(\frac{1}{T^3}\right) - \frac{1}{T^2} (\Delta \lambda) (1, \vec{r}/T)
\end{align}

Furthermore, in order for $\lambda$ to be a smooth function, its value at the boundary of $i^+$ (which is a 2-sphere of radius $|\vec{r}| = T$) must be equal to its value on $\mathcal{I}^+_+$, which we have denoted $\lambda^+(z^A)$. Because $\lambda$ is harmonic, its values on the whole of $i^+$ are completely determined by its values on the boundary sphere.

Actually, $\lambda$ is not the only function completely determined on $i^+$. The EM fields are completely determined on $i^+$ as well, in the sense that they are Coulombic and therefore depend only on the locations of the charged particles. Aside from an overall scaling of $T$, timelike infinity is strangely ``timeless'' and contains no electromagnetic degrees of freedom.

Let us now finally show that the soft charge $Q^{i^+,S}_\lambda=0$ at $i^+$. By linearity, it is enough to show this holds for a single particle. By Lorentz symmetry, we can take this particle to have a velocity of $\vec{0}$. We then calculate
\begin{align}
    Q^{i^+,S}_\lambda &= - \int_{i^+} d \Sigma_\mu F^{\mu \nu} \nabla_\nu \lambda \\
    &= - \int_0^T r^2 dr \int_{S^2}  F^{ti} \partial_i \lambda \; dS \\
    &= - \int_0^T r^2 dr \int_{S^2} \frac{e}{4 \pi r^2} \partial_r \lambda \; dS\\
    &= e \big( \lambda(\vec{0}) - \lambda_{\rm avg}(T) \big)
\end{align}
where $\lambda_{\rm avg}(r)$ is the average value of $\lambda$ on a sphere of radius $r$ centered around $\vec{0}$. (Note that $dS$ is the unit 2-sphere area element.)
\begin{equation}
    \lambda_{\rm avg}(r) \equiv \int_{S^2(\vec{0}, r)} \frac{dS}{4 \pi} \; \lambda 
\end{equation}
Crucially, harmonic functions satisfy the ``mean value property,'' which says that their value in the center of any sphere is equal to their average value on the surface of the sphere. Therefore, $Q^{i^+,S}_\lambda =0$ in Lorenz gauge, which is exactly what we wanted to show!

By the way, the mean value property can be proven quickly by differentiating $\lambda_{\rm avg}(r)$ with respect to $r$ and showing it is 0 using the divergence theorem:
\begin{align}
    \frac{d}{dr} \lambda_{\rm avg}(r) &= \int_{S^2(\vec{0}, r)} \frac{dS}{4 \pi} \; \partial_r \lambda \\
    &= \int_{B^3(\vec{0}, r)} \frac{d^3 x}{4 \pi} \Delta \lambda \\
    &= 0.
\end{align}

We have now brought all parts of infinity into our story of soft charges. $\mathcal{I}^\pm$ is the place where radiation goes, making it our main object of study. The behavior at $i^0$ is completely determined by the even/oddness properties of early time Coulombic fields and requires antipodal matching for $\lambda$ so that soft charge to be conserved. Finally, the behavior at $i^\pm$ is also completely determined by the early/late time Coulombic fields and up to a scaling all motion completely ``freezes out'' there. No radiation is present and the soft part of the soft charge is 0 at $i^\pm$.

\subsection{Soft charge conservation \& memory}

Let's write out the soft charge conservation equation explicitly when we take 
\begin{equation}
    \lambda(x) = G(x, \hat q)
\end{equation}
for some unit vector $\hat q$. Because $G \propto \delta^2(\hat n - \hat q)$ at $\mathcal{I}^+$, this will give us all the possible information we can extract from soft charge conservation.

The equation
\begin{equation}
    Q_\lambda^{+,H} - Q_\lambda^{-,H} = - Q_\lambda^{+,S} + Q_\lambda^{-,S}
\end{equation}
becomes
\begin{align}\label{memoryu0}
    \sum_k \frac{e_k}{4 \pi} \left( \frac{1 - \beta_{k,f}^2}{(1 - \hat q \cdot \vec{\beta}_{k,f})^2} - \frac{1 - \beta_{k,i}^2}{(1 - \hat q \cdot \vec{\beta}_{k,i})^2} \right) = D^A \left( \int_{\mathcal{I}^+_{\hat q}} du F_{uA} - \int_{\mathcal{I}^-_{-\hat q}} dv F_{vA} \right).
\end{align}
Both sides are functions of $\hat q$. The first term on the RHS is an integral over $\mathcal{I}^+$ restricted the to sphere angle $\hat q$, and the second term is an integral over $\mathcal{I}^-$ at $-\hat q$. This equation means that the overall amount of radiation that leaves null infinity at each angle is constrained by the initial and final velocities of the charged particles. Incidentally, if one of the charged particles is massless, then on the LHS one should replace
\begin{equation}
    \frac{e_k}{4 \pi} \frac{1 - \beta_{k,f}^2}{(1 - \hat q \cdot \vec{\beta}_{k,f})^2} \mapsto \frac{e_k}{\sqrt{\gamma}} \delta^2(z^A - q^A)
\end{equation}
where $z^A$ gives the direction of the particle's velocity.

Interestingly, the RHS of \eqref{memoryu0} can be simplified further by performing the $u$- and $v$-integrals over $\mathcal{I}^\pm$. For instance,
\begin{align}
    \int_{\mathcal{I}^+_{\hat q}} du F_{uA} &= \int_{\mathcal{I}^+_{\hat q}} du( \partial_u \underbrace{A_A}_{\mathcal{O}(1)} - \partial_A \underbrace{A_u}_{\mathcal{O}(1/r)} ) = \eval{A_A(\hat q)}_{\mathcal{I}^+_-}^{\mathcal{I}^+_+}.
\end{align}
Therefore, \eqref{memoryu0} becomes
\begin{align}\label{memoryu}
    \sum_k \frac{e_k}{4 \pi} \left( \frac{1 - \beta_{k,f}^2}{(1 - \hat q \cdot \vec{\beta}_{k,f})^2} - \frac{1 - \beta_{k,i}^2}{(1 - \hat q \cdot \vec{\beta}_{k,i})^2} \right) = D^A \left( \eval{A_A(\hat q) }_{\mathcal{I}^+_-}^{\mathcal{I}^+_+} - \eval{A_A(- \hat q) }_{\mathcal{I}^-_-}^{\mathcal{I}^-_+} \right).
\end{align}
If we assume that no radiation enters from $\mathcal{I}^-$ then the second term on the RHS will be 0. This equation then means that $A_A$ will necessarily undergo some net change depending on the initial and final velocities of the charged particles. Therefore, we can see how integrating over null infinity has turned our soft charge conservation equation into a ``memory effect'' equation. We will say more about the memory effect in section \ref{sec_memory}.

The equation \eqref{memoryu} represents and infinite number of equations, one for each $\hat q$. It can also be understood as a single \textit{differential} equation. One might wonder if this differential equation can be expressed at the equation of motion of some Euclidean action defined on $S^2$. This is indeed the case---such an action was given in \cite{kapec2021shadows} (see also \cite{nguyen2021celestial,Nguyen:2020hot}). However, we will not comment any further on this infrared boundary action.

Actually, \eqref{memoryu} can be simplified again by noting that both sides are total divergences. This is manifestly true on the RHS. On the LHS, one can use the identity
\begin{align}
    \frac{1 - \beta^2}{(1- \hat q \cdot \vec{\beta})^2} &= D^A D_A \log(1 - \hat q \cdot \vec{\beta}) + 1 \\
    &= D^A \left( \frac{- \partial_A \hat q \cdot \vec{\beta}}{1 - \hat q \cdot \vec{\beta}} \right) + 1. \label{id2}
\end{align}
This formula is easiest to prove in polar coordinates with $\hat q \cdot \vec{\beta} = |\vec{\beta}| \cos \theta$, where is boils down to checking
\begin{align}
    D^A D_A \log(1 - \hat q \cdot \vec{\beta} ) &= \frac{1}{\sin \theta} \partial_\theta \left( \sin \theta \partial_\theta \log(1 - |\vec{\beta}| \cos \theta) \right) \\
    &= -1 + \frac{1 - \beta^2}{(1 - |\vec{\beta}| \cos \theta )^2}.
\end{align}
Using \eqref{id2}, we can simply ``pick off'' the $D^A$ on both sides\footnote{In \cite{strominger2018lectures} the derivatives were `picked off' by choosing $\lambda = \tfrac{1}{w - w_0}$ in complex sphere coordinates and using the residue theorem} of \eqref{memoryu} and get
\begin{equation}\label{memoryu1}
    \sum_k \frac{e_k}{4 \pi} \left( - \frac{\partial_A \hat q \cdot \vec{\beta}_{kf} }{1 - \hat q \cdot \vec{\beta}_{k,f}} + \frac{\partial_A \hat q \cdot \vec{\beta}_{k,i} }{1 - \hat q \cdot \vec{\beta}_{k,i}} \right) = \eval{A_A(\hat q)}^{\mathcal{I}^+_+}_{\mathcal{I}^+_-} - \eval{A_A(-\hat q)}^{\mathcal{I}^-_+}_{\mathcal{I}^-_-}
\end{equation}
We can therefore see that the conservation of soft charge is such a powerful constraint on the evolution of the EM field that it almost completely solves for $A_A$ at $\mathcal{I}^\pm_\pm$.

Notice that if we set
\begin{align}
    p^\mu &= m \gamma(1, \vec{\beta}) \\
    \vep^\mu_A &= \partial_A(1, \hat q)
\end{align}
then \eqref{memoryu1} becomes
\begin{equation}\label{classical_soft_photon}
    \sum_k \frac{e_k}{4 \pi} \left( \frac{ \vep_A \cdot p_{k,f}}{q \cdot p_{k,f}} - \frac{\vep_A \cdot p_{k,i}}{q \cdot p_{k,i}} \right) = \int_{\mathcal{I}^+_{\hat q}} du F_{uA} -  \int_{\mathcal{I}^-_{-\hat q}} dv F_{vA}.
\end{equation}
This equations bears a striking resemblance to Low's soft photon theorem \cite{low1958bremsstrahlung,weinberg1965infrared}. In fact, upon quantization, it literally is Low's soft photon theorem \cite{he2014new}. $F_{uA}$ on the RHS will break up into a sum of terms like $\hat a^\dagger + \hat a$ where $\hat a^\dagger$ $(\hat a)$ creates (annihilates) a 0 energy photon with polarization $\vep^\mu_A$ and a four-momentum proportional to $(1,\hat q)$. The soft photon theorem is then given by the Ward identity
\begin{equation}
    0 = \bra{f} [ Q_\lambda, \mathcal{S}] \ket{i} = \bra{f} Q_\lambda^+ \mathcal{S} \ket{i} - \bra{f}\mathcal{S}  Q_\lambda^-  \ket{i}.
\end{equation}

\section{Intro to Covariant Phase Space Formalism}

It is often said that in order to study the phase space of field theories, one must single out a time direction and break manifest Lorentz symmetry. This is false. The ``covariant phase space formalism'' gives a way to construct symplectic forms, Poisson brackets, and canonical charges all while preserving manifest covariance \cite{crnkovic1987covariant}. It will come in handy when studying the symmetry algebra of large gauge transformations so we review the formalism here.

\subsection{The symplectic form}

First let's review what a symplectic form is. Say you have a classical phase space in which each point $P$ corresponds to a physical state. The symplectic form $\omega$ takes in as input two tangent vectors $v_1,v_2$ based at a point $P$ and outputs a real number.

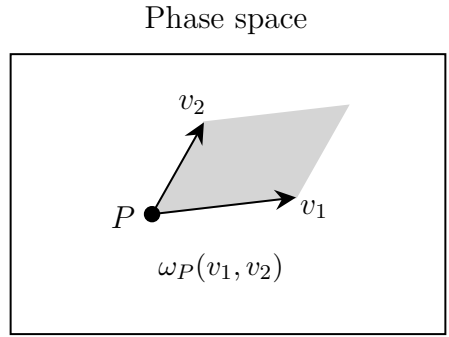
\begin{figure}[h]
\begin{center}

\tikzset{every picture/.style={line width=0.75pt}} 

\begin{tikzpicture}[x=0.75pt,y=0.75pt,yscale=-1,xscale=1]

\draw   (4,32) -- (223.26,32) -- (223.26,173.3) -- (4,173.3) -- cycle ;
\draw  [color={rgb, 255:red, 213; green, 213; blue, 213 }  ,draw opacity=1 ][fill={rgb, 255:red, 213; green, 213; blue, 213 }  ,fill opacity=1 ] (101.48,66.44) -- (174.11,58) -- (147.95,104.32) -- (75.32,112.76) -- cycle ;
\draw  [fill={rgb, 255:red, 0; green, 0; blue, 0 }  ,fill opacity=1 ] (71.69,112.76) .. controls (71.69,110.76) and (73.31,109.13) .. (75.32,109.13) .. controls (77.32,109.13) and (78.95,110.76) .. (78.95,112.76) .. controls (78.95,114.77) and (77.32,116.39) .. (75.32,116.39) .. controls (73.31,116.39) and (71.69,114.77) .. (71.69,112.76) -- cycle ;
\draw    (75.32,112.76) -- (144.97,104.67) ;
\draw [shift={(147.95,104.32)}, rotate = 533.37] [fill={rgb, 255:red, 0; green, 0; blue, 0 }  ][line width=0.08]  [draw opacity=0] (10.72,-5.15) -- (0,0) -- (10.72,5.15) -- (7.12,0) -- cycle    ;
\draw    (75.32,112.76) -- (100.01,69.05) ;
\draw [shift={(101.48,66.44)}, rotate = 479.46] [fill={rgb, 255:red, 0; green, 0; blue, 0 }  ][line width=0.08]  [draw opacity=0] (10.72,-5.15) -- (0,0) -- (10.72,5.15) -- (7.12,0) -- cycle    ;

\draw (70,6) node [anchor=north west][inner sep=0.75pt]   [align=left] {Phase space};
\draw (110.17,139.63) node  [font=\small]  {$\omega _{P}( v_{1} ,v_{2})$};
\draw (68.28,114.45) node [anchor=east] [inner sep=0.75pt]    {$P$};
\draw (148.61,104.05) node [anchor=north west][inner sep=0.75pt]    {$v_{1}$};
\draw (95.59,57.37) node    {$v_{2}$};

\end{tikzpicture}

\end{center}
\caption{\label{fig:phasespace_v} Abstract picture of the symplectic form $\omega$ as area of parallelogram spanned by tangent vectors $v_1, v_2$ based at point $P$.}
\end{figure}

As a very basic example, consider the phase space $\mathbb{R}^2$, parameterized by $P = (q,p)$. The symplectic form will be the 2-form
\begin{equation}
    \omega = \mathrm{d} q \wedge \mathrm{d} p.
\end{equation}
Actually, from here on out, let's use the symbol ``$\delta$'' instead of ``$\mathrm{d}$''. In this line of work, it is helpful to use ``$\mathrm{d}$'' for the exterior derivative on \textit{spacetime} and use ``$\delta$'' for the exterior derivative on \textit{phase space}. So,
\begin{equation}
    \omega = \delta q \wedge \delta p.
\end{equation}
As a reminder on how 2-forms work, we can expand out the wedge product as
\begin{equation}
    \omega = \delta_1 q \, \delta_2 p - \delta_2 q \, \delta_1 p.
\end{equation}
If we write out the two tangent vectors $v_1$ and $v_2$ in components as
\begin{equation}
    v_1 = v_1^q \frac{\partial}{\partial q} + v_1^p \frac{\partial}{\partial p} \hspace{1 cm} v_2 = v_2^q \frac{\partial}{\partial q} + v_2^p \frac{\partial}{\partial p}
\end{equation}
then
\begin{equation}
\begin{split}
    \omega(v_1, v_2) = v_1^q v_2^p - v_2^q v_1^p.
\end{split}
\end{equation}
Notice that this expression is just the area of the parallelogram spanned by $v_1$ and $v_2$! This is the geometrical interpretation of the symplectic form, as discussed in \cite{mcduff2010symplectic}.

If our phase space is $\mathbb{R}^{2N}$, parameterized by $P = (q_i, p_i)_{i = 1\ldots N}$, the the symplectic form is
\begin{equation}
    \omega = \sum_{i=1}^N \delta q_i \wedge \delta p_i.
\end{equation}
$\omega(v_1, v_2)$ is then the sum of the areas of the parallelograms spanned by $v_1$ and $v_2$ when projected down into the $N$ different $(q_i, p_i)$ planes. Morally speaking, we can see that $\omega$ ``pairs'' $q$'s with $p$'s in some preferred way that it won't pair $q$'s with other $q$'s or $p$'s with other $p$'s.\footnote{Having said that, there is no invariant way to specify the difference between a $q$ and a $p$ because they can be mixed into each other by canonical transformations.}

One important use of $\omega$ is that it can be wedged together with itself $N$ times to get the volume form on the phase space.
\begin{align}\label{vol}
    \mathrm{vol} &\equiv \underbrace{\omega \wedge \omega \wedge \ldots \wedge \omega}_{N \text{ times}} \\
    &= \delta q_1 \wedge \delta p_1 \wedge \delta q_2 \wedge \delta p_2 \wedge \ldots \wedge \delta q_N \wedge \delta p_N.
\end{align}

Another very important job of $\omega$ is that it is used to define ``Hamiltonian vector fields'' on phase space. If $f$ is some function on phase space, then its Hamiltonian vector field $X_f$ is defined by the equation
\begin{equation}\label{Xfdef}
\begin{split}
    \omega(X_f, v ) = \delta f (v)
\end{split}
\end{equation}
where $v$ is an arbitrary input tangent vector. Let's unwrap this definition for our simple $\mathbb{R}^2$ example. We'll write out $X_f$ and $v$ in components as
\begin{align}
    X_f &= X_f^q \frac{\partial}{\partial q} + X_f^p \frac{\partial}{\partial p} \\
    v &= v^q \frac{\partial}{\partial q} + v^p \frac{\partial}{\partial p}.
\end{align}
Then the \eqref{Xfdef} becomes becomes
\begin{align}
\omega(X_f, v ) &= \left( \frac{\partial f}{\partial q} \delta q + \frac{\partial f}{\partial p} \delta p \right) (v) \\
    X_f^q v^p - X_f^p v^q &= \frac{\partial f}{\partial q} v^q + \frac{\partial f}{\partial p} v^p
\end{align}
and we can extract the $X_f^q$ and $X_f^p$ components to get
\begin{equation}
    (X_f^q, X_f^p) = \left( \frac{\partial f}{\partial p},- \frac{\partial f}{\partial q} \right).
\end{equation}
Notice how the above vector field differs from the gradient of $f$
\begin{equation}
    \nabla f = \left( \frac{\partial f}{\partial q}, \frac{\partial f}{\partial p} \right).
\end{equation}
Interestingly, we can see that $X_f$ is just $\nabla f$ rotated by $90^\circ$! If we draw lines of constant $f$ on phase space, $\nabla f$ will point perpendicular to these lines while $X_f$ will point along them. This is the geometric picture for the vector field $X_f$ that a function $f$ generates.

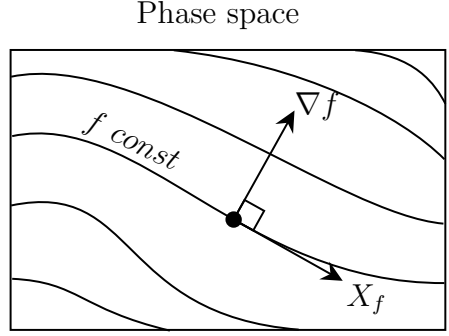
\begin{figure}[H]
\begin{center}

\tikzset{every picture/.style={line width=0.75pt}} 

\begin{tikzpicture}[x=0.75pt,y=0.75pt,yscale=-1,xscale=1]

\draw   (237,32) -- (456.26,32) -- (456.26,173.3) -- (237,173.3) -- cycle ;
\draw  [fill={rgb, 255:red, 0; green, 0; blue, 0 }  ,fill opacity=1 ] (345.86,117.6) .. controls (345.86,115.59) and (347.48,113.96) .. (349.49,113.96) .. controls (351.49,113.96) and (353.12,115.59) .. (353.12,117.6) .. controls (353.12,119.6) and (351.49,121.23) .. (349.49,121.23) .. controls (347.48,121.23) and (345.86,119.6) .. (345.86,117.6) -- cycle ;
\draw    (237.8,75.5) .. controls (301.8,62.16) and (354.3,150.16) .. (455.8,149.66) ;
\draw    (237.3,45.5) .. controls (301.3,32.16) and (402.8,115.16) .. (455.3,119.66) ;
\draw    (319.3,32.16) .. controls (384.31,39.31) and (433.8,65.16) .. (455.8,87.16) ;
\draw    (410.8,32.16) .. controls (441.3,34.16) and (451.3,49.66) .. (456.3,59.16) ;
\draw    (237.3,110.5) .. controls (301.3,97.16) and (289.8,165.66) .. (382.3,173.16) ;
\draw    (237.8,147.5) .. controls (271.8,149.16) and (271.3,162.66) .. (317.3,173.66) ;
\draw    (349.49,117.6) -- (378.73,65.16) ;
\draw [shift={(380.19,62.54)}, rotate = 479.15] [fill={rgb, 255:red, 0; green, 0; blue, 0 }  ][line width=0.08]  [draw opacity=0] (10.72,-5.15) -- (0,0) -- (10.72,5.15) -- (7.12,0) -- cycle    ;
\draw    (349.49,117.6) -- (401.91,147.07) ;
\draw [shift={(404.53,148.54)}, rotate = 209.35] [fill={rgb, 255:red, 0; green, 0; blue, 0 }  ][line width=0.08]  [draw opacity=0] (10.72,-5.15) -- (0,0) -- (10.72,5.15) -- (7.12,0) -- cycle    ;
\draw   (354.99,107.7) -- (364.88,113.21) -- (359.38,123.1) -- (349.49,117.6) -- cycle ;

\draw (299,6) node [anchor=north west][inner sep=0.75pt]   [align=left] {Phase space};
\draw (378.85,51.59) node [anchor=north west][inner sep=0.75pt]    {$\nabla f$};
\draw (404.85,148.59) node [anchor=north west][inner sep=0.75pt]    {$X_{f}$};
\draw (294.41,86.08) node [anchor=south] [inner sep=0.75pt]  [rotate=-25.87]  {$f\ const$};

\end{tikzpicture}

\end{center}
\caption{\label{fig:90deg}Lines of constant $f$ are drawn on phase space. $X_f$ is a 90 degree rotation of $\nabla f$.}
\end{figure}

\subsection{The Poisson bracket}

The definition of the Poisson bracket is
\begin{equation}\label{poisson_def}
    \{ f, g \} \equiv \omega(X_f, X_g).
\end{equation}
Let's unwrap this equation so we can arrive at a more explicit formula. If we write the symplectic form with generalized coordinates $P^I$ as
\begin{equation}
    \omega = \frac{1}{2} \omega_{IJ} \delta P^I \wedge \delta P^J
\end{equation}
where $\omega_{IJ} = -\omega_{JI}$, then we can write $X_f$ and an arbitrary vector field $v$ as
\begin{equation}
    X_f = X_f^I \partial_I \hspace{1.2 cm} v = v^I \partial_I.
\end{equation}
This implies
\begin{align}
    \omega(X_f, v) &= \delta f (v) \\
    \omega_{IJ} X_f^I v^J &= (\partial_I f) v^I
\end{align}
and assuming $\omega_{IJ}$ is non-degenerate we can solve for $X_f^I$ as
\begin{equation}\label{XfI}
    X_f^I = -\omega^{II'} \partial_{I'} f
\end{equation}
where $\omega^{IJ}$ is the inverse matrix of $\omega_{IJ}$. Plugging this into \eqref{poisson_def} we get
\begin{align}
    \{ f, g \} &= \omega_{IJ} X_f^I X_g^J \\
    &= \omega_{IJ} (\omega^{II'} \partial_{I'} f)( \omega^{JJ'} \partial_{J'} g) \\
    &= -\omega^{IJ} (\partial_I f) (\partial_J g). \label{poissonfinal}
\end{align}
This is a more useful equation than \eqref{poisson_def}. In particular,
\begin{equation}
\begin{split}\label{poisson_useful}
    \{ P^I, P^J \} = -\omega^{IJ}.
\end{split}
\end{equation}
Finally, comparing \eqref{XfI} and \eqref{poissonfinal}, we see that
\begin{equation}
\begin{split}
    X_f(g) = \{ g, f \}.
\end{split}
\end{equation}
Therefore, the Poisson bracket $\{g, f\}$ can be interpreted as the change in $g$ after some infinitesimal flow along the vector field $X_f$.

\subsection{Field theory \& fall-offs}

How can we describe the phase space of a field theory in a manifestly covariant way, without choosing a foliation of time slices? The answer is: each \textit{solution} to the equations of motion is one \textit{point} in our phase space. For example, in Maxwell EM, every on-shell function $A_\mu(x)$ on our 4D spacetime is a single point (modulo small gauge transformations, but we'll get to that later). ``Tangent vectors'' $\delta A$ are field variations that satisfy the linearized equations of motion. (Here we are abusing notation in a standard way by referring to phase space tangent vectors, i.e. field variations, with the symbol $\delta$ which we have also used to represent the phase space exterior derivative.)

\begin{figure}[H]
\begin{center}

\tikzset{every picture/.style={line width=0.75pt}} 

\begin{tikzpicture}[x=0.75pt,y=0.75pt,yscale=-1,xscale=1]

\draw   (237,32) -- (456.26,32) -- (456.26,173.3) -- (237,173.3) -- cycle ;
\draw  [color={rgb, 255:red, 213; green, 213; blue, 213 }  ,draw opacity=1 ][fill={rgb, 255:red, 213; green, 213; blue, 213 }  ,fill opacity=1 ] (334.48,66.44) -- (407.11,58) -- (380.95,104.32) -- (308.32,112.76) -- cycle ;
\draw  [fill={rgb, 255:red, 0; green, 0; blue, 0 }  ,fill opacity=1 ] (304.69,112.76) .. controls (304.69,110.76) and (306.31,109.13) .. (308.32,109.13) .. controls (310.32,109.13) and (311.95,110.76) .. (311.95,112.76) .. controls (311.95,114.77) and (310.32,116.39) .. (308.32,116.39) .. controls (306.31,116.39) and (304.69,114.77) .. (304.69,112.76) -- cycle ;
\draw    (308.32,112.76) -- (377.97,104.67) ;
\draw [shift={(380.95,104.32)}, rotate = 533.37] [fill={rgb, 255:red, 0; green, 0; blue, 0 }  ][line width=0.08]  [draw opacity=0] (10.72,-5.15) -- (0,0) -- (10.72,5.15) -- (7.12,0) -- cycle    ;
\draw    (308.32,112.76) -- (333.01,69.05) ;
\draw [shift={(334.48,66.44)}, rotate = 479.46] [fill={rgb, 255:red, 0; green, 0; blue, 0 }  ][line width=0.08]  [draw opacity=0] (10.72,-5.15) -- (0,0) -- (10.72,5.15) -- (7.12,0) -- cycle    ;

\draw (303,6) node [anchor=north west][inner sep=0.75pt]   [align=left] {Phase space};
\draw (292.87,138.05) node [anchor=north west][inner sep=0.75pt]  [font=\small]  {$\omega [ A;\delta _{1} A,\delta _{2} A]$};
\draw (290.83,108.68) node [anchor=north west][inner sep=0.75pt]    {$A$};
\draw (381.61,104.05) node [anchor=north west][inner sep=0.75pt]    {$\delta _{1} A$};
\draw (307.61,46.72) node [anchor=north west][inner sep=0.75pt]    {$\delta _{2} A$};

\end{tikzpicture}

\end{center}
\caption{\label{fig:phase_A} Each on-shell solution $A$ is a single point in our phase space. A tangent vector $\delta A$ is a variation such that $A+ \delta A$ is also on-shell to linear order.}
\end{figure}
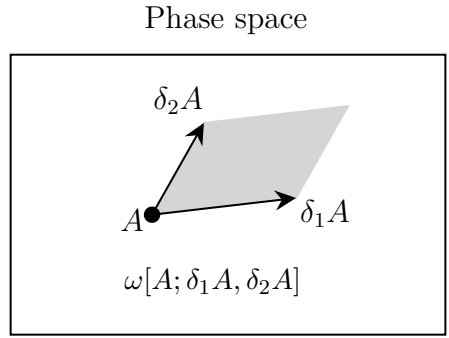

Now, how exactly do we find $\omega?$ Interestingly, it turns out that it can be derived from the field's Lagrangian! Notice, for a moment, that in Lagrangian mechanics you only need a single function $L$ in order to determine the e.o.m.. However, in Hamiltonian mechanics, you need both a function $H$ as well as a symplectic form in order to determine the equations of motion. The symplectic form is required in order to define $X_H$--equivalently, you need the symplectic form in order to single out $q$'s and $p$'s with which to define Hamilton's equations.
\begin{align*}
\text{Lagrangian mechanics: }&\text{ need $L$ to get e.o.m.} \\ \text{Hamiltonian mechanics: }&\text{ need $H$ and $\omega$ to get e.o.m.}
\end{align*}
This means that $L$ is somehow equivalent to the pair $(H, \omega)$.
\begin{equation*}
    L \longleftrightarrow (H, \omega).
\end{equation*}
It is commonly discussed how $H$ can be obtained from $L$. But how do you obtain $\omega$? Well, let's start with the covariant EM action
\begin{equation}
    S = \int_V d^4 x \sg \left( - \frac{1}{4} F^{\mu \nu} F_{\mu \nu} - j^\nu A_\nu \right).
\end{equation}
Here, $j^\mu$ is an external current source satisfying $\nabla_\mu j^\mu = 0$ and $V$ is the volume of our spacetime. If we vary this action, we get
\begin{align}
    \delta S &= \int_V d^4 x \sg \left( - F^{\mu \nu} \partial_\mu \delta A_\nu - j^\nu \delta A_\nu \right) \\
    &= \int_V d^4 x \sg \left( - \nabla_\mu (F^{\mu \nu} \delta A_\nu) + (\nabla_\mu F^{\mu \nu}  - j^\nu) \delta A_\nu \right).
\end{align}
The second term is the e.o.m., which we are well familiar with, and the first term is a total derivative. We are used to ignoring the total derivative term, but we shouldn't be so wasteful. Every part of the Lagrangian is important! In fact, the quantity inside the derivative $\nabla_\mu( \ldots )$ is a very important object called the ``presymplectic potential density'' $\Theta^\mu[A; \delta A]$.\footnote{If one adds a total derivative to the Lagrangian, $\Theta^\mu$ will change by a boundary term. One must in general determine what the ``best'' boundary term to add is. In EM we can just use the naive $\Theta^\mu$, but in GR there are multiple options which have different properties \cite{compere2019advanced}. }
\begin{equation} \label{deltaL_full}
    \delta \mathcal{L} = \sg \nabla_\mu\underbrace{( - F^{\mu \nu} \delta A_\nu )}_{\Theta^\mu[A; \delta A]} + \sg \underbrace{(\nabla_\mu F^{\mu \nu}  - j^\nu )}_{\text{e.o.m.}} \delta A_\nu
\end{equation}
Now, on-shell, e.o.m. is 0. For this section alone, we denote this with
\begin{equation}
    \text{e.o.m.} \approx 0.
\end{equation}
Here ``$\approx$'' is the ``weak equality'' symbol which is used for equations which only hold when evaluated on solutions to the equations of motion. Therefore, \eqref{deltaL_full} becomes
\begin{equation}
    \sg \nabla_\mu \Theta^\mu[A; \delta A] \approx \delta \mathcal{L}.
\end{equation}
Notice that the symbol ``$\delta$'' on the RHS is the exterior derivative on phase space. Therefore, if we act $\delta$ on both sides, using $\delta^2 = 0$ we have
\begin{equation}\label{deltatheta}
    \nabla_\mu \left( \delta \Theta^\mu \right) \approx 0.
\end{equation}
The exterior derivative of the ``presymplectic potential density'' is a very important thing---it's called the ``symplectic density,'' denoted by $\omega^\mu$ or $\omega^\mu[A; \delta_1 A, \delta_2 A]$.
\begin{align}
    \omega^\mu &\equiv \delta \Theta^\mu \\
    \omega^\mu[A; \delta_1 A, \delta_2 A]&= \delta_2 \Theta^\mu[A; \delta_1 A] - \delta_1 \Theta^\mu[A; \delta_2 A].
\end{align}
In the second line, the exterior derivative $\delta$ has been written out more explicitly. Notice that $\Theta^\mu[A; \delta A]$ is a 1-form on phase space while $\omega^\mu[A; \delta_1 A, \delta_2 A]$ is a 2-form on phase space. As a bit of explication, when we vary, say, $\delta_1 \Theta^\mu[A; \delta_2 A]$, we are varying the basepoint $A$ by $\delta_1 A$ and leaving $\delta_2 A$ untouched.

From \eqref{deltatheta}, we have
\begin{equation}
    \nabla_\mu \omega^\mu[A; \delta_1 A, \delta_2 A] \approx 0
\end{equation}
meaning that, on solutions to the equations of motion, $\omega^\mu$ is conserved. Therefore, if $\omega^\mu$ is integrated over a 3D spacetime surface $\Sigma$, the result will stay the same as long as you keep $\partial \Sigma$ fixed.

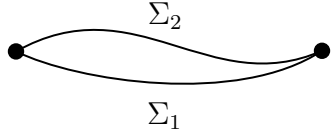
\begin{figure}[H]
\begin{center}

\tikzset{every picture/.style={line width=0.75pt}} 

\begin{tikzpicture}[x=0.75pt,y=0.75pt,yscale=-1,xscale=1]

\draw  [fill={rgb, 255:red, 0; green, 0; blue, 0 }  ,fill opacity=1 ] (28.36,266.75) .. controls (28.36,264.75) and (29.98,263.12) .. (31.99,263.12) .. controls (33.99,263.12) and (35.62,264.75) .. (35.62,266.75) .. controls (35.62,268.76) and (33.99,270.38) .. (31.99,270.38) .. controls (29.98,270.38) and (28.36,268.76) .. (28.36,266.75) -- cycle ;
\draw  [fill={rgb, 255:red, 0; green, 0; blue, 0 }  ,fill opacity=1 ] (182.69,266.42) .. controls (182.69,264.41) and (184.31,262.79) .. (186.32,262.79) .. controls (188.32,262.79) and (189.95,264.41) .. (189.95,266.42) .. controls (189.95,268.42) and (188.32,270.05) .. (186.32,270.05) .. controls (184.31,270.05) and (182.69,268.42) .. (182.69,266.42) -- cycle ;
\draw    (31.99,266.75) .. controls (92.43,232.8) and (129.09,291.47) .. (186.32,266.42) ;
\draw    (31.99,266.75) .. controls (62.76,281.8) and (143.76,294.47) .. (186.32,266.42) ;

\draw (96.67,290.85) node [anchor=north west][inner sep=0.75pt]    {$\Sigma _{1}$};
\draw (97,240.52) node [anchor=north west][inner sep=0.75pt]    {$\Sigma _{2}$};

\end{tikzpicture}

\end{center}
\caption{\label{fig:sigma12vary} $\int_{\Sigma_1} d \Sigma_\mu \omega^\mu = \int_{\Sigma_2} d \Sigma_\mu \omega^\mu$ because $\nabla_\mu \omega^\mu \approx 0$.}
\end{figure}

We finally have all the ingredients necessary to define the symplectic form in the covariant phase space formalism---it is just
\begin{align}
    \omega &\equiv \int_\Sigma d \Sigma_\mu \omega^\mu \\
    &= -\int_\Sigma d \Sigma_\mu \delta F^{\mu \nu} \wedge \delta A_\nu \label{omega_em}
\end{align}
which we can also write as
\begin{equation}
    \omega[A; \delta_1 A, \delta_2 A] = - \int_\Sigma d \Sigma_\mu \left( \delta_1 F^{\mu \nu} \delta_2 A_\nu - \delta_2 F^{\mu \nu} \delta_1 A_\nu \right).
\end{equation}
Here, $\Sigma$ is a Cauchy slice for the spacetime. Notice that while we need to choose a Cauchy slice in order to compute $\omega$, it shouldn't matter \textit{which} slice we choose. The final number will be the same no matter what.

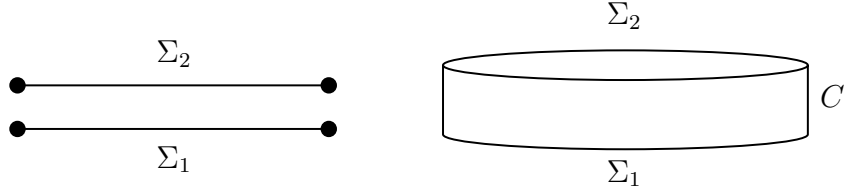
\begin{figure}
\begin{center}

\tikzset{every picture/.style={line width=0.75pt}} 

\begin{tikzpicture}[x=0.75pt,y=0.75pt,yscale=-1,xscale=1]

\draw  [fill={rgb, 255:red, 0; green, 0; blue, 0 }  ,fill opacity=1 ] (135.02,50.75) .. controls (135.02,48.75) and (136.65,47.12) .. (138.65,47.12) .. controls (140.66,47.12) and (142.28,48.75) .. (142.28,50.75) .. controls (142.28,52.76) and (140.66,54.38) .. (138.65,54.38) .. controls (136.65,54.38) and (135.02,52.76) .. (135.02,50.75) -- cycle ;
\draw  [fill={rgb, 255:red, 0; green, 0; blue, 0 }  ,fill opacity=1 ] (292.02,50.75) .. controls (292.02,48.75) and (293.65,47.12) .. (295.65,47.12) .. controls (297.66,47.12) and (299.28,48.75) .. (299.28,50.75) .. controls (299.28,52.76) and (297.66,54.38) .. (295.65,54.38) .. controls (293.65,54.38) and (292.02,52.76) .. (292.02,50.75) -- cycle ;
\draw    (138.65,50.75) -- (295.65,50.75) ;
\draw  [fill={rgb, 255:red, 0; green, 0; blue, 0 }  ,fill opacity=1 ] (135.02,72.75) .. controls (135.02,70.75) and (136.65,69.12) .. (138.65,69.12) .. controls (140.66,69.12) and (142.28,70.75) .. (142.28,72.75) .. controls (142.28,74.76) and (140.66,76.38) .. (138.65,76.38) .. controls (136.65,76.38) and (135.02,74.76) .. (135.02,72.75) -- cycle ;
\draw    (138.65,72.75) -- (295.65,72.75) ;
\draw  [fill={rgb, 255:red, 0; green, 0; blue, 0 }  ,fill opacity=1 ] (292.02,72.75) .. controls (292.02,70.75) and (293.65,69.12) .. (295.65,69.12) .. controls (297.66,69.12) and (299.28,70.75) .. (299.28,72.75) .. controls (299.28,74.76) and (297.66,76.38) .. (295.65,76.38) .. controls (293.65,76.38) and (292.02,74.76) .. (292.02,72.75) -- cycle ;
\draw   (537.26,40.57) -- (537.26,75.51) .. controls (537.26,79.65) and (496.07,83) .. (445.26,83) .. controls (394.45,83) and (353.26,79.65) .. (353.26,75.51) -- (353.26,40.57) .. controls (353.26,36.43) and (394.45,33.08) .. (445.26,33.08) .. controls (496.07,33.08) and (537.26,36.43) .. (537.26,40.57) .. controls (537.26,44.7) and (496.07,48.06) .. (445.26,48.06) .. controls (394.45,48.06) and (353.26,44.7) .. (353.26,40.57) ;

\draw (207.33,79.85) node [anchor=north west][inner sep=0.75pt]    {$\Sigma _{1}$};
\draw (207.33,26.52) node [anchor=north west][inner sep=0.75pt]    {$\Sigma _{2}$};
\draw (434.33,7.52) node [anchor=north west][inner sep=0.75pt]    {$\Sigma _{2}$};
\draw (434.33,86.85) node [anchor=north west][inner sep=0.75pt]    {$\Sigma _{1}$};
\draw (542,49.4) node [anchor=north west][inner sep=0.75pt]    {$C$};

\end{tikzpicture}

\end{center}
\caption{\label{fig:sigmacylinder} Cauchy slices which are $t = const$ surfaces. $C \cong S^2 \times [0,1]$ is a 3-surface which connects $\Sigma_1$ and $\Sigma_2$.}
\end{figure}

Let's look at a specific example in detail. We'll take our Cauchy slices to be $t = const$ surfaces. The first thing we should check is that our symplectic form is finite. Explicitly, it is
\begin{equation}\label{omega_large_r}
    \omega = -\int_0^\infty r^2 dr \; \int_{S^2} d^2 z \sqrt{\gamma} \left( \delta_1 F^{t i} \delta_2 A_i - \delta_2 F^{t i} \delta_1 A_i  \right).
\end{equation}
Given that $F^{ti} \sim \mathcal{O}(1/r^2)$ and $A_i \sim \mathcal{O}(1/r)$, it seems as though $\omega$ diverges as $\log(r)$. However, we must remember to use even/oddness conditions at large $r$. Recall that the field of a charged particle which moves at a constant velocity and passes through the origin is
\begin{equation}
    A_\mu(t, \vec{r}) = \frac{e}{4 \pi} \frac{(1, -\vec{\beta})}{\sqrt{ (\vec{r} - \vec{\beta} t)^2 + (\vec{r} \cdot \vec{\beta} )^2 - \beta^2 r^2 }}.
\end{equation}
This field is prototypical of the fields we want to consider (as constant shifts in $\vec{r}$ will not affect the large $r$ behavior). If we hold $t$ fixed and take $r \to \infty$, we have the even/oddness conditions
\begin{equation}
    \begin{matrix} A_\mu(t, -\vec{r}) = + A_\mu(t, \vec{r}) \\  F_{\mu \nu}(t, - \vec{r}) = - F_{\mu \nu}(t, \vec{r}) \end{matrix} \hspace{1 cm} (\text{for } t \text{ fixed, }r\to \infty.)
\end{equation}
Therefore, at large $r$ the $\int_{S^2}$ integral will vanish due to the cancellation of antipodal points and $\omega$ is finite!\footnote{If one considers a variation of the form $\delta_2 A_i = \partial_i \lambda$ where $\lambda = \lambda(z^A)$ then it can be shown using the divergence theorem that $\omega$ remains finite for arbitrary $\lambda$. In particular, odd $\lambda$'s will render $\omega = 0$ and even $\lambda$'s will render $\omega$ finite. This implies that we should only demand that $A_i$ is even at $i^0$ up to a pure gauge piece, as discussed in \cite{henneaux2018asymptotic}. It turns out that only the even $\lambda$'s should be thought of as `physical' gauge transformations at spatial infinity, which is consistent with the story we saw at null infinity. }

The final thing to show is that $\omega$ doesn't depend on which timeslice we evaluate it. To show this, we must check that the integral of $\omega^\mu$ over the cylindrical surface $C$, which connects $\Sigma_1$ and $\Sigma_2$, vanishes:

\begin{align}
    \int_C d\Sigma_\mu \omega^\mu &= -\lim_{r \to \infty} \int_{t_1}^{t_2} dt \, r^2 \int_{S^2} d^2 z \sqrt{\gamma} \Big( \underbrace{\delta_1 F^{r \mu} }_{\mathcal{O}(1/r^2)} \underbrace{\delta_2 A_\mu}_{\mathcal{O}(1/r)} - \underbrace{\delta_2 F^{r \mu}}_{\mathcal{O}(1/r^2)} \underbrace{ \delta_1 A_\mu}_{\mathcal{O}(1/r)}  \Big) \\
    &= 0.
\end{align}
For more thorough treatments of the symplectic structure of EM at $i^0$, see \cite{henneaux2018asymptotic,campiglia2017asymptotic,Prabhu:2018gzs}.

\newpage
\subsection{When is the covariant phase space formalism useful?}

The covariant phase space formalism is admittedly a little funny. Usually, Hamiltonian mechanics is presented as a way to express the equations of motion, with the Hamiltonian function $H$ generating a set of first-order differential equations to be solved. However, in the covariant phase space formalism, every point in the phase space is already a full solution to the equations of motion. There is still a Hamiltonian function $H$, but its interpretation is different. Instead of giving the equations of motion, it will instead generate a transformation which translates the \textit{whole} solution forwards in time.

The thing about the covariant phase space formalism is that it's very well-suited for studying how symmetry transformations act on full solutions to the equations of motion, but very poorly-suited for actually helping one solve the equations of motion to begin with. This means that it's best to use when one already has a good idea what the space of solutions looks like. This is the very situation we find ourselves in in this note.

\subsection{Why are large gauge transformations physical?}

We are told in textbooks that gauge transformations are not true ``physical'' symmetry transformations, but are instead mere redundancies in our mathematical description. However, in this note it has been claimed repeatedly that large gauge transformations \textit{are} physical symmetries while small gauge transformations are not. It is time to confront the issue head on. What makes a transformation ``physical?''

Consider, for a moment, the phase space of the universe. Total momentum conservation follows from translational symmetry. However, a translation moves every object in the universe together, keeping their relative positions fixed. Is this transformation physical? A philosopher would probably say `no.' Due to the fact that all observers are a part of the universe, there is no experiment they could do to detect the difference between the universe and the universe shifted 10 feet to the left.

However, the philosopher is wrong. The only true criterion which a transformation must satisfy in order to count as ``physical'' is that it must correspond to a \textit{non-degenerate direction} with respect to the symplectic form. A degenerate direction is any tangent vector $\delta_{\rm deg} P$ based at $P$ such that
\begin{equation}
\begin{split}
    \omega_P(\delta_{\rm deg} P, \delta P) = 0 \hspace{0.5 cm} \text{ for all } \delta P.
\end{split}
\end{equation}

\begin{figure}[H]
\begin{center}

\tikzset{every picture/.style={line width=0.75pt}} 

\begin{tikzpicture}[x=0.75pt,y=0.75pt,yscale=-1,xscale=1]

\draw   (69.5,42) -- (188.5,42) -- (160,129) -- (41,129) -- cycle ;
\draw [line width=2.25]    (41,129) -- (67.94,46.75) ;
\draw [shift={(69.5,42)}, rotate = 468.14] [fill={rgb, 255:red, 0; green, 0; blue, 0 }  ][line width=0.08]  [draw opacity=0] (16.07,-7.72) -- (0,0) -- (16.07,7.72) -- (10.67,0) -- cycle    ;
\draw  [dash pattern={on 4.5pt off 4.5pt}]  (91,97) -- (119.5,10) ;
\draw    (119.5,10) -- (238.5,10) ;
\draw    (210,97) -- (238.5,10) ;
\draw  [dash pattern={on 4.5pt off 4.5pt}]  (91,97) -- (210,97) ;
\draw    (160,129) -- (210,97) ;
\draw    (188.5,42) -- (238.5,10) ;
\draw    (69.5,42) -- (119.5,10) ;
\draw [line width=2.25]    (41,129) -- (155,129) ;
\draw [shift={(160,129)}, rotate = 180] [fill={rgb, 255:red, 0; green, 0; blue, 0 }  ][line width=0.08]  [draw opacity=0] (16.07,-7.72) -- (0,0) -- (16.07,7.72) -- (10.67,0) -- cycle    ;
\draw [line width=2.25]    (41,129) -- (86.79,99.7) ;
\draw [shift={(91,97)}, rotate = 507.38] [fill={rgb, 255:red, 0; green, 0; blue, 0 }  ][line width=0.08]  [draw opacity=0] (16.07,-7.72) -- (0,0) -- (16.07,7.72) -- (10.67,0) -- cycle    ;

\draw (65.5,27.57) node [anchor=east] [inner sep=0.75pt]  [font=\large]  {$\delta _{\mathrm{deg}} P$};
\draw (26.11,132.69) node  [font=\Large]  {$P$};
\draw (365.07,60) node [anchor=east] [inner sep=0.75pt]   [align=left] {\begin{minipage}[lt]{60.57pt}\setlength\topsep{0pt}
\begin{center}
phase space\\volume
\end{center}

\end{minipage}};
\draw (372,57.2) node [anchor=west] [inner sep=0.75pt]    {$=\ 0$};

\end{tikzpicture}

\end{center}
\caption{\label{fig:parallelpiped} A parallelpiped in phase space with one leg spanned by a degenerate direction $\delta_{\rm deg} P$ has zero phase space volume.}
\end{figure}
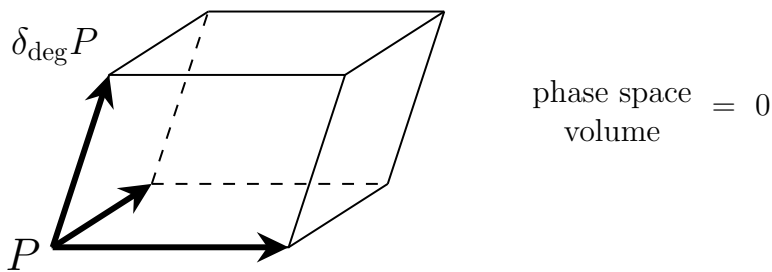

Recall, for instance, that the phase-space volume form comes from repeatedly wedging the symplectic form with itself, as in \eqref{vol}. This means that the volume of a parallelpiped in phase space which has as one leg a degenerate direction $\delta_{\rm deg} P$ will be zero, as is shown in Figure \ref{fig:parallelpiped}. This implies that the point $P$ and $P + \delta_{\rm deg} P$ are physically equivalent. The orbits generated by flowing along these degenerate directions are called ``gauge orbits.'' A whole gauge orbit counts as a single physical state.

Note that one must invert the symplectic form in order to compute Poisson brackets, and in order for the symplectic form to be invertible it cannot be degenerate. One must therefore choose a slice which cuts through each gauge orbit exactly once.

\begin{figure}[H]
\begin{center}

\tikzset{every picture/.style={line width=0.75pt}} 

\begin{tikzpicture}[x=0.75pt,y=0.75pt,yscale=-1,xscale=1]

\draw   (59,36) -- (278.26,36) -- (278.26,177.3) -- (59,177.3) -- cycle ;
\draw    (187.6,36.27) -- (139.6,176.87) ;
\draw    (147.6,36.27) -- (99.6,176.87) ;
\draw    (107.6,36.27) -- (59.6,176.87) ;
\draw    (227.6,36.27) -- (179.6,176.87) ;
\draw    (267.6,36.27) -- (219.6,176.87) ;
\draw    (67.6,36.27) -- (58.6,62.67) ;
\draw    (278,123.67) -- (259.6,176.87) ;
\draw    (59.2,136) .. controls (128.8,101.07) and (199.8,180.27) .. (278.6,139.87) ;
\draw  [fill={rgb, 255:red, 0; green, 0; blue, 0 }  ,fill opacity=1 ] (149.63,136.82) .. controls (149.63,134.85) and (151.22,133.25) .. (153.2,133.25) .. controls (155.17,133.25) and (156.77,134.85) .. (156.77,136.82) .. controls (156.77,138.79) and (155.17,140.39) .. (153.2,140.39) .. controls (151.22,140.39) and (149.63,138.79) .. (149.63,136.82) -- cycle ;
\draw [line width=1.5]    (153.2,135.82) -- (169.3,89.56) ;
\draw [shift={(170.61,85.78)}, rotate = 469.19] [fill={rgb, 255:red, 0; green, 0; blue, 0 }  ][line width=0.08]  [draw opacity=0] (13.4,-6.43) -- (0,0) -- (13.4,6.44) -- (8.9,0) -- cycle    ;

\draw (125,10) node [anchor=north west][inner sep=0.75pt]   [align=left] {Phase space};
\draw (236.58,90.7) node  [font=\normalsize,rotate=-289] [align=left] {gauge orbit};
\draw (166.44,150.64) node  [rotate=-11.69] [align=left] {gauge fixed slice};
\draw (153.82,131.13) node [anchor=south east] [inner sep=0.75pt]    {$P$};
\draw (153.93,76.53) node  [font=\normalsize]  {$\delta _{\mathrm{deg}} P$};

\end{tikzpicture}

\end{center}
\caption{\label{fig:gauge_phase} Degenerate tangent vectors point along gauge orbits. Each gauge orbit corresponds to a single physical state. Fixing a gauge slice is necessary to make the symplectic form non-degenerate and therefore invertible.}
\end{figure}
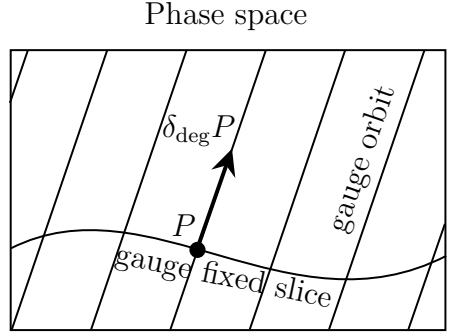

So, our universe-wide translation counts as a physical symmetry simply because the phase-space-vector-field it corresponds to is non-degenerate with respect to the symplectic form. Is this the case for large gauge transformations as well?

Yes. We can verify explicitly that large gauge transformations do not correspond to degenerate tangent vectors while small gauge transformations do. Showing this simply requires is an integration by parts. Let's plug $\delta_2 A_\mu = \partial_\mu \lambda$ into the symplectic form. Note that $\delta_2 F_{\mu \nu} = 0$, so
\begin{align}
    \omega[A; \delta_1 A, \partial \lambda] &= - \int_\Sigma d \Sigma_\mu (\delta_1 F^{\mu \nu}) \nabla_\mu \lambda \\
    &= - \int_S dS_{\mu \nu} \; \lambda \, \delta_1 F^{\mu \nu}
\end{align}
where $S = \partial \Sigma$. Here we used $\nabla_\mu F^{\mu \nu} \approx j^\nu$ which implies $\nabla_\mu \delta F^{\mu \nu} \approx \delta j^\nu = 0$ because $j^\mu$ is a background current. If $\lambda$ is compact, meaning it vanishes on $S$, then $\omega = 0$. However, if $\lambda$ is a large gauge transformation, then $\omega \neq 0$. This is exactly what we wanted to show.

\section{Canonical Edge Modes and Vacuum Transitions}

\subsection{Vacuum transitions}

\begin{figure}[H]
\begin{center}

\tikzset{every picture/.style={line width=0.75pt}} 

\begin{tikzpicture}[x=0.75pt,y=0.75pt,yscale=-1,xscale=1]

\draw  [color={rgb, 255:red, 187; green, 187; blue, 187 }  ,draw opacity=1 ][fill={rgb, 255:red, 187; green, 187; blue, 187 }  ,fill opacity=1 ][line width=1.5]  (55.6,187.07) -- (70.38,201.84) -- (62.71,209.08) -- (61.13,207.25) -- (56.63,199.25) -- (55.63,195.83) -- cycle ;
\draw  [color={rgb, 255:red, 187; green, 187; blue, 187 }  ,draw opacity=1 ][fill={rgb, 255:red, 187; green, 187; blue, 187 }  ,fill opacity=1 ] (54.93,186.4) -- (155.73,85.6) -- (171.46,101.33) -- (70.66,202.13) -- cycle ;
\draw  [color={rgb, 255:red, 187; green, 187; blue, 187 }  ,draw opacity=1 ][fill={rgb, 255:red, 187; green, 187; blue, 187 }  ,fill opacity=1 ][line width=1.5]  (81.13,86.75) -- (94.63,97.5) -- (53.21,138.58) -- (52.38,129.75) -- (50.71,124.08) -- (49.21,119.08) -- cycle ;
\draw [color={rgb, 255:red, 187; green, 187; blue, 187 }  ,draw opacity=1 ]   (54.67,186.67) -- (156,85.33) ;
\draw [color={rgb, 255:red, 187; green, 187; blue, 187 }  ,draw opacity=1 ]   (62.52,210.75) -- (172.22,101.04) ;
\draw [color={rgb, 255:red, 187; green, 187; blue, 187 }  ,draw opacity=1 ]   (47.83,119.5) -- (119.33,48) ;
\draw [color={rgb, 255:red, 187; green, 187; blue, 187 }  ,draw opacity=1 ]   (52.67,140) -- (131.33,61.33) ;
\draw [line width=1.5]    (87.63,251.75) -- (62.63,210.75) ;
\draw [shift={(75.13,231.25)}, rotate = 418.63] [fill={rgb, 255:red, 0; green, 0; blue, 0 }  ][line width=0.08]  [draw opacity=0] (11.61,-5.58) -- (0,0) -- (11.61,5.58) -- cycle    ;
\draw [line width=1.5]    (62.63,210.75) .. controls (53.76,195.02) and (55.33,200) .. (54.67,186.67) ;
\draw [shift={(56.22,199.4)}, rotate = 421.03999999999996] [fill={rgb, 255:red, 0; green, 0; blue, 0 }  ][line width=0.08]  [draw opacity=0] (11.61,-5.58) -- (0,0) -- (11.61,5.58) -- cycle    ;
\draw [line width=1.5]    (54.67,186.67) -- (52.67,140) ;
\draw [shift={(53.67,163.33)}, rotate = 447.55] [fill={rgb, 255:red, 0; green, 0; blue, 0 }  ][line width=0.08]  [draw opacity=0] (11.61,-5.58) -- (0,0) -- (11.61,5.58) -- cycle    ;
\draw  [color={rgb, 255:red, 187; green, 187; blue, 187 }  ,draw opacity=1 ][fill={rgb, 255:red, 187; green, 187; blue, 187 }  ,fill opacity=1 ] (56.88,110.83) -- (118.85,48.85) -- (131.19,61.19) -- (69.22,123.16) -- cycle ;
\draw [line width=1.5]    (82.38,11.72) -- (229.62,158.95) ;
\draw [line width=1.5]    (52.67,140) .. controls (52.38,137.75) and (48.63,121.25) .. (46.42,116.92) ;
\draw [shift={(49.99,128.2)}, rotate = 436.28] [fill={rgb, 255:red, 0; green, 0; blue, 0 }  ][line width=0.08]  [draw opacity=0] (11.61,-5.58) -- (0,0) -- (11.61,5.58) -- cycle    ;
\draw [line width=1.5]    (46.42,116.92) -- (28.63,83.75) ;
\draw [shift={(37.52,100.33)}, rotate = 421.78999999999996] [fill={rgb, 255:red, 0; green, 0; blue, 0 }  ][line width=0.08]  [draw opacity=0] (11.61,-5.58) -- (0,0) -- (11.61,5.58) -- cycle    ;

\draw (200.03,126.5) node [anchor=south west] [inner sep=0.75pt]  [font=\small]  {$A^{(0)}_{A} =\partial _{A} \Phi _{1} \ $};
\draw (146.03,73.44) node [anchor=south west] [inner sep=0.75pt]  [font=\small]  {$A^{(0)}_{A} =\partial _{A} \Phi _{2} \ $};
\draw (102.27,29.3) node [anchor=south west] [inner sep=0.75pt]  [font=\small]  {$A^{(0)}_{A} =\partial _{A} \Phi _{3} \ $};
\draw (88.04,91.82) node  [rotate=-315] [align=left] {\textbf{radiation}};
\draw (111.2,145.86) node  [rotate=-315] [align=left] {\textbf{radiation}};
\draw (231.62,162.35) node [anchor=north west][inner sep=0.75pt]  [font=\Large]  {$\mathcal{I}^{+}$};

\end{tikzpicture}

\end{center}
\caption{\label{fig:vacuum_transition} When charged particles change their velocity, they release radiation. At $\mathcal{I}^+$, this induces a vacuum transition. While $A^{(0)}_A$ will be pure gauge before and after the radiation hits $\mathcal{I}^+$, with $A^{(0)}_A = \partial_A \Phi$ for some angle dependent function $\Phi$, the function $\Phi$ itself will change and $\Delta \Phi$ will only depend on the particle's initial and final velocity.}
\end{figure}
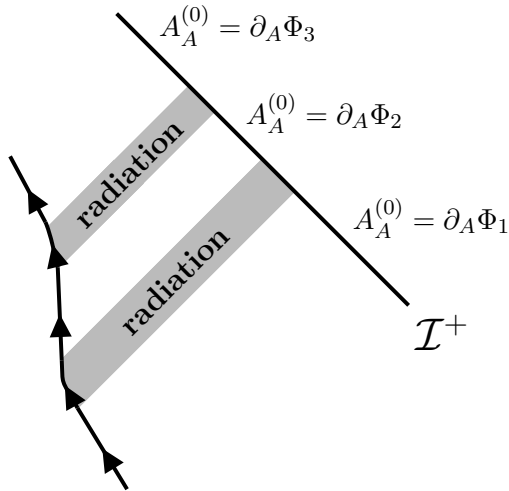

From equation \eqref{AA_orig}, we can see that when the bulk contains a collection of particles, labelled by $k = 1 \ldots N$ with charges $e_k$ and constant velocities $\vec{\beta}_k$, $A_A$ at large $r$ is
\begin{equation}\label{AA_vel}
    A_A^{(0)} = \partial_A \sum_{k} \frac{e_k}{4 \pi} \log( 1 - \vec{\beta}_k   \cdot \hat{n}(z^A) ).
\end{equation}
(The other components $A_u$ and $A_r$ die off at large $r$.) We can see an two interesting features of $A^{(0)}_A$: (i) it is pure gauge, meaning $A^{(0)}_A = \partial_A \Phi$ for some $\Phi$, and (ii) $\Phi$ only depends of the velocities of the charged particles. If the particles accelerate and change their velocity, then $\Phi$ will change. This is shown in Figure \ref{fig:vacuum_transition}.

Actually, we should be careful because \eqref{AA_vel} is not a gauge-invariant equation. $\Phi$ can always be changed if we perform a large gauge transformation which sends $\Phi \mapsto \Phi + \lambda$. However, because $\lambda$ must be $u$-independent, the \textit{difference} between $\Phi$ at different values of $u$ will be gauge invariant.

\begin{figure}
\input{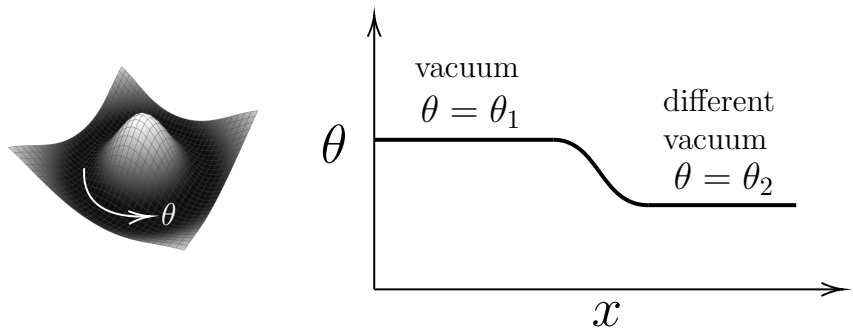}
\caption{\label{fig:potential} In the well-known example of spontaneous $U(1)$ symmetry breaking, it is possible to be in different vacua at different regions of space.}
\end{figure}

This behavior is therefore exactly the same as the kind seen in spontaneous symmetry breaking. By analogy, consider a basic ``Mexican hat'' potential with a $U(1)$ circle of vacua labelled by an angle $\theta$. It is possible to have field configurations which are in different vacua at different locations in space, as in Figure \ref{fig:potential}.

\begin{figure}
\input{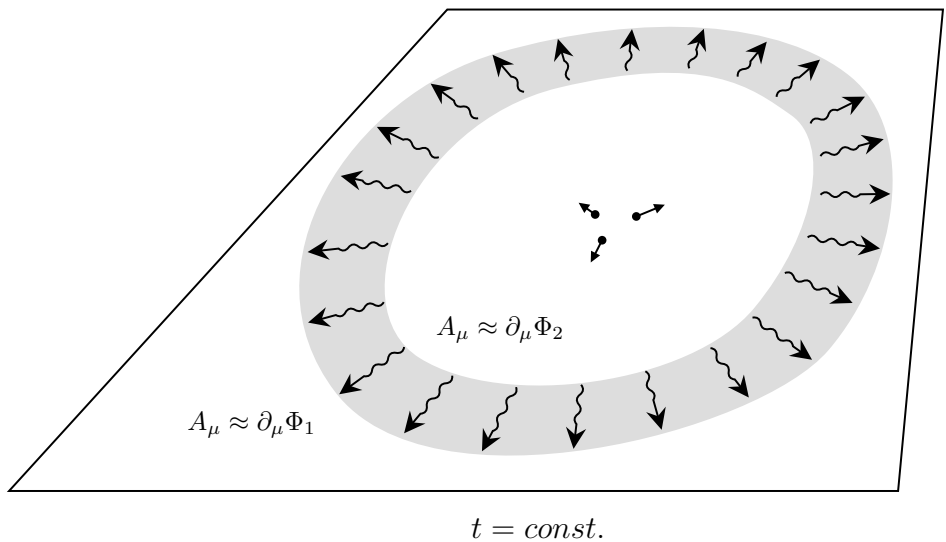}
\caption{\label{fig:timeslice_vacuum} A spherical shell of radiation expands outwards at the speed of light. On either side of this shell, the fields are Coulombic and approximately pure gauge. The shell represents a domain wall between spontaneously broken large gauge symmetry vacua.}
\end{figure}

Returning to electromagnetism, one can consider a picture like that of Figure \ref{fig:timeslice_vacuum} which depicts a set of charged particles which scattered long ago. A spherical shell of radiation created by the scattering event expands outward at the speed of light. On either side of the expanding front, the fields are Coulombic and rapidly approaching a vanishing $F_{\mu \nu}$. Roughly speaking, $A_\mu$ will be pure gauge in these Coulombic regions, meaning it will be in a vacuum state. However, it will be in different vacua on either side of the radiation. Therefore, radiation comprises a domain wall separating spontaneously broken large gauge vacua.

\subsection{The soft photon mode $N_A$}

Let us finally explore the connection between what we have called the ``soft charge'' $Q_\lambda$ and actual long wavelength soft radiation. At $\mathcal{I}^+$, the soft-part of the soft charge is
\begin{align}
    Q_\lambda^{+,S} &= \int_{\mathcal{I}^+} \measnor F_{uA} D^A \lambda
\end{align}
Notice that the only term in the integrand which actually depends on $u$ is $F_{uA}$. Therefore, if we define the ``soft photon mode'' $N_A$ as
\begin{equation}
    N_A \equiv \int_{\mathcal{I}^+_{\hat n}} du F_{uA}
\end{equation}
which is an integral over $u$ at a fixed angle, we can re-express $Q_\lambda^{+,S}$ as the 2-sphere integral
\begin{align}
    Q_\lambda^{+,S} = \int_{S^2} d^2z \sqrt{\gamma} N_A D^A \lambda.
\end{align}

Now, why did we name $N_A$ the ``soft photon mode,'' exactly? It's because it's a measure of how many soft photons hit $\mathcal{I}^+$ per angle over its entire history! Recall that in Lorenz gauge, solutions to the vacuum equations of motion can be Fourier decomposed into linear combinations of plane waves of the form
\begin{align}
    A_\mu &= \vep_\mu e^{-ikx} \\
    F_{\mu \nu} &= -i (k_\mu \vep_\nu - k_\nu \vep_\mu) e^{-ikx}
\end{align}
where $\vep_\mu$ is a polarization vector. $F_{uA}$ in particular is
\begin{equation}
    F_{uA} = -i (\partial_A \hat n \cdot \vec{\vep})\,  k_0 \, e^{-ikx}
\end{equation}
where $\vec{\vep}$ is the spatial component of $\vep_\mu$.

In section \ref{sec_resid}, we argued that plane waves on $\mathcal{I}^+$ can be expressed (in a distributional sense) as
\begin{equation}
    \eval{e^{-ikx}}_{\mathcal{I}^+} = \begin{cases} e^{-ik_0 u } & \text{ if } \hat n = \hat{k} \\
    0 & \text{ for any other angle}
    \end{cases}.
\end{equation} \vskip .2cm
\noindent Therefore, we can measure the amount of radiation which hits $\mathcal{I}^+$ at an angle $\hat n$, frequency $\omega$, polarization vector $\vec{\vep} = \partial_A \hat n$ as

\begin{equation}
    \begin{matrix} \text{``amount'' of radiation emitted} \\ \text{in the $\hat n$-direction with} \\
    \text{frequency $\omega$ polarized along $\partial_A$} \end{matrix} =  \int_{\mathcal{I}^+_{\hat n}} du \; e^{i \omega u} F_{u A}.
\end{equation} \vskip 0.3cm

\noindent $N_A$ is just the $\omega \to 0$ limit of the above expression, confirming that it does indeed measure the ``amount'' of soft radiation emitted per angle!

One might wonder ``where'' exactly soft photons are located in the spacetime. Note that if one wants to find the Fourier components of $A_\mu$, one must compute the Fourier transform of $A_\mu$ which is an integral over the full 4-dimensional spacetime. This is a highly ``non-local'' computation, depending on the full history of $A_\mu$. So, while $A_\mu$ itself evolves in a manifestly local way, the designation of how many ``photons'' have been emitted in a spacetime is a non-local notion. What's more, infinite wavelength soft photons are the most non-local of all. So, while $F_{uA}$ itself is only non-zero at large $r$ when the news that a particle has accelerated finally reaches $\mathcal{I}^+$, a soft-photon should be understood as being de-localized over all of null infinity.

\begin{figure}[H]
\centering
\includegraphics[width=0.6\textwidth]{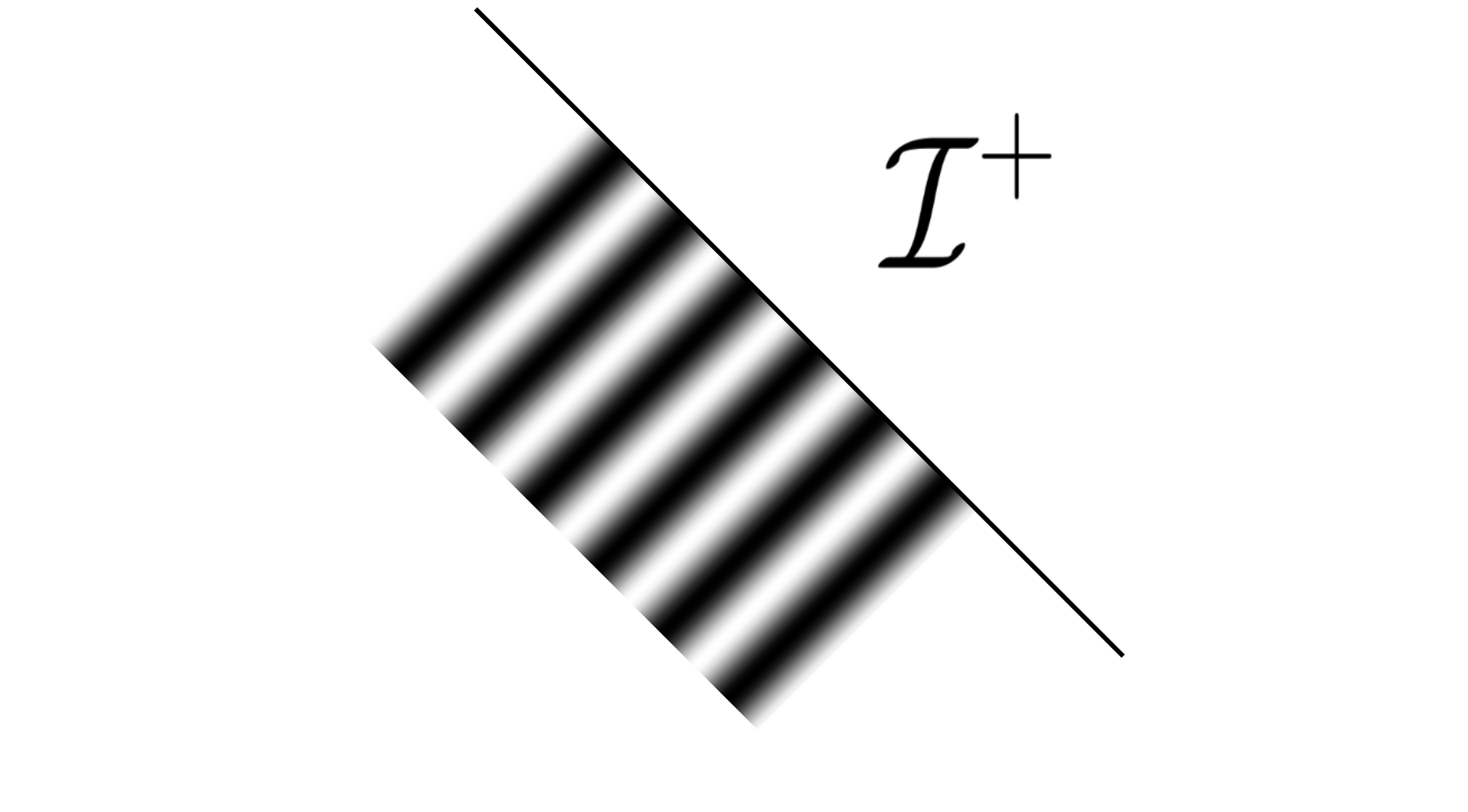}
\caption{\label{fig:wave_penrose} A plane wave hitting future null infinity. The shade of gray represents the phase of the wave. A soft photon can be pictured by taking the infinite wavelength limit.}
\end{figure}

\noindent In particular, it is possible for accelerating charged particles to both create \textit{and destroy} soft photons. For instance, if a charge starts travelling at velocity $\vec{\beta}_1$, then accelerates to $\vec{\beta}_2$, and then accelerates back to its original velocity $\vec{\beta}_1$, no net soft photons will have been produced despite the fact that radiation will have been released during both acceleration events.

Because we know that the amount of soft radiation released only depends on the initial and final velocities of the charged particles, we might expect that $N_A$ can be expressed in terms of of quantities at $\mathcal{I}^+_\pm$. This is indeed the case. Recalling the definitions of $\phi_+$ and $\phi_-$
\begin{equation}
    \eval{A_A}_{\mathcal{I}^+_+} = \partial_A \phi_+, \hspace{0.5 cm} \eval{A_A}_{\mathcal{I}^+_-} = \partial_A \phi_- 
\end{equation}
we can see that
\begin{align}
    N_A &= \int_{\mathcal{I}^+_{\hat n}} du F_{uA} = \int_{\mathcal{I}^+_{\hat n}} du ( \partial_u \underbrace{A_A}_{O(1)} - \partial_A \underbrace{A_u}_{O(1/r)} ) \\
    &= A_A^{(0)}(u = \infty) - A_A^{(0)}(u = -\infty) \\
    &= \partial_A (\phi_+ - \phi_-) \\
    &= \partial_A N
\end{align}
where
\begin{equation}
    N \equiv \phi_+ - \phi_-.
\end{equation}
Amazingly we have found that $N_A$ is the total derivative of a single function $N$ which is equal to the difference in the large gauge vacuum modes between early and late times!

Interestingly, $Q_\lambda^+$, which a priori is an integral over a 3-dimensional Cauchy surface $\mathcal{I}^+$, only \textit{truly} depends on quantities defined at $\mathcal{I}^+_\pm$. This can be seen as an manifestation of Noether's 1.5th theorem.

\subsection{Poisson bracket algebra at null infinity}

Let us compute the algebra of Poisson brackets of our phase space at null infinity. The symplectic form becomes, using \eqref{omega_em},
\begin{align}
    \omega &= \int_{\mathcal{I}^+} d \Sigma_\mu \omega^\mu = \int_{\mathcal{I}^+} \meas \omega_u \\
    &= \int_{\mathcal{I}^+} \measnor \gamma^{AB} (\delta F_{uA} \wedge \delta A_B )\\
    &= \int_{\mathcal{I}^+} \measnor \gamma^{AB} \big(\delta (\partial_u A_A^{(0)})\wedge \delta A_B^{(0)} \big) \label{omega_almost}
\end{align}
One may be concerned that we have excluded timelike infinity from our surface of integration. However, at $i^+$ the field is Coulombic and completely specified by locations of the charged particles via the equations of motion. Therefore, the on-shell variations $\delta A_\mu$ are 0 at timelike infinity and do not contribute to the symplectic form. The only true degrees of freedom of the electromagnetic field are the radiative ones which hit $\mathcal{I}$.\footnote{In Lorenz gauge, the e.o.m. becomes $\partial^2 A^\mu = j^\mu$ which is an inhomogeneous linear differential equation. One can always add a solution to the homogeneous vacuum equation to $A_\mu$ which represents adding pure plane-wave radiation to the spacetime. In particular, the \textit{only} possible on-shell variations $\delta A_\mu$ are radiative.}

Amazingly, at null infinity the symplectic structure only depends on $A_A^{(0)}$ and its $u$-derivative. In order to clarify what role $N_A$ plays in the symplectic structure, let us separate $A_A^{(0)}$ into a ``large gauge'' $u$-independent piece and a $u$-dependent piece $\hat A_A$ as follows:
\begin{align}
    \phi \equiv \frac{1}{2}(\phi_+ + \phi_-), \hspace{1.5 cm}
    A^{(0)}_A(u,z) = \hat A_A(u,z) + \partial_A \phi(z).
\end{align}
Plugging the above expression for $A^{(0)}_A$ into \eqref{omega_almost}, we arrive at our final expression of
\begin{equation}\label{edge_sym_form}
    \omega = \int_{\mathcal{I}^+} \measnor \gamma^{AB} \delta( \partial_u \hat A_A) \wedge \delta \hat A_B + \int_{S^2} d^2 z \sqrt{\gamma} \gamma^{AB} \delta N_A \wedge \delta ( \partial_B \phi)
\end{equation}
Apparently, $\phi$ is the symplectic partner of the soft photon mode $N_A$.

We can now find the Poisson brackets by inverting the symplectic form \eqref{edge_sym_form} and using \eqref{poisson_useful}. To begin, $\hat A_A$ has the Poisson brackets of
\begin{align}
    \Big\{ \partial_u \hat A_A(u_1, z_1), \hat A_B(u_2, z_2) \Big\} &= - \frac{\gamma_{AB}}{\sqrt{\gamma}} \delta^2(z_1 - z_2) \, \delta(u_1 - u_2)\\
    \Big\{ \hat A_A(u_1, z_1), \hat A_B(u_2, z_2) \Big\} &= - \frac{\gamma_{AB}}{\sqrt{\gamma}} \delta^2(z_1 - z_2) \, \frac{1}{2} \mathrm{sign}(u_1 - u_2).
\end{align}
To get the second line, we integrated the first line with respect to $u$ and found the constant of integration using the fact that Poisson brackets must be anti-symmetric.

The Poisson bracket of $N_A$ and $\partial_B\phi$ is given by
\begin{align}\label{pb_na_phib}
    \Big\{ N_A(z_1), \partial_B \phi(z_2) \Big\} &= -\frac{\gamma_{AB}}{\sqrt{\gamma}} \delta^2(z_1 - z_2)
\end{align}
and the following Poisson brackets vanish:
\begin{align}
    \Big\{ \partial_u \hat A_A(u_1, z_1), N_B(z_2) \Big\} &= 0 \\ \Big\{ \partial_u \hat A_A(u_1, z_1), \partial_B \phi(z_2) \Big\} &= 0.
\end{align}
The soft charge is equal to
\begin{align}
    Q^+_\lambda &= Q^{+,H}_\lambda + \int_{S^2} d^2 z \sqrt{\gamma} N_A D^A\lambda^+.
\end{align}
The hard part of the soft charge only depends on the background charge current so its Poisson brackets with EM field variables is zero.
\begin{equation}
    \Big\{ Q_\lambda^{+,H}, \hat A_A(u,z) \Big\} = \Big\{ Q_\lambda^{+,H}, N_A(z) \Big\} = \Big\{ Q_\lambda^{+,H}, \partial_A \phi(z) \Big\} = 0.
\end{equation}
Therefore, we can immediately see that
\begin{equation}
    \Big\{ Q_\lambda^+, A_A^{(0)}(u,z) \Big\} = \Big\{ Q_\lambda^+, \partial_A \phi(z) \Big\} = -\partial_A \lambda^+(z)
\end{equation}
and the soft charge generates large gauge transformations exactly as one would hope! Keep in mind that this large gauge transformation acts on the entire spacetime solution of $A_\mu$ as a whole.

As a brief aside, it is interesting to note that the soft charge is only constructed out of $N_A$ which itself only depends on the difference $\phi_+ - \phi_-$. Meanwhile, the goldstone mode $\phi$ only depends on the sum $\phi_+ + \phi_-$. If we use the intuition presented in Figure \ref{fig:90deg}, we can see why this is natural. A large $u$-independent gauge transformation preserves the difference of $\phi_+$ and $\phi_-$ while changing the sum. Therefore, if we flow along the Hamiltonian vector field $X_{Q_\lambda}$, the difference $\phi_+ - \phi_-$ will remain constant along the flow while the sum $\phi_+ + \phi_-$ will not. 

While the hard part $Q^{+,H}_\lambda$ did not play a role in the Poisson algebra of our EM field, it would play a role if our charge current $j^\mu$ was the global $U(1)$ Noether current of a dynamical matter field $\psi$ with Poisson bracket
\begin{equation}
    \Big\{j^0(t, \vec{x}), \psi(t, \vec{y}) \Big\} = i e \delta^3(\vec{x} - \vec{y}) \psi(t, \vec{y}).
\end{equation}
The hard part would therefore act on $\psi(x)$ as
\begin{equation}
    \Big\{ Q^{H}_\lambda, \psi(x) \Big\} = ie \lambda(x) \psi(x).
\end{equation}
This means that $Q^{+,H}_\lambda$ generates the action of large gauge transformations on the matter field just as $Q^{+,S}_\lambda$ generates the action of large gauge transformations on the EM field.

What about the Poisson bracket of $N$ and $\phi$, namely $\{ N, \phi \}$? To find it, we can plug $N_A = \partial_A N$ into the symplectic form \eqref{edge_sym_form}, integrate by parts and read off
\begin{equation}
    D^2 \Big\{ N(z_1), \phi(z_2) \Big\} = \frac{1}{\sqrt{\gamma}} \delta^2(z_1 - z_2).
\end{equation}
Solving this equation would give us $\{ N, \phi \}$.
The only issue is that this equation cannot actually be solved. This can be seen by integrating $z_1$ over the 2-sphere and using the divergence theorem---the LHS will be 0 but the RHS will be 1. Having said that, the modified equation
\begin{equation}
    D^2 G_{S^2}(z_1 - z_2) = \frac{1}{\sqrt{\gamma}} \delta^2(z_1 - z_2) - \frac{1}{4 \pi}
\end{equation}
can be solved, and its solution is
\begin{equation}
    G_{S^2}(z_1 - z_2) = \frac{1}{4 \pi} \log\big(|\hat n(z_1) - \hat n(z_2) |^2\big).
\end{equation}
So what is going on? What is $\{ N, \phi \}$?

The resolution involves the fact that $\phi$ is really only defined up to a constant, so $\phi \sim \phi + c$. Therefore, the value of $\phi$ at a point $z$ is not actually a valid function on the phase space. However, the difference of $\phi$ at two different points, is. This would imply
\begin{equation}
    \Big\{ N(z_1) , \phi(z_2) - \phi(z_3) \Big\} = \frac{1}{4 \pi} \log\big(|\hat n(z_1) - \hat n(z_2)|^2\big) - \frac{1}{4 \pi} \log\big(|\hat n(z_1) - \hat n(z_3)|^2\big)
\end{equation}
resolving the apparent paradox. We can see that the derivative of $\phi$ has a well-defined Poisson brackets because it is a limit of differences of $\phi$ at two points.

Anyway, aside from this thorny issue of addition by a constant, one may get the feeling that the phase space of electromagnetism is a little too simple. The complete phase space is given by only four functions (and two of them don't even depend on $u$!)
\begin{equation}
\begin{split}\label{cauchy_plus}
    (\text{Cauchy data on }\mathcal{I}^+) = \big( \hat A_A(u, z), N(z), \phi(z) \big).
\end{split}
\end{equation}
However, actually finding the full solution to the equations of motion based on this Cauchy data is a non-trivial task. Therefore, all the complication resides in the map
\begin{equation}
\begin{split}
    (\text{Cauchy data on }\mathcal{I}^-) \longrightarrow (\text{Cauchy data on }\mathcal{I}^+)
\end{split}
\end{equation}
which depends on the exact knowledge of the locations of the charged particles in the bulk at all times.

To conclude this section, let me point out that we were able to invert the symplectic form and solve for the Poisson brackets without ever choosing a gauge slice. Shouldn't this have been impossible, because without doing so the symplectic form isn't be invertible? Interestingly, this issue was avoided because we pushed our Cauchy slice to $\mathcal{I}^+$ where trivial gauge symmetries do not act and the symplectic form really is invertible.

\section{Electromagnetic Memory}\label{sec_memory}

\begin{figure}[H]
\begin{center}

\begin{tikzpicture}[x=0.75pt,y=0.75pt,yscale=-1,xscale=1]

\draw  [fill={rgb, 255:red, 187; green, 187; blue, 187 }  ,fill opacity=1 ] (16.83,30.67) -- (40.83,6.67) -- (106.83,6.67) -- (106.83,62.67) -- (82.83,86.67) -- (16.83,86.67) -- cycle ; \draw   (106.83,6.67) -- (82.83,30.67) -- (16.83,30.67) ; \draw   (82.83,30.67) -- (82.83,86.67) ;
\draw    (40.83,6.67) -- (40.83,62.67) ;
\draw    (40.83,62.67) -- (106.83,62.67) ;
\draw    (40.83,62.67) -- (16.83,86.67) ;
\draw  [fill={rgb, 255:red, 0; green, 0; blue, 0 }  ,fill opacity=1 ] (58,49.53) .. controls (58,48.01) and (59.23,46.78) .. (60.75,46.78) .. controls (62.27,46.78) and (63.5,48.01) .. (63.5,49.53) .. controls (63.5,51.05) and (62.27,52.28) .. (60.75,52.28) .. controls (59.23,52.28) and (58,51.05) .. (58,49.53) -- cycle ;

\end{tikzpicture}

\end{center}
\caption{\label{fig:memory_vat} A test charge in a tank of viscous fluid.}
\end{figure}
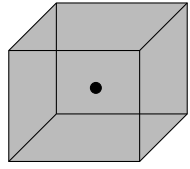

We will now describe a thought experiment, developed in \cite{pasterski2017asymptotic,bieri2013electromagnetic}, which can detect large gauge vacuum transitions. The basic measurement device, depicted in Figure \ref{fig:memory_vat}, is a test charge immersed in a tank of viscous fluid.
\noindent Imagine this fluid supplies a linear damping force (i.e. Stokes' drag) such that the equation of motion for the position of the particle is
\begin{equation}\label{newton2}
    m \frac{d^2}{dt^2} \vec{x} = \vec{F}_{\rm ext} - b \frac{d}{dt} \vec{x}
\end{equation}
where $\vec{F}_{\rm ext}$ is the external force on the particle and $b$ is a proportionality constant which sets the strength of the damping force. Integrating both sides of \eqref{newton2}, we get
\begin{equation}
    m \Delta \vec{v} = \int_{-\infty}^{\infty} dt \vec{F}_{\rm ext} - b \Delta \vec{x}.
\end{equation}
Assume our particle starts at rest and $\vec{F}_{\rm ext}$ is only non-zero for some finite length of time. Then as long as $b > 0$, the particle will eventually come to rest once again. This implies that $\Delta \vec{v} = 0$ and
\begin{equation}
    \Delta \vec{x} = \frac{1}{b} \int_{- \infty}^\infty dt \vec{F}_{\rm ext}.
\end{equation}
Therefore, the final displacement of the particle in the tank only depends on the time integral of the external force and $b$.

Now, imagine arranging a large number of these paritcle-in-a-vat apparatuses in a very big sphere which surrounds some charged bulk particles which are scattering/accelerating/radiating. If we assume our test charges move slowly then they'll only feel a force from the electric field and not the magnetic field. If we write the tangential component of the tiny displacement vector $\Delta \vec{x}$ in spherical coordinates, we find
\begin{align}
    \Delta x_A &= \frac{e}{b} \int_{-\infty}^\infty du F_{uA} \\
    &= \frac{e}{b} \Delta A_A^{(0)}\\
    &= \frac{e}{b} N_A
\end{align}
Thus we see that the change in the tangential position of the test charge is proportional to the soft photon mode $N_A$! Note that because $\Delta x_A = r \partial_A \hat n \cdot \Delta \vec{x}$, the overall displacement shrinks as $1/r$.

It should be emphasized that the fact that the test particles have some overall displacement is not particularly interesting. What \textit{is} interesting is that these displacements depend only on the initial and final velocities of the bulk particles, which is because that is all $N_A$ depends on. Hence the name ``memory.''

\section{Is this holography? What information does $Q_\lambda$ hold?}

Noether's 1.5th theorem implies that gauge theories have intrinsically holographic properties. Because the Noether current of a gauge symmetry satisfies
\begin{equation}
    J^\mu_\lambda = \partial_\nu K^{\mu \nu}_\lambda
\end{equation}
the soft charge $Q_\lambda$ is really a surface charge. Can we somehow use this fact to construct a dual ``boundary theory'' to electromagnetism?

Nobody knows the answer to this question, but its certainly worth speculating on. One of the first distinctions we must make is between ``trivial'' holography and true holography. By pushing a Cauchy surface up to $\mathcal{I}^+$, can we automatically say that we have collected all of the information of the system onto a spacetime of one fewer dimension? Not really. A true holographic principle would be more like AdS/CFT, where each spatial slice in the bulk corresponds to a spatial slice in the boundary. In our case, while points on $\mathcal{I}^+$ with different values of $u$ really just live on a single Cauchy slice, it is also not unnatural to consider $u$ as a time coordinate under which a 2D celestial sphere ``evolves.''

\begin{figure}[h]
\input{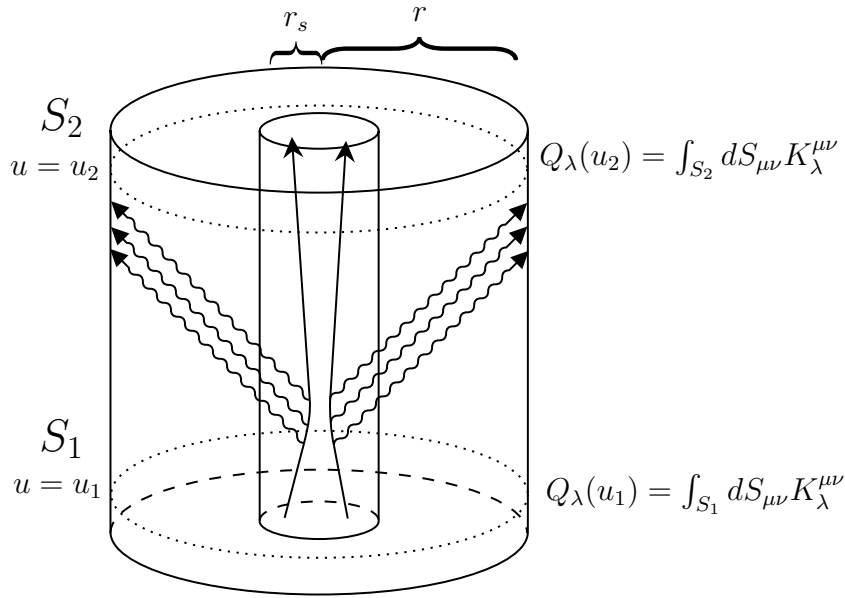}
\caption{\label{fig:hologram} A large holographic cylinder surrounding a scattering event. Imagine a boundary observer had access to $Q_\lambda(u)$ measured at each sphere of constant $u$.}
\end{figure}

Let's see if we can ascertain exactly what bulk information $Q_\lambda$ does and does not have access to, in hopes of assessing the possibility of ``bulk reconstruction.'' Consider the set up drawn in Figure \ref{fig:hologram}. We have a collection of charged particles scattering in the bulk which are confined to be in a sphere of radius $r_s$. Far away from these particles we have another large sphere of radius $r$, with $r \gg r_s$. Imagine we have access to the soft charges $Q_\lambda(u)$, as measured on the sphere of constant $u$, for all $\lambda$ and $u$. When the particles in the bulk radiate, soft charge will flow out of the cylinder and $Q_\lambda(u)$ will change with time.

If one knows $Q_\lambda$ for all $\lambda$, one knows the leading $1/r^2$ component of the radial electric field $E_r$ at all angles. If the particle is travelling at a constant velocity this will only depend on the velocity of the particle. Specifically, in the $u$ fixed large $r$ limit,
\begin{equation}
    E_r = \frac{e}{4 \pi r^2} \frac{1 - \beta^2}{(1 - \vec \beta \cdot \hat n)} + \mathcal{O}\left( \frac{r_s}{r} \right).
\end{equation}
So, at first glance it seems as though $Q_\lambda$ only only holds information about the velocity of charges particles but not the positions of charged particles. The positions of the particles only affect $Q_\lambda$ at sub-leading order in $1/r$.

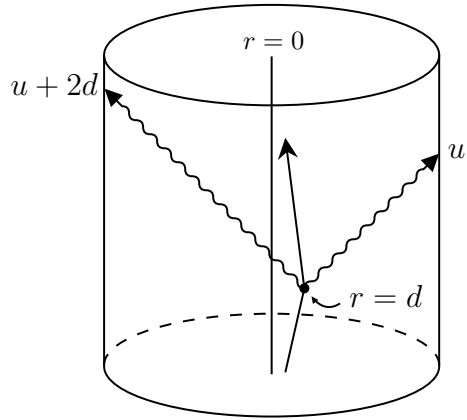
\begin{figure}[H]
\begin{center}

\tikzset{every picture/.style={line width=0.75pt}} 

\begin{tikzpicture}[x=0.75pt,y=0.75pt,yscale=-1,xscale=1]

\draw   (231,47.55) -- (231,204.63) .. controls (231,218.63) and (193.17,229.98) .. (146.5,229.98) .. controls (99.83,229.98) and (62,218.63) .. (62,204.63) -- (62,47.55) .. controls (62,33.55) and (99.83,22.2) .. (146.5,22.2) .. controls (193.17,22.2) and (231,33.55) .. (231,47.55) .. controls (231,61.55) and (193.17,72.9) .. (146.5,72.9) .. controls (99.83,72.9) and (62,61.55) .. (62,47.55) ;
\draw    (146.5,48.2) -- (146.5,208.64) ;
\draw  [fill={rgb, 255:red, 0; green, 0; blue, 0 }  ,fill opacity=1 ] (160.97,165.27) .. controls (160.97,164.14) and (161.88,163.23) .. (163,163.23) .. controls (164.12,163.23) and (165.03,164.14) .. (165.03,165.27) .. controls (165.03,166.39) and (164.12,167.3) .. (163,167.3) .. controls (161.88,167.3) and (160.97,166.39) .. (160.97,165.27) -- cycle ;
\draw    (163,165.27) -- (153.4,207.67) ;
\draw    (163,165.27) -- (153.78,92.24) ;
\draw [shift={(153.4,89.27)}, rotate = 442.8] [fill={rgb, 255:red, 0; green, 0; blue, 0 }  ][line width=0.08]  [draw opacity=0] (10.72,-5.15) -- (0,0) -- (10.72,5.15) -- (7.12,0) -- cycle    ;
\draw    (163,165.27) .. controls (163,162.91) and (164.18,161.73) .. (166.54,161.73) .. controls (168.89,161.73) and (170.07,160.55) .. (170.07,158.2) .. controls (170.07,155.84) and (171.25,154.66) .. (173.61,154.66) .. controls (175.97,154.66) and (177.15,153.48) .. (177.14,151.12) .. controls (177.14,148.76) and (178.32,147.58) .. (180.68,147.59) .. controls (183.04,147.59) and (184.22,146.41) .. (184.21,144.05) .. controls (184.21,141.69) and (185.39,140.51) .. (187.75,140.52) .. controls (190.11,140.52) and (191.29,139.34) .. (191.28,136.98) .. controls (191.28,134.62) and (192.46,133.44) .. (194.82,133.45) .. controls (197.18,133.45) and (198.36,132.27) .. (198.36,129.91) .. controls (198.36,127.56) and (199.54,126.38) .. (201.89,126.38) .. controls (204.25,126.38) and (205.43,125.2) .. (205.43,122.84) .. controls (205.42,120.48) and (206.6,119.3) .. (208.96,119.3) .. controls (211.32,119.31) and (212.5,118.13) .. (212.5,115.77) .. controls (212.49,113.41) and (213.67,112.23) .. (216.03,112.23) .. controls (218.39,112.24) and (219.57,111.06) .. (219.57,108.7) .. controls (219.56,106.34) and (220.74,105.16) .. (223.1,105.16) -- (223.22,105.04) -- (228.88,99.39) ;
\draw [shift={(231,97.27)}, rotate = 495] [fill={rgb, 255:red, 0; green, 0; blue, 0 }  ][line width=0.08]  [draw opacity=0] (8.93,-4.29) -- (0,0) -- (8.93,4.29) -- cycle    ;
\draw    (163,165.27) .. controls (160.64,165.27) and (159.46,164.09) .. (159.46,161.73) .. controls (159.46,159.38) and (158.28,158.2) .. (155.93,158.2) .. controls (153.57,158.2) and (152.39,157.02) .. (152.39,154.66) .. controls (152.4,152.3) and (151.22,151.12) .. (148.86,151.12) .. controls (146.5,151.13) and (145.32,149.95) .. (145.32,147.59) .. controls (145.33,145.23) and (144.15,144.05) .. (141.79,144.05) .. controls (139.43,144.06) and (138.25,142.88) .. (138.25,140.52) .. controls (138.26,138.16) and (137.08,136.98) .. (134.72,136.98) .. controls (132.36,136.99) and (131.18,135.81) .. (131.18,133.45) .. controls (131.18,131.09) and (130,129.91) .. (127.64,129.91) .. controls (125.29,129.91) and (124.11,128.73) .. (124.11,126.38) .. controls (124.11,124.02) and (122.93,122.84) .. (120.57,122.84) .. controls (118.21,122.84) and (117.03,121.66) .. (117.04,119.3) .. controls (117.04,116.94) and (115.86,115.76) .. (113.5,115.77) .. controls (111.14,115.77) and (109.96,114.59) .. (109.97,112.23) .. controls (109.97,109.87) and (108.79,108.69) .. (106.43,108.7) .. controls (104.07,108.7) and (102.89,107.52) .. (102.9,105.16) .. controls (102.9,102.8) and (101.72,101.62) .. (99.36,101.63) .. controls (97,101.63) and (95.82,100.45) .. (95.82,98.09) .. controls (95.82,95.74) and (94.64,94.56) .. (92.29,94.56) .. controls (89.93,94.56) and (88.75,93.38) .. (88.75,91.02) .. controls (88.76,88.66) and (87.58,87.48) .. (85.22,87.48) .. controls (82.86,87.49) and (81.68,86.31) .. (81.68,83.95) .. controls (81.69,81.59) and (80.51,80.41) .. (78.15,80.41) .. controls (75.79,80.42) and (74.61,79.24) .. (74.61,76.88) .. controls (74.62,74.52) and (73.44,73.34) .. (71.08,73.34) -- (69.98,72.24) -- (64.32,66.59) ;
\draw [shift={(62.2,64.47)}, rotate = 405] [fill={rgb, 255:red, 0; green, 0; blue, 0 }  ][line width=0.08]  [draw opacity=0] (8.93,-4.29) -- (0,0) -- (8.93,4.29) -- cycle    ;
\draw  [draw opacity=0][dash pattern={on 4.5pt off 4.5pt}] (62,204.07) .. controls (62,204.07) and (62,204.07) .. (62,204.07) .. controls (62,189.71) and (99.83,178.07) .. (146.5,178.07) .. controls (193.17,178.07) and (231,189.71) .. (231,204.07) -- (146.5,204.07) -- cycle ; \draw  [dash pattern={on 4.5pt off 4.5pt}] (62,204.07) .. controls (62,204.07) and (62,204.07) .. (62,204.07) .. controls (62,189.71) and (99.83,178.07) .. (146.5,178.07) .. controls (193.17,178.07) and (231,189.71) .. (231,204.07) ;
\draw    (181,172.87) .. controls (176.87,175.62) and (172.45,175.41) .. (168.49,171.48) ;
\draw [shift={(166.6,169.27)}, rotate = 413.75] [fill={rgb, 255:red, 0; green, 0; blue, 0 }  ][line width=0.08]  [draw opacity=0] (3.57,-1.72) -- (0,0) -- (3.57,1.72) -- cycle    ;

\draw (147.6,40.08) node  [font=\footnotesize]  {$r=0$};
\draw (239.82,95.83) node    {$u$};
\draw (36.88,62.63) node    {$u+2d$};
\draw (184.2,169.23) node [anchor=west] [inner sep=0.75pt]    {$r=d$};

\end{tikzpicture}

\end{center}
\caption{\label{fig:hologram2} A charged particle accelerates at a distance of $d$ from the origin. News of this event first reaches the large cylinder at one particular angle at $u$. However, news of the event reaches the antipodal angle at $u+2d$.}
\end{figure}

However, if we truly know $Q_\lambda(u)$ for all $u$, this conclusion is too hasty. Consider the scenario in Figure \ref{fig:hologram2}. A charged particle accelerates when it is at a distance of $d$ from the origin. The news of this event will not reach all points of the boundary at the same moment in time. If the point on the boundary closest to the particle gets the news at $u$, then the antipodal point will get the news at $u + 2d$. Therefore, if one can cross-correlate $Q_\lambda(u)$ measurements at different times, one should be able to extract detailed information about the positions of the charged particles when they undergo scattering.

In fact, if one knows $Q_\lambda(u)$ for all $u$, one is essentially measuring the soft flux which leaves the sphere at all times. This is tantamount to knowing $F_{uA} = \partial_u A_A$ for all angles and $u$. Therefore, knowledge of $Q_\lambda(u)$ allows one to reconstruct $A_A^{(0)}$ (up to a large gauge transformation) which is essentially all of the Cauchy data at $\mathcal{I}^+$ by \eqref{cauchy_plus}. If we assume that no radiation enters from $\mathcal{I}^-$, then knowledge of the radiative Cauchy data at both $\mathcal{I}^+$ and $\mathcal{I}^-$ is enough to allow us reconstruct the precise path of the particle in the bulk. In fact, it's not even particularly hard. From \eqref{AA_vel0}, essentially reproduced below,
\begin{equation}
    A^{(0)}_A = -\frac{e}{4 \pi} \frac{ \partial_A \hat n \cdot \vec{\beta}}{1 - \vec{\beta} \cdot \hat n} + \partial_A \Phi_{\rm arbitrary}
\end{equation}
we can see how $\vec{\beta}$ can be read off of $A^{(0)}_A$, up to a large gauge transformation. The presence of the arbitrary gauge piece does not invalidate the overall procedure because it is $u$-independent, meaning that if the particle accelerates then one can extract $\vec{\beta}_i$ and $\vec{\beta}_f$ using the difference $\Delta A_A^{(0)}$. What's more, one is always able to find $\vec{\beta}_i$ using an initial soft charge measurement.

If we have multiple particles in the bulk instead of just one, the situation complicates. Soft charges cannot distinguish between one particle of charge $2e$ and two particles of charge $e$ if they have equal velocities. Having said that, if one keeps track of the positions of the individual particles using the timing method discussed previously, the two particles will generically be able to be resolved.

Therefore, the prospects of bulk reconstruction in classical electromagnetism are actually very promising: almost all of the information about the path of the particle ends up imprinted on $\mathcal{I}^+$. But, at the end of the day, one has to wonder if this really amounts to `true' holography or if this is just `trivial' holography with extra steps. In particular, is the bulk theory equivalent to a boundary theory which has its own intrinsic and non-trivial mathematical formulation? Who knows!

If one graduates from classical mechanics to quantum mechanics, the ability to perform this thought experiment and execute bulk reconstruction encounters major complications, as discussed originally in \cite{Bousso:2017xyo}. If one really wanted to measure $E_r$ at large $r$, because $E_r \sim 1/r^2$ this measurement would have to be extremely sensitive. However, quantum field fluctuations would preclude one from accurately measuring $\langle E_r \rangle$ in some small region of space without waiting a very long time. In fact, the period of time one would have to wait to measure $\langle E_r \rangle$ with any accuracy will increase without bound as $r$ increases. Quantum fluctuations make it impossible to measure asymptotic charges in a finite length of time.

Having said that, statistical uncertainty which precludes the efficient measurement of quantum systems is not usually considered to be an essential obstacle in the formulation of theories. (For instance, the black hole information paradox is usually formulated without considering whether it is statistically feasible to determine if Hawking radiation is truly unitary.) The seeming ease with which the bulk can be reconstructed classically could be interpreted as an encouraging sign that it may indeed be possible.

\section*{Acknowledgements}

The author has greatly benefited from enlightening discussions with Alex Atanasov, Adam Ball, Erin Crawley, Alfredo Guevara, Elizabeth Himwich, Daniel Kapec, Sruthi Narayanan, Andrew Strominger, and especially Ana-Maria Raclariu throughout the conception and writing of this note. This work was supported by NSF GRFP grant DGE1745303.

\newpage

\appendix

\section{Covariant Integration}

\subsection{Helpful formulae}

We begin with the handy identity
\begin{equation}\label{varsqrtdet}
    \delta  \sg = \frac{1}{2} \sg g^{\mu \nu} \delta g_{\mu \nu}.
\end{equation}
Using the fact that $g^{\alpha \beta}$ is symmetric, we can see that
\begin{align}
    \frac{1}{\sg} \partial_\nu \sg &= \frac{1}{2} g^{\alpha \beta} \partial_\nu g_{\alpha \beta} \\
    &= \frac{1}{2} g^{\alpha \beta} (\partial_\nu g_{\alpha \beta} + \partial_\beta g_{\alpha \nu} - \partial_\alpha g_{\beta \nu}) \\
    &= g^{\alpha \beta} \Gamma_{\alpha \beta \nu} \\
    &= \Gamma^\mu_{\; \mu \nu}. \label{dersqgchristoffel}
\end{align}
We can use this to derive the indispensable formula
\begin{align}
    \nabla_{\mu} J^{\mu} &= \partial_{\mu} J^{\mu} + \Gamma^{\mu}_{\; \mu \nu} J^\nu \\
    &= \frac{1}{\sg} \partial_\mu (\sg J^\mu). \label{wonderfuldiv}
\end{align}
This formula allows for one to integrate by parts with $\nabla_\mu$ and the covariant measure $\sg d^4 x$ just as one one would with $\partial_\mu$ and $d^4 x$ in flat space.

We can also derive a similar formula for an anti-symmetric rank-2 tensor $K^{\mu \nu}$.
\begin{align}
    \nabla_\mu K^{\mu \nu} &= \partial_\mu K^{\mu \nu} + \Gamma^\mu_{\; \mu \rho} K^{\rho \nu} + \Gamma^\nu_{\; \mu \rho} K^{\mu \rho} \\
    &= \partial_\mu K^{\mu \nu} + \Gamma^\mu_{\; \mu \rho} K^{\rho \nu} \\
    &= \frac{1}{\sg} \partial_\mu ( \sg K^{\mu \nu}). \label{wonderfuldiv2form}
\end{align}
This formula immediately generalizes to anti-symmetric tensors of any rank. It allows one to switch back and forth between equations like $J^\mu = \partial_\nu K^{\mu \nu}$ and $J^\mu = \nabla_\nu K^{\mu \nu}$ at the small cost of redefining $J^\mu$ and $K^{\mu \nu}$ by factors of $\sg$.

\subsection{Surface elements and Stokes' theorem}

For a complete treatment of covariant integration see chapter 2 of \cite{poisson2004relativist}. Say you have a $3$-surface $\Sigma$ parameterized by some variables $y^a$. The surface is given by $x^\mu = x^\mu(y)$, where $\mu = 0,1,2,3$ and $a = 1,2,3$. The tangent space of $\Sigma$ is spanned by the three vectors $e^\mu_a$ defined by
\begin{equation}
    e^\mu_a \partial_\mu = \pdv{x^\mu}{y^a} \partial_\mu.
\end{equation}
You can restrict the metric $g_{\mu \nu}$ to $\Sigma$ to get the induced metric $h_{ab}$. We have
\begin{align}
    ds^2_\Sigma &= g_{\mu \nu} dx^\mu dx^\nu \\
    &= g_{\mu \nu} \Big( \pdv{x^\mu}{y^a} dy^a \Big) \Big( \pdv{x^\nu}{y^b} dy^b \Big) \\
    &= h_{ab} dy^a dy^b
\end{align}
where we have defined
\begin{equation}
    h_{ab} \equiv g_{\mu \nu} \frac{\partial x^\mu}{\partial y^a} \frac{\partial x^\nu}{\partial y^b}=  g_{\mu \nu} e^\mu_a e^\nu_b.
\end{equation}
When we want to integrate over this 3-surface, we use the volume measure
\begin{equation}
    d \Sigma \equiv | h|^{1/2} d^3 y
\end{equation}
where $h = \det(h_{ab})$. Let us now introduce the outward-pointing normal vector $n^\mu$ which is orthogonal to all $e^\mu_a$ and has norm $n^\mu n_\mu = +1 (-1)$ if $n^\mu$ is timelike (spacelike). We can define the directed surface element
\begin{align}\label{dSigma_def1}
    d\Sigma^\mu &\equiv \pm n^\mu d \Sigma \\
    d\Sigma_\mu &= \pm n_\mu d \Sigma.
\end{align}
where the sign choice of $\pm$ is $+$ $(-)$ if $n^\mu$ is timelike (spacelike)\footnote{If you don't understand the need for the sign, consider that in $(t,r,z^A)$ coordinates, the outward pointing normal to an $r=const$ surface is given by $n^\mu = (0,1,0,0)$. Upon lowering, we get $n_\mu = (0,-1,0,0)$.}. The case when $n^\mu$ is null must be handled separately.

We now have the ingredients to covariantly express the divergence theorem. Say $V$ is a 4-volume and $\Sigma = \partial V$. Then,
\begin{align}
    \int_V d^4 x \sg \nabla_\mu J^\mu &= \int_\Sigma d \Sigma_\mu J^\mu \\
    &= \int_\Sigma d^3 y \sqrt{|h|} (\pm n_\mu) J^\mu.
\end{align}
The above expression can be easily proven if one chooses a coordinate system in which $\Sigma$ is the $x^0=1$ surface, $n^\mu = (1,0,0,0)$, and the metric factorizes on $\Sigma$ as
\begin{equation}
    \eval{g_{\mu \nu}}_\Sigma = \begin{pmatrix} \pm 1 & 0 \\ 0 & h_{ab} \end{pmatrix}.
\end{equation}

\begin{figure}[H]
\begin{center}

\tikzset{every picture/.style={line width=0.75pt}} 

\begin{tikzpicture}[x=0.75pt,y=0.75pt,yscale=-1,xscale=1]

\draw   (103.5,98) .. controls (123.5,88) and (168.5,104) .. (188,113) .. controls (207.5,122) and (231.5,132) .. (188,173) .. controls (144.5,214) and (118,203) .. (98,173) .. controls (78,143) and (83.5,108) .. (103.5,98) -- cycle ;
\draw   (300.5,98) .. controls (320.5,88) and (365.5,104) .. (385,113) .. controls (404.5,122) and (428.5,132) .. (385,173) .. controls (341.5,214) and (315,203) .. (295,173) .. controls (275,143) and (280.5,108) .. (300.5,98) -- cycle ;
\draw  [dash pattern={on 4.5pt off 4.5pt}] (306.25,106.83) .. controls (321.75,97.83) and (368.75,118.83) .. (383.25,125.33) .. controls (397.75,131.83) and (395.75,146.33) .. (375.25,166.83) .. controls (354.75,187.33) and (326.5,193.67) .. (310.75,175.33) .. controls (295,157) and (290.75,115.83) .. (306.25,106.83) -- cycle ;
\draw  [dash pattern={on 4.5pt off 4.5pt}] (321.25,122.33) .. controls (336.75,113.33) and (365.25,132.83) .. (366.75,139.33) .. controls (368.25,145.83) and (366.75,153.33) .. (357.25,163.83) .. controls (347.75,174.33) and (332.25,175.33) .. (320.75,167.83) .. controls (309.25,160.33) and (305.75,131.33) .. (321.25,122.33) -- cycle ;
\draw  [dash pattern={on 4.5pt off 4.5pt}] (326.75,147.33) .. controls (326.75,142.09) and (331,137.83) .. (336.25,137.83) .. controls (341.5,137.83) and (345.75,142.09) .. (345.75,147.33) .. controls (345.75,152.58) and (341.5,156.83) .. (336.25,156.83) .. controls (331,156.83) and (326.75,152.58) .. (326.75,147.33) -- cycle ;
\draw    (295,173) -- (276.75,185.17) ;
\draw [shift={(274.25,186.83)}, rotate = 326.31] [fill={rgb, 255:red, 0; green, 0; blue, 0 }  ][line width=0.08]  [draw opacity=0] (10.72,-5.15) -- (0,0) -- (10.72,5.15) -- (7.12,0) -- cycle    ;

\draw (135.13,152.67) node    {$V$};
\draw (163,98.93) node [anchor=south west] [inner sep=0.75pt]    {$\Sigma =\partial V$};
\draw (273.44,182.23) node [anchor=north east] [inner sep=0.75pt]    {$n^{\mu }$};
\draw (369.62,96.09) node  [font=\footnotesize,rotate=-21.53]  {$x^{0} =1$};
\draw (351.02,119.97) node  [font=\tiny,rotate=-21.53]  {$x^{0} =\text{const}$};

\end{tikzpicture}

\end{center}
\caption{\label{fig:stokes} Diagram of coordinate system in which the divergence theorem is easy to prove.}
\end{figure}
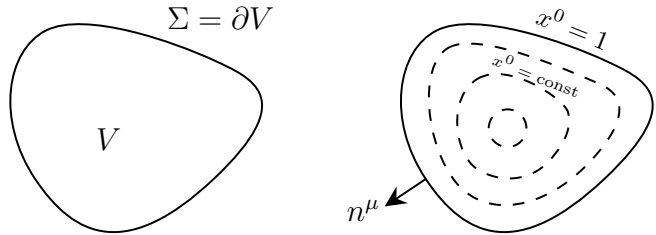

\begin{figure}[H]
\begin{center}

\tikzset{every picture/.style={line width=0.75pt}} 

\begin{tikzpicture}[x=0.75pt,y=0.75pt,yscale=-1,xscale=1]

\draw    (66,33) .. controls (123.5,-38) and (226.5,44) .. (113.5,104) ;
\draw    (66,33) .. controls (56.5,71) and (88.5,107) .. (113.5,104) ;
\draw  [dash pattern={on 4.5pt off 4.5pt}]  (66,33) .. controls (94.5,24) and (133.5,101) .. (113.5,104) ;
\draw    (137,89) -- (148.83,103.2) ;
\draw [shift={(150.75,105.5)}, rotate = 230.19] [fill={rgb, 255:red, 0; green, 0; blue, 0 }  ][line width=0.08]  [draw opacity=0] (10.72,-5.15) -- (0,0) -- (10.72,5.15) -- (7.12,0) -- cycle    ;
\draw    (76,82.5) -- (61.21,92.79) ;
\draw [shift={(58.75,94.5)}, rotate = 325.18] [fill={rgb, 255:red, 0; green, 0; blue, 0 }  ][line width=0.08]  [draw opacity=0] (10.72,-5.15) -- (0,0) -- (10.72,5.15) -- (7.12,0) -- cycle    ;

\draw (139.82,29.17) node    {$\Sigma $};
\draw (59.52,46.9) node [anchor=north east] [inner sep=0.75pt]    {$S=\partial \Sigma $};
\draw (151,102.9) node [anchor=north west][inner sep=0.75pt]    {$n^{\mu }$};
\draw (55.23,95.4) node [anchor=north] [inner sep=0.75pt]    {$m^{\mu }$};

\end{tikzpicture}

\end{center}
\caption{\label{fig:stokes2} $m^\mu$ must be defined as the outward pointing normal on $\partial \Sigma$.}
\end{figure}
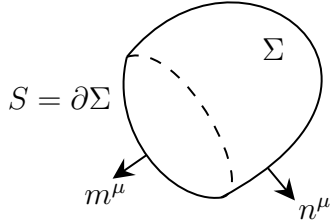

We can define an analogous surface element for a 2-surface $S$. Say you parametize $S$ by some variables $z^A$. The surface is given by $x^\mu = x^\mu(z)$ where $A=1,2$. The induced metric on $S$ is given by
\begin{equation}
    \gamma_{AB} \equiv g_{\mu \nu} \frac{\partial x^\mu}{\partial z^A} \frac{\partial x^\nu}{\partial z^B}.
\end{equation}
We can define the surface element by
\begin{equation}
    d S \equiv \sqrt{\gamma} d^2 z.
\end{equation}
Say $n^\mu$ and $m^\mu$ are two unit vectors which are normal to $S$. Note that exactly one of $n^\mu$ and $m^\mu$ will timelike and the other will be spacelike. We define the directed surface element by
\begin{equation}
    d S_{\mu \nu} \equiv -(n_\mu m_\nu - n_\mu m_\nu) d S.
\end{equation}
Say $S = \partial \Sigma$. We can covariantly express Stokes' theorem as
\begin{align}
    \int_\Sigma d\Sigma_\mu \nabla_\nu K^{\mu \nu} &= \frac{1}{2} \int_S d S_{\mu \nu} K^{\mu \nu} \\
    &= \int_S d^2 z \sqrt{\gamma} (-n_\mu m_\nu) K^{\mu \nu}.
\end{align}
Actually, for the above formula to be true, we must choose for $m^\mu$ to be outward pointing from $\Sigma$, as shown in Figure \ref{fig:stokes2}.

\subsection{How to integrate over $\mathcal{I}^+$}\label{app_null_integrate}

In this note, we frequently use the equation
\begin{equation}\label{Ipint0}
    \int_{\mathcal{I}^+} d \Sigma_\mu J^\mu = \int_{\mathcal{I}^+} \meas J_u.
\end{equation}
The purpose of this section is to derive the above formula, as its proof is a bit subtle. Recall that in $(u,r,z^A)$ coordinates, the metric is
\begin{equation}
    ds^2 = du^2 + 2 du dr - r^2 \gamma_{AB} d z^A d z^B.
\end{equation}
A point on $\mathcal{I}^+$ is reached by holding $u$ and $z^A$ fixed and sending $r \to \infty$. Therefore, $\mathcal{I}^+$ is defined as the $r = const$ surface where that constant is going to infinity.\footnote{One may be confused because an $r=const$ suface has signature $(1,2)$ while $\mathcal{I}^+$ should be a null surface with signature $(0,2)$. Crucially, $\mathcal{I}^+$ should really be understood as a null surface in the \textit{conformal compactification} where one multiplies the metric by $1/r^2$.} The outward pointing normal of this $r=const$ surface in $(u,r,z^A)$ coordinates is
\begin{align}
    n^\mu &= (-1, 1, 0, 0) \\
    n_\mu &= (0, -1, 0, 0).
\end{align}
This implies that
\begin{align}
    \int_{\mathcal{I}^+} d \Sigma_\mu J^\mu &= \int_{\mathcal{I}^+} \meas J^r \label{Ipint1} \\
    &= \int_{\mathcal{I}^+} \meas (J_u - J_r)  \label{Ipint2}
\end{align}
which does not quite look like \eqref{Ipint0} due to the presence of $J_r$. We will now argue that for the flux integral of a typical conserved current $J^\mu$, the $J_r$ integral will be subleading and vanish as $r \to \infty$, leaving only the $J_u$ part.

We begin by assuming that the total charge $\int d^3 x J^0$ on a $t=const$ time slice is finite and that a finite amount of flux will pass through $\mathcal{I}^+$, which from \eqref{Ipint1} implies
\begin{equation}
    J^r \sim \mathcal{O}(1/r^2).
\end{equation}
The current conservation equation then reads
\begin{equation}
    0 = \nabla_\mu J^\mu = \underbrace{\frac{1}{r^2} \partial_r ( r^2 J^r)}_{\mathcal{O}(1/r^4)} + \partial_u J^u + \nabla_A J^A.
\end{equation}
Using
\begin{align}
    J^u &= J_r \\
    J^r &= J_u - J_r
\end{align}
we find that
\begin{equation}
    J_r = \nabla_A (...)^A + \mathcal{O}(1/r^4)
\end{equation}
meaning $J_r$ is a total derivative on the sphere plus a piece which shrinks as $1/r^4$. Therefore, the integral of $J_r$ in \eqref{Ipint2} will go to $0$ as $r \to \infty$. This proves \eqref{Ipint0}.

While we have given a satisfactory proof of \eqref{Ipint0}, it is still reassuring to check that the $J_r$ piece really is subleading in some key examples. We will now turn to two such examples. One will be where $J^\mu$ is the EM Noether's theorem gauge transformation current $J^\mu_\lambda$, and the other will be where $J^\mu$ is the charge current $j^\mu$ of a massless particle.

\subsubsection{Example 1: Integrating the Noether current $J^\mu_\lambda$}
The Noether current of gauge symmetry (defined in \eqref{Jlambda_def}) in EM is
\begin{equation}
    (J_\lambda)_\mu = -F_{\mu \nu} \nabla^\nu \lambda+ \lambda j_\mu.
\end{equation}
We take the gauge parameter to only depend on the sphere coordinates, so $\lambda = \lambda(z^A)$. From the previous section, we want to ensure that the $(J_\lambda)_r$ integral is $0$, which we now do:
\begin{align}
    \int_{\mathcal{I}^+} \meas (J_\lambda)_r &= \int_{\mathcal{I}^+} \meas (- F_{r A} \nabla^A \lambda + \lambda j_r) \\
    &= \int_{\mathcal{I}^+} \meas \, \lambda(\nabla^A F_{rA} + j_r) \\
    &= \int_{\mathcal{I}^+} \meas \, \lambda (-\nabla^u F_{r u} ) \\
    &= \int_{\mathcal{I}^+} \meas \, \lambda \underbrace{(-\nabla_r F_{r u} )}_{\mathcal{O}(1/r^4)} \\
    &= 0.
\end{align}
In the first line we integrated by parts on the 2-sphere. In the second line we used the equations of motion $\nabla^\mu F_{\mu \nu} = j_\nu$. In the third line we used that $F_{ru} \sim \mathcal{O}(1/r^2)$ from \eqref{falloff2}. 
We have now proved that
\begin{align}
    \int_{\mathcal{I}^+} d \Sigma_\mu J^\mu_\lambda &= \int_{\mathcal{I}^+} \meas (J_\lambda)_u \\
    &= \int_{\mathcal{I}^+} \meas (-F_{u A} \nabla^A \lambda + \lambda j_u).
\end{align}

\subsubsection{Example 2: Integrating the charge flux of a massless particle}

If a massless particle of charge $e$ travels on a path parameterized by affine parameter $\tau$ as
\begin{align}
    u &= u_0 \\
    r &= \tau \\
    z^A &= z^A_0
\end{align}
then its charge current will be given by
\begin{align}
    j^u &= 0 \\
    j^r &= \frac{e}{r^2} \delta(u - u_0) \frac{1}{\sqrt{\gamma}} \delta^2(z^A - z^A_0) \\
    j^A &= 0.
\end{align}
Note that $j^u = 0$ because $j^\mu$ is the charge flux passing through an $x^\mu = const$ surface, and our particle will not pass through any $u = const$ surfaces. In any case,
\begin{align}
    j_u &= j^u + j^r = \frac{e}{r^2} \delta(u - u_0) \frac{1}{\sqrt{\gamma}} \delta^2(z^A - z^A_0) \\
    j_r &= j^u = 0
\end{align}
and the full flux integral is
\begin{equation}
    \int_{\mathcal{I}^+} d \Sigma_\mu j^\mu = \int_{\mathcal{I}^+} \meas (j_u - j_r) = e.
\end{equation}

\section{Noether's Theorems}\label{app_noether}

For this section we will reintroduce the ``weak equality'' symbol ``$\approx$''. We use weak equality for equations which only hold on solutions to the equations of motion (on-shell), so
\begin{equation}
    \text{e.o.m.} \approx 0.
\end{equation}
We use strong equality ``$=$'' for mathematical identities which are always true (off-shell) such as $\partial_\mu \partial_\nu F^{\mu \nu} = 0$. Much of the following is based on \cite{avery2016noether}.

\subsection{Noether's 1st Theorem}

Say you have field $\phi_i$ and an action
\begin{equation}
    S = \int d^4 x \, \mathcal{L}.
\end{equation}
If you have a symmetry transformation
\begin{equation}
    \delta \phi_i = \eps \delta_{\rm sym} \phi_i
\end{equation}
where $\eps$ is an infinitesimal constant, then the action will change by the total derivative of some vector $\mathcal{J}^\mu$.
\begin{equation}\label{noether1sym0}
    \delta S = \int d^4 x \; \eps \, \partial_\mu \mathcal{J}^\mu
\end{equation}
If we now make $\eps$ a spacetime dependent function, i.e. $\eps = \eps(x)$, and change the symmetry variation to
\begin{equation}\label{noether1sym}
    \delta \phi_i = \eps(x) \delta_{\rm sym} \phi_i
\end{equation}
then the variation of the action must be of the form
\begin{equation}
    \delta S = \int d^4 x \big(\eps \, \partial_\mu \mathcal{J}^\mu + (\partial_\mu \eps) \mathcal{K}^\mu \big).
\end{equation}
This is because the above equation must reduce to \eqref{noether1sym0} when $\eps = const$. In principle, if $S$ depends on higher derivatives of $\phi_i$, there could also be terms which depend on $\partial_\mu \partial_\nu \eps$, $\partial_\mu \partial_\nu \partial_\rho \eps$, etc., but for brevity we will assume this is not the case. If we take $\eps$ to have compact support, we can integrate by parts to find
\begin{equation}
    \delta S = \int d^4 x \; \eps \, \partial_\mu \big(\mathcal{J}^\mu - \mathcal{K}^\mu \big).
\end{equation}
Finally, recall that a solution to the equations of motion will make the action stationary, and $\delta S \approx 0$ for \textit{any} variation, including our symmetry variation \eqref{noether1sym}. Allowing $\eps$ to be any function implies that the current
\begin{equation}
    J_\mu \equiv \mathcal{J}^\mu - \mathcal{K}^\mu
\end{equation}
is conserved
\begin{equation}
    \partial_\mu J^\mu \approx 0.
\end{equation}
This is Noether's 1st theorem; a symmetry which leaves the action invariant up to a total derivative can be used to find a current which is conserved on-shell.

The proof we have given also provides us an effective method to find these conserved currents. If you have a symmetry variation, simply multiply it by a spacetime dependent infinitesimal parameter $\eps(x)$ and deduce the equation which must hold to have $\delta S \approx 0$ on-shell. This equation will be guaranteed to be a conservation equation of the form $\partial_\mu J^\mu \approx 0$ by our proof above.

\subsection{Noether's 2nd Theorem}

Noether's 2nd theorem states that every gauge symmetry implies there is an \textit{off-shell} relationship between the equations of motion. In particular, it means that the equations of motion are not all independent. We will define a ``gauge symmetry'' as an infinite dimensional symmetry which is locally parametized by a spacetime dependent function.

The proof is perhaps easiest to understand using a specific example rather than formulating it in complete generality. Say we have the gauge field $A_\mu$, the external current $j^\mu$, and the action
\begin{equation}
    S = \int d^4 x \left( - \frac{1}{4} F^{\mu \nu} F_{\mu \nu} - j^\nu A_\nu \right).
\end{equation}
For any variation $\delta A_\mu$, the action will vary as
\begin{align}
    S &= \int d^4 x \big( - F^{\mu \nu} \partial_\mu \delta A_\nu - j^\nu \delta A_\nu \big) \\
    &= \int d^4 x \big( - \partial_\mu( F^{\mu \nu} \delta A_\nu ) + (\partial_\mu F^{\mu \nu} - j^\nu) \delta A_\nu \big)
\end{align}
which yields the equations of motion $E^\nu \approx 0$ where
\begin{equation}
    E^\nu \equiv \partial_\mu F^{\mu \nu} - j^\nu.
\end{equation}

This action has a gauge symmetry
\begin{equation}
    \delta A_\mu = \partial_\mu \lambda
\end{equation}
under which $\delta S = 0$ \textit{off-shell}. Therefore, if we choose $\lambda$ to be compact, we have
\begin{align}
    0 = \delta S &= \int d^4 x \, E^\nu \partial_\nu \lambda \\
    &= -\int d^4 x \, \lambda \partial_\nu E^\nu.
\end{align}
This can only hold for arbitary $\lambda$ if we have the off-shell equation
\begin{align}
    0 &= \partial_\nu E^\nu \\
    &= \partial_\nu \partial_\mu F^{\mu \nu} - \partial_\nu j^\nu. \label{noether2}
\end{align}
Therefore, we see that gauge symmetries can be used to find off-shell relationships between the equations of motion, completing our proof of Noether's 2nd theorem.

Let us reflect briefly on how this proof compares to that of Noether's 1st theorem. In both proofs, one needs some sort of varying arbitrary function in order get the final result. In Noether's 1st theorem, it was $\eps(x)$; in Noether's 2nd theorem, it was $\lambda(x)$. The difference is that in Noether's 1st theorem, the transformation that depended on $\eps(x)$ was not really a symmetry of the action (even though it reduced to one when $\eps = const$) and we only had $\delta S \approx 0$ on-shell, meaning our conservation equation only held on-shell. However, in Noether's 2nd theorem, we had a gauge symmetry transformation which left $\delta S = 0$ off-shell and already depended on an arbitrary function $\lambda(x)$ without the need of $\eps(x)$. Because $\delta S = 0$ off-shell, we could find an equation which held off-shell.

Incidentally, Noether's 2nd theorem should not really come as a surprise. If you have a Lagrangian with a gauge symmetry, you can take a solution to the equations of motion and gauge transform it by a function $\lambda$ whose value and derivatives all vanish on the $t = 0$ slice, giving you two solutions to the equations of motion with the exact same initial conditions. This means that the equations of motion cannot be ``predictive,'' in the sense that initial conditions cannot determine unique solutions.\footnote{By declaring that two solutions are ``equivalent'' if they differ by a gauge transformation, we restore the predictivity of the equations.} This is exactly the behavior which results from the equations of motion not all being independent.

This might make Noether's 2nd theorem seem trivial and uninteresting. However, that is far from the truth. Let's take a closer look at \eqref{noether2}. We see that Noether's 2nd theorem implies that $\partial_\mu j^\mu = 0$. This, however, is a strange conclusion, because $j^\mu$ is an external current which a priori could be completely arbitrary. What we have seen is that the equations of motion are literally inconsistent unless this external current is constrained to satisfy $\partial_\mu j^\mu = 0$. This leads us to conclude that fields with gauge symmetries can only couple consistently to currents which are conserved.

This can be seen as a reason why gauge theories are desirable in the first place. If one is trying to concoct a theory in which a field couples to a conserved current, using a field with a gauge symmetry is a great way to accomplish this task! 

Noether's 2nd theorem also illuminates the common procedure where one ``promotes'' a global symmetry to a gauge symmetry, coupling a matter field to a newly defined gauge field in the process. For instance, the free Dirac field $\psi$ has a global $U(1)$ symmetry $\psi \mapsto e^{-i \theta} \psi$. If one substitutes $\theta \to \lambda(x)$, promotes this to a gauge symmetry, and tries to make sure the overall action is invariant under this gauge symmetry, one will find that the Dirac field has coupled to a gauge field $A_\mu$. This procedure is essentially just running Noether's 2nd theorem backwards, as the original global $U(1)$ symmetry (which still exists as a special case of the local symmetry) gives rise to a conserved dynamical charge current $j^\mu$ which couples to $A_\mu$. In other words, by creating a gauge symmetry out of thin air, one has created a new field which manifestly couples to our global $U(1)$ current.

The history behind Noether's 2nd theorem, described for instance in \cite{brading2005note, brading2012hilbert, kosmann2020noether}, is both interesting and instructive. Right after the birth of general relativity, Albert Einstein and David Hilbert were puzzled as to the status of energy conservation in the new theory. The field equations read
\begin{equation}
    G^{\mu \nu} \approx 8 \pi G T^{\mu \nu}
\end{equation}
and the off-shell Bianchi identity implied that $\nabla_\mu G^{\mu \nu} = 0$. This meant that $\nabla_\mu T^{\mu \nu} \approx 0$ had to hold automatically. At the time, this came as a surprise because Hilbert expected that energy-momentum conservation either would have to be a constraint imposed on the theory or would end up being be violated altogether. In 1915, he asked Emmy Noether, a colleague he held in high regard, to see if she could satisfactorily understand the situation.

Her resulting 1918 paper (english translation given in \cite{noether1971invariant}) proved both of her theorems as well as their converses. She observed that in general relativity, the global group of translations was embedded into the diffeomorphism gauge symmetry. In section 6 of her paper, titled ``A Hilbertian Assertion," she wrote \cite{kosmann2020noether}:
\begin{quote}
    \textit{``Given $[S]$ invariant under a group of translations, then the energy relations are improper if and only if $[S]$ is invariant under an infinite group which contains the group of translations as a subgroup.''}
\end{quote}
Read: Energy momentum conservation $\nabla_\mu T^{\mu \nu} \approx 0$ will hold as a consequence of an off-shell identity like $\nabla_\mu G^{\mu \nu} = 0$ if the group of translations is embedded in a larger infinite dimensional group of diffeomorphism gauge symmetries.

Let's see for ourselves how we can derive $\nabla_\mu G^{\mu \nu} = 0$ from diffeomorphism invariance. The Einstein Hilbert action for pure gravity is given by
\begin{equation}
    S =  \frac{-1}{16 \pi G} \int d^4 x \sg R
\end{equation}
and its variation (up to boundary terms) is
\begin{equation}
    \delta S =  \frac{1}{16 \pi G} \int d^4 x \sg G^{\mu \nu}  \delta g_{\mu \nu}.
\end{equation}
If one then uses the variation of the metric corresponding to an infinitesimal diffeomorphism along the vector field $\xi$
\begin{equation}
    \delta g_{\mu \nu} = \mathcal{L}_\xi g_{\mu \nu} = \nabla_\mu \xi_\nu + \nabla_\nu \xi_\mu
\end{equation}
under which $\delta S = 0$. Assuming $\xi_\nu$ is compact, we have
\begin{align}
0 = \delta S &= \frac{1}{8 \pi G} \int d^4 x \, \sg G^{\mu \nu} \nabla_\mu \xi_\nu \\
&= \frac{-1}{8 \pi G} \int d^4 x \, \sg \nabla_\mu G^{\mu \nu} \xi_\nu.
\end{align}
Allowing $\xi_\nu$ to freely vary, we see that
\begin{equation}
    \nabla_\mu G^{\mu \nu} = 0
\end{equation}
off-shell. Therefore, the conservation of energy momentum has the same status in gravity that the conservation of charge has in electromagnetism: they are consistency conditions required by gauge symmetry. 

As a brief aside, one might wonder how an equation like $\partial_\nu(\partial_\mu F^{\mu \nu} - j^\nu)=0$ could possibly hold off-shell when $j^\nu$ is not a background source but rather depends on a dynamical field, in which case we would have $\partial_\mu j^\mu \approx 0$ but not $\partial_\mu j^\mu = 0$. Take for instance the QED Lagrangian
\begin{equation}
    \mathcal{L}_{\rm QED} = - \frac{1}{4} F^{\mu \nu} F_{\mu \nu} + \overline{\psi}( i \gamma^\mu(\partial_\mu + i e A_\mu) - m )\psi
\end{equation}
which has the gauge symmetry
\begin{align} \label{vary_g_A}
    A_\mu &\mapsto A_\mu + \partial_\mu \lambda \\
    \psi &\mapsto e^{-i \lambda} \psi. \label{vary_g_psi}
\end{align}
The Noether current from the global $\lambda = const$ symmetry is given by
\begin{equation}
    j^\mu = e \overline{\psi} \gamma^\mu \psi.
\end{equation}
If one computes the Noether's-2nd-theorem off-shell identity resulting from $\mathcal{L}_{\rm QED}$, one will find that it is just $\partial_\mu \partial_\nu F^{\mu \nu} = 0$. $j^\mu$ doesn't appear in this equation, as it couldn't. We can still easily derive that $j^\mu$ is conserved from $\partial_\mu \partial_\nu F^{\mu \nu} = 0$; it just takes one extra line of algebra, namely $\partial_\mu F^{\mu \nu} \approx j^\nu$. To sum up: Noether's 2nd theorem can only be used to impose $\partial_\mu j^\mu = 0$ off-shell when $j^\mu$ is an external current, which is very reasonable. Otherwise, it will give you $\partial_\mu j^\mu \approx 0$. Interestingly, we could also derive this result by using a ``half'' gauge transformation where one just varies $A_\mu$ as $\delta A_\mu = \partial_\mu \lambda$ but doesn't vary $\psi$. As this is not a full gauge transformation which would render $\delta S = 0$ off-shell, we can't derive an off-shell equation. However, we can still evoke $\delta S \approx 0$ under this half gauge transformation to derive $\partial_\nu(\partial_\mu F^{\mu \nu} - j^\nu)\approx 0$.

\subsection{Noether's 1.5th Theorem}

An astute observer may notice that a one-parameter family of symmetries generated by an arbitrary infinitesimal gauge transformation can be thought of as a global symmetry. One might wonder: how does the corresponding conserved current derived from Noether's 1st theorem behave? This question is addressed by a result we will call ``Noether's 1.5th theorem,'' as it exists in the middle ground between her 1st and 2nd theorems.\footnote{This is a well-known fact \cite{Karatas:1989mt} but to my knowledge had not been given a specific name in the literature. Having said that, a closely related result is called the `Generalized Noether Theorem' \cite{compere2019advanced}.} 

To be specific, say that the infinitesimal gauge symmetries are parameterized by a spacetime function $f$. The conserved current $J^\mu_f$ will satisfy $\partial_\mu J^\mu_f \approx 0$. This means we can always find an anti-symmetric 2-form $K^{\mu \nu}_f$ such that $J^\mu_f \approx \partial_\nu K^{\mu \nu}_f$ if our spacetime has trivial de Rham cohomology. Actually, this $K^{\mu \nu}_f$ can only be well defined up to the addition of a total derivative term of the form
\begin{equation}
    K_f^{\mu \nu} \to K_f^{\mu \nu} + \partial_\rho M^{\mu \nu \rho}
\end{equation}
where $M^{\mu \nu \rho}$ is any completely anti-symmetric rank-3 tensor.

Noether's 1.5th theorem states that there is some choice of $K^{\mu \nu}_f$ which is \textit{locally constructed} out of the fields and gauge parameter $f$. 

Note that the addition of a term $\partial_\rho M^{\mu \nu \rho}$ will not affect integrals of the form $\int_S dS_{\mu \nu} K^{\mu \nu}$ when $S$ is a closed surface. Therefore, to prove the theorem, it is sufficient to show that for any surface $S$, the integral $\int_S dS_{\mu \nu} K^{\mu \nu}$ only depends on the values and derivatives of the fields and gauge parameter $f$ on the surface itself.

To examine why this ``locally constructed'' clause is important, let's quickly review how $K^{\mu \nu}$ can be found in general when $\partial_\nu K^{\mu \nu} = J^\mu$ and $\partial_\mu J^\mu = 0$. It can be found using the integral
\begin{equation}
    K^{\mu \nu}(x) = \int_{x_0}^x (J^\mu  dx^\nu - J^\nu dx^\mu)
\end{equation}
where we integrate along an arbitrary path connecting a basepoint $x_0$ to $x$. If the path is deformed, $K^{\mu \nu}$ will change by a total derivative. To see this, one can consider a small square path in the $x^\alpha$-$x^\beta$ plane of side length $\eps$, which we can denote as ``$\square_{\alpha \beta}$'', and calculate that
\begin{equation}
    \int_{\square_{\alpha \beta}} (J^\mu dx^\nu - J^\nu dx^\mu) = \eps^2 \partial_\rho M_{\alpha \beta }^{\mu \nu \rho}
\end{equation}
where
\begin{equation}
    M_{\alpha \beta }^{\mu \nu \rho} = -\delta^{[ \mu}_\alpha \delta^\nu_\beta J^{\rho]}.
\end{equation}
In any case, a generic $K^{\mu \nu}$ defined by such an integral will be intrinsically non-local, as its value at $x$ depends on the value $J$ at other spacetime points as well.

We will now prove Noether's 1.5th theorem, using a modified version of the proof presented in \cite{avery2016noether}. The proof is non-constructive, meaning in practice one must just observe it holds for whatever example one is interested in.

We will crucially use the fact that $J^\mu_f$ is itself guaranteed to be locally constructed out of the fields and $f$ by its standard method of construction used in the proof of Noether's 1st theorem. Furthermore, $J^\mu_f(x)$ will be $0$ at any spacetime points $x$ where $f(x) = 0$ and its derivatives $\partial_\mu f(x) = 0$.

\begin{figure}[H]
\begin{center}

\tikzset{every picture/.style={line width=0.75pt}} 

\begin{tikzpicture}[x=0.75pt,y=0.75pt,yscale=-1,xscale=1]

\draw [line width=1.5]    (111.5,99) .. controls (93.5,97) and (83.5,26) .. (111.5,22) ;
\draw    (111.5,22) .. controls (24.5,0) and (24.5,119) .. (111.5,99) ;
\draw [line width=1.5]  [dash pattern={on 5.63pt off 4.5pt}]  (111.5,99) .. controls (134.5,90) and (128.5,33) .. (111.5,22) ;
\draw    (111.5,22) .. controls (209.5,-17) and (222.5,132) .. (111.5,99) ;

\draw (146.13,60.17) node    {$V$};
\draw (45.09,61.05) node [anchor=east] [inner sep=0.75pt]    {$\Sigma _{1}$};
\draw (205.54,59.05) node    {$\Sigma _{2}$};
\draw (91.45,70.93) node [anchor=south east] [inner sep=0.75pt]    {$S$};

\end{tikzpicture}

\end{center}
\caption{\label{fig:noether15} $\partial V = \Sigma_1 + \Sigma_2$ and $\partial \Sigma_1 = - \partial \Sigma_2 = S$.}
\end{figure}
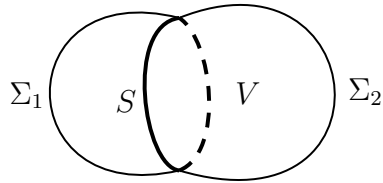

First, choose an arbitrary 4-volume $V$. Denote its boundary as $\partial V = \Sigma_1 + \Sigma_2$ where $\Sigma_1$ and $\Sigma_2$ share a common boundary $\partial \Sigma_1 = -\partial \Sigma_2 = S$ (with a sign for orientation). A diagram is provided in Figure \ref{fig:noether15}.

Notice that
\begin{align}
    0 &\approx \int_V d^4 x \sg \nabla_\mu J^\mu_f \\
    &= \int_{\Sigma_1} d\Sigma_\mu J^\mu_f + \int_{\Sigma_2} d\Sigma_\mu J^\mu_f
\end{align}
which can be rewritten as
\begin{equation}
    \int_{\Sigma_1} d\Sigma_\mu J^\mu_f \approx - \int_{\Sigma_2} d\Sigma_\mu J^\mu_f.
\end{equation}
Let's now choose an $f$ such that $f(x) = 0$ when $x \in \Sigma_2$ and $\partial_\mu f(x) = 0$ when $x \in S$. The RHS of the above equation will be $0$, proving that
\begin{equation}
    \int_{\Sigma_1} d\Sigma_\mu J^\mu_f \approx \int_S dS_{\mu \nu} K^{\mu \nu}_f \approx 0 \hspace{0.5 cm} \text{if $f$ and its derivatives are $0$ on $S$.}
\end{equation}
Because $J^\mu_f$ is derived by performing an infinitesimal gauge transformation on our fields, it is guaranteed to be linear in $f$, meaning that
\begin{align}
    J^\mu_{f_1} + J^\mu_{f_2} &= J^\mu_{f_1+f_2} \\K^{\mu\nu}_{f_1} + K^{\mu\nu}_{f_2} &= K^{\mu\nu}_{f_1+f_2}.
\end{align}
Therefore, we have just shown that
\begin{equation}
    \int_S dS_{\mu \nu} K^{\mu \nu}_{f_1} \approx \int_S dS_{\mu \nu} K^{\mu \nu}_{f_2} \hspace{0.5 cm} \text{if $f_1$, $f_2$ and their derivatives are equal on $S$.}
\end{equation}
We have therefore proven that $K^{\mu \nu}_f$ must depend locally on the gauge parameter $f$. To complete the proof, we must also show that it depends locally on the field variables as well.

Because $K^{\mu \nu}_f$ depends locally on $f$, we know that the equation
\begin{equation}
    \int_S dS_{\mu \nu} K^{\mu \nu}_{f} = \int_{\Sigma_1} d \Sigma_\mu J^\mu_f
\end{equation}
will still hold if on the RHS we replace $f$ with $f'$ such that $f$ and its derivatives are equal to $f'$ and its derivatives on $S$. We can then take $f'$ to be 0 outside of an arbitrarily small tubular neighborhood of $S$. However, the RHS of the above equation would then only depend on field variables arbitrarily close to $S$, as $J_{f'}^\mu$ will be 0 wherever $f'$ and its derivatives are $0$. Therefore, the LHS can also only depend on field variables arbitrarily close to $S$ as well. This completes the proof of Noether's 1.5th theorem.

A curious reader might wonder what the Noether's-1.5th-theorem diffeomorphism 2-form is for pure GR. Given an arbitrary vector $\xi$, it is
\begin{equation}
    K_\xi^{\mu \nu} = \frac{1}{16 \pi G} (\nabla^\mu \xi^\nu - \nabla^\nu \xi^\mu).
\end{equation}
Note that the $J^\mu_\xi \approx \nabla_\nu K_\xi^{\mu \nu} = 0$ if $\xi$ is a Killing vector of a particular solution $g_{\mu \nu}$. This is because such a transformation represents an ``empty'' transformation $\delta g_{\mu \nu}= \mathcal{L}_\xi g_{\mu \nu} = 0$ on that solution, meaning the Noether current must vanish.\footnote{This is the same thing that happens in EM when $\lambda = 1$ and we recover Gauss's law.} This means that the surface of integration of $K^{\mu \nu}_\xi$ can be deformed freely without changing the value of the integral. This is a very powerful observation when coupled with the fact that $K^{\mu \nu}_\xi$ depends locally on $\xi$ and the metric. For instance, if one has an arbitrary spacetime which asymptotically approaches some fixed background metric $\overline{g}_{\mu \nu}$ at large distances, one can use a Killing vector $\xi$ of $\overline{g}_{\mu \nu}$ to define surface charges $Q_\xi = \int_S dS_{\mu \nu} K^{\mu \nu}_\xi$ where $S$ is topologically a large 2-sphere at infinity. $Q_\xi$ will not change if $S$ is deformed as long as $S$ remains sufficiently far away from the interior of the spacetime where $g_{\mu \nu}$ can deviate from $\overline{g}_{\mu \nu}$ and $\xi$ will no longer satisfy Killing's equation. These surface charges will give, for instance, the Komar mass and Komar angular momentum when one takes $\xi = \partial_t$ or $\partial_\phi$ for asymptotically Kerr spacetimes.

Actually, some of these Komar charges will be off by relative factors of 2. These discrepancies will be resolved if one considers more intricate constructions of diffeomorphism surface charges using the covariant phase space formalism. This was done, for instance, in Wald's classic paper \cite{wald1993black} where he gave a Noetherian proof of the first law of black hole thermodynamics. However, the most general ``best'' proscription for defining these surface charges is still a matter of debate, with different approaches used by different practitioners in different contexts \cite{compere2019advanced, harlow2020covariant, Wald:1999wa, iyer1994some, abbott1982stability}. 

\section{Results From Classical Electromagnetism}\label{app_EM}

\subsection{Green's functions}

Say you have a massless scalar field satisfying the inhomogeneous wave equation
\begin{equation}\label{inhomo_wave}
\partial^\mu \partial_\mu \phi = j(x)
\end{equation}
where $j(x)$ is an external source. We are interested in solving for $\phi$ (given some boundary conditions) for any $j(x)$. To do this let's find the Green's function $G(x)$ which satisfies
\begin{equation} \label{G_eq}
    \partial^\mu \partial_\mu G(x) = \delta^4(x).
\end{equation}
We can use the Fourier transforms
\begin{align}
    G(x) &= \frac{1}{(2 \pi)^4} \int d^4 k \; e^{-i k x} \widetilde{G}(k) \\
    \delta^4(x) &= \frac{1}{(2 \pi)^4} \int d^4 k \; e^{-i k x}
\end{align}
to rewrite \eqref{G_eq} as
\begin{equation}
    - k^2 \widetilde{G}(k) = 1.
\end{equation}
Now, the above equation does not specify a unique solution for $\widetilde{G}(k)$ due to the fact that one cannot divide by $k^2$ when it is 0. This non-uniqueness was actually inevitable. Note that if you have a function $F$ that satisfies the homogeneous wave equation $\partial^2 F = 0$, then $G + F$ will still solve \eqref{G_eq}. So the non-invertibility of $\partial^2$ is due to the fact that there are many possible solutions which differ by solutions to the homogeneous equation. In particular, these different solutions will all have different boundary conditions.

Two possible Green's functions are the ``retarded'' and ``advanced'' solutions
\begin{align}
    \widetilde{G}_{\substack{\rm ret \\ \rm adv} }(k) &= \frac{-1}{(k_0 \pm i \eps) ^2 - \bk^2 } \\
    G_{\substack{\rm ret \\ \rm adv} }(x) &= \frac{-1}{(2 \pi)^4} \int d^4 k \frac{e^{- i k x}}{(k_0 \pm i \eps) ^2 - \bk^2 }. \label{G_retadv_fourier}
\end{align}
Note that the integrands have two poles in the complex $k_0$ plane.
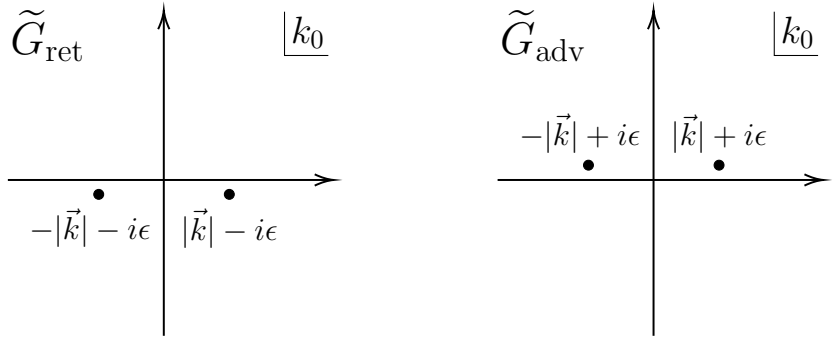
\begin{figure}[H]
\begin{center}   

\begin{tikzpicture}[x=0.65pt,y=0.65pt,yscale=-1,xscale=1]

\draw [line width=0.75]    (17,109) -- (204.5,109) ;
\draw [shift={(206.5,109)}, rotate = 180] [color={rgb, 255:red, 0; green, 0; blue, 0 }  ][line width=0.75]    (10.93,-3.29) .. controls (6.95,-1.4) and (3.31,-0.3) .. (0,0) .. controls (3.31,0.3) and (6.95,1.4) .. (10.93,3.29)   ;
\draw [line width=0.75]    (107.75,199.75) -- (107.75,12.25) ;
\draw [shift={(107.75,10.25)}, rotate = 450] [color={rgb, 255:red, 0; green, 0; blue, 0 }  ][line width=0.75]    (10.93,-3.29) .. controls (6.95,-1.4) and (3.31,-0.3) .. (0,0) .. controls (3.31,0.3) and (6.95,1.4) .. (10.93,3.29)   ;
\draw   (203.5,36.5) -- (178,36.5) -- (178,11) ;
\draw  [fill={rgb, 255:red, 0; green, 0; blue, 0 }  ,fill opacity=1 ] (143,117.38) .. controls (143,115.79) and (144.29,114.5) .. (145.88,114.5) .. controls (147.46,114.5) and (148.75,115.79) .. (148.75,117.38) .. controls (148.75,118.96) and (147.46,120.25) .. (145.88,120.25) .. controls (144.29,120.25) and (143,118.96) .. (143,117.38) -- cycle ;
\draw  [fill={rgb, 255:red, 0; green, 0; blue, 0 }  ,fill opacity=1 ] (67,117.38) .. controls (67,115.79) and (68.29,114.5) .. (69.88,114.5) .. controls (71.46,114.5) and (72.75,115.79) .. (72.75,117.38) .. controls (72.75,118.96) and (71.46,120.25) .. (69.88,120.25) .. controls (68.29,120.25) and (67,118.96) .. (67,117.38) -- cycle ;
\draw [line width=0.75]    (302,109) -- (489.5,109) ;
\draw [shift={(491.5,109)}, rotate = 180] [color={rgb, 255:red, 0; green, 0; blue, 0 }  ][line width=0.75]    (10.93,-3.29) .. controls (6.95,-1.4) and (3.31,-0.3) .. (0,0) .. controls (3.31,0.3) and (6.95,1.4) .. (10.93,3.29)   ;
\draw [line width=0.75]    (392.75,199.75) -- (392.75,12.25) ;
\draw [shift={(392.75,10.25)}, rotate = 450] [color={rgb, 255:red, 0; green, 0; blue, 0 }  ][line width=0.75]    (10.93,-3.29) .. controls (6.95,-1.4) and (3.31,-0.3) .. (0,0) .. controls (3.31,0.3) and (6.95,1.4) .. (10.93,3.29)   ;
\draw   (488.5,36.5) -- (463,36.5) -- (463,11) ;
\draw  [fill={rgb, 255:red, 0; green, 0; blue, 0 }  ,fill opacity=1 ] (428,100.38) .. controls (428,98.79) and (429.29,97.5) .. (430.88,97.5) .. controls (432.46,97.5) and (433.75,98.79) .. (433.75,100.38) .. controls (433.75,101.96) and (432.46,103.25) .. (430.88,103.25) .. controls (429.29,103.25) and (428,101.96) .. (428,100.38) -- cycle ;
\draw  [fill={rgb, 255:red, 0; green, 0; blue, 0 }  ,fill opacity=1 ] (352,100.38) .. controls (352,98.79) and (353.29,97.5) .. (354.88,97.5) .. controls (356.46,97.5) and (357.75,98.79) .. (357.75,100.38) .. controls (357.75,101.96) and (356.46,103.25) .. (354.88,103.25) .. controls (353.29,103.25) and (352,101.96) .. (352,100.38) -- cycle ;

\draw (191.71,23.05) node  [font=\large]  {$k_{0}$};
\draw (17,8.4) node [anchor=north west][inner sep=0.75pt]  [font=\Large]  {$\widetilde{G}_{\mathrm{ret}}$};
\draw (64.84,136.67) node  [font=\normalsize]  {$-|\vec{k} |-i\epsilon $};
\draw (145.84,137.17) node  [font=\normalsize]  {$|\vec{k} |-i\epsilon $};
\draw (476.71,23.05) node  [font=\large]  {$k_{0}$};
\draw (302,8.4) node [anchor=north west][inner sep=0.75pt]  [font=\Large]  {$\widetilde{G}_{\mathrm{adv}}$};
\draw (349.84,81.67) node  [font=\normalsize]  {$-|\vec{k} |+i\epsilon $};
\draw (430.84,82.17) node  [font=\normalsize]  {$|\vec{k} |+i\epsilon $};

\end{tikzpicture}

\end{center}
\caption{\label{fig:Gretadv_poles} Pole locations for retarded and advanced Green's functions.}
\end{figure}
Let's evaluate, say, $G_{\rm ret}(x)$. Note that in the numerator of \eqref{G_retadv_fourier} we have $e^{- i k_0 x_0}$. If $x_0 < 0$, this dampens for $\Im(k_0) > 0$ and explodes for $\Im(k_0) < 0$. If $x_0 > 0$, this dampens for $\Im(k_0) < 0$ and explodes for $\Im(k_0) > 0$.

Consider a large semi-circular contour like $\cap$ or $\cup$. Because the overall power of the $\widetilde{G}$ integral is roughly $ \int \tfrac{d k_0}{k_0^2} \sim 1/k_0$, this will go to zero for large radius as long as the exponential factor $e^{-ik_0 x_0}$ doesn't explode. So for $x_0 < 0$ the upper semi circular integral is 0, while for $x_0 > 0$ the lower semi circular integral is 0.

\begin{figure}[H]
\begin{center}

\begin{tikzpicture}[x=0.65pt,y=0.65pt,yscale=-1,xscale=1]

\draw [line width=0.75]    (17,109) -- (204.5,109) ;
\draw [shift={(206.5,109)}, rotate = 180] [color={rgb, 255:red, 0; green, 0; blue, 0 }  ][line width=0.75]    (10.93,-3.29) .. controls (6.95,-1.4) and (3.31,-0.3) .. (0,0) .. controls (3.31,0.3) and (6.95,1.4) .. (10.93,3.29)   ;
\draw [line width=0.75]    (107.75,199.75) -- (107.75,12.25) ;
\draw [shift={(107.75,10.25)}, rotate = 450] [color={rgb, 255:red, 0; green, 0; blue, 0 }  ][line width=0.75]    (10.93,-3.29) .. controls (6.95,-1.4) and (3.31,-0.3) .. (0,0) .. controls (3.31,0.3) and (6.95,1.4) .. (10.93,3.29)   ;
\draw   (203.5,36.5) -- (178,36.5) -- (178,11) ;
\draw  [fill={rgb, 255:red, 0; green, 0; blue, 0 }  ,fill opacity=1 ] (143,117.38) .. controls (143,115.79) and (144.29,114.5) .. (145.88,114.5) .. controls (147.46,114.5) and (148.75,115.79) .. (148.75,117.38) .. controls (148.75,118.96) and (147.46,120.25) .. (145.88,120.25) .. controls (144.29,120.25) and (143,118.96) .. (143,117.38) -- cycle ;
\draw  [fill={rgb, 255:red, 0; green, 0; blue, 0 }  ,fill opacity=1 ] (67,117.38) .. controls (67,115.79) and (68.29,114.5) .. (69.88,114.5) .. controls (71.46,114.5) and (72.75,115.79) .. (72.75,117.38) .. controls (72.75,118.96) and (71.46,120.25) .. (69.88,120.25) .. controls (68.29,120.25) and (67,118.96) .. (67,117.38) -- cycle ;
\draw [line width=0.75]    (302,109) -- (489.5,109) ;
\draw [shift={(491.5,109)}, rotate = 180] [color={rgb, 255:red, 0; green, 0; blue, 0 }  ][line width=0.75]    (10.93,-3.29) .. controls (6.95,-1.4) and (3.31,-0.3) .. (0,0) .. controls (3.31,0.3) and (6.95,1.4) .. (10.93,3.29)   ;
\draw [line width=0.75]    (392.75,199.75) -- (392.75,12.25) ;
\draw [shift={(392.75,10.25)}, rotate = 450] [color={rgb, 255:red, 0; green, 0; blue, 0 }  ][line width=0.75]    (10.93,-3.29) .. controls (6.95,-1.4) and (3.31,-0.3) .. (0,0) .. controls (3.31,0.3) and (6.95,1.4) .. (10.93,3.29)   ;
\draw   (488.5,36.5) -- (463,36.5) -- (463,11) ;
\draw  [fill={rgb, 255:red, 0; green, 0; blue, 0 }  ,fill opacity=1 ] (428,117.38) .. controls (428,115.79) and (429.29,114.5) .. (430.88,114.5) .. controls (432.46,114.5) and (433.75,115.79) .. (433.75,117.38) .. controls (433.75,118.96) and (432.46,120.25) .. (430.88,120.25) .. controls (429.29,120.25) and (428,118.96) .. (428,117.38) -- cycle ;
\draw  [fill={rgb, 255:red, 0; green, 0; blue, 0 }  ,fill opacity=1 ] (352,117.38) .. controls (352,115.79) and (353.29,114.5) .. (354.88,114.5) .. controls (356.46,114.5) and (357.75,115.79) .. (357.75,117.38) .. controls (357.75,118.96) and (356.46,120.25) .. (354.88,120.25) .. controls (353.29,120.25) and (352,118.96) .. (352,117.38) -- cycle ;
\draw  [draw opacity=0][line width=1.5]  (36.4,108.6) .. controls (36.4,108.6) and (36.4,108.6) .. (36.4,108.6) .. controls (36.4,69.17) and (68.37,37.2) .. (107.8,37.2) .. controls (147.23,37.2) and (179.2,69.17) .. (179.2,108.6) -- (107.8,108.6) -- cycle ; \draw  [color={rgb, 255:red, 82; green, 198; blue, 22 }  ,draw opacity=1 ][line width=1.5]  (36.4,108.6) .. controls (36.4,108.6) and (36.4,108.6) .. (36.4,108.6) .. controls (36.4,69.17) and (68.37,37.2) .. (107.8,37.2) .. controls (147.23,37.2) and (179.2,69.17) .. (179.2,108.6) ;
\draw  [draw opacity=0][line width=1.5]  (321.27,109.35) .. controls (321.27,109.35) and (321.27,109.35) .. (321.27,109.35) .. controls (321.27,148.78) and (353.23,180.75) .. (392.67,180.75) .. controls (432.1,180.75) and (464.07,148.78) .. (464.07,109.35) -- (392.67,109.35) -- cycle ; \draw  [color={rgb, 255:red, 82; green, 198; blue, 22 }  ,draw opacity=1 ][line width=1.5]  (321.27,109.35) .. controls (321.27,109.35) and (321.27,109.35) .. (321.27,109.35) .. controls (321.27,148.78) and (353.23,180.75) .. (392.67,180.75) .. controls (432.1,180.75) and (464.07,148.78) .. (464.07,109.35) ;
\draw  [color={rgb, 255:red, 82; green, 198; blue, 22 }  ,draw opacity=1 ][fill={rgb, 255:red, 82; green, 198; blue, 22 }  ,fill opacity=1 ] (161.62,67.26) -- (155.63,55.27) -- (166.85,62.6) -- (159.93,60.1) -- cycle ;
\draw  [color={rgb, 255:red, 82; green, 198; blue, 22 }  ,draw opacity=1 ][fill={rgb, 255:red, 82; green, 198; blue, 22 }  ,fill opacity=1 ] (60.98,58.41) -- (49.96,66.04) -- (55.63,53.9) -- (54.13,61.1) -- cycle ;

\draw  [color={rgb, 255:red, 82; green, 198; blue, 22 }  ,draw opacity=1 ][fill={rgb, 255:red, 82; green, 198; blue, 22 }  ,fill opacity=1 ] (448.16,148.81) -- (442.16,160.79) -- (453.38,153.47) -- (446.47,155.97) -- cycle ;
\draw  [color={rgb, 255:red, 82; green, 198; blue, 22 }  ,draw opacity=1 ][fill={rgb, 255:red, 82; green, 198; blue, 22 }  ,fill opacity=1 ] (342.9,155.28) -- (331.92,147.61) -- (337.53,159.77) -- (336.07,152.57) -- cycle ;

\draw (191.71,23.05) node  [font=\large]  {$k_{0}$};
\draw (17,8.4) node [anchor=north west][inner sep=0.75pt]  [font=\Large]  {$\widetilde{G}_{\mathrm{ret}}$};
\draw (476.71,23.05) node  [font=\large]  {$k_{0}$};
\draw (302,8.4) node [anchor=north west][inner sep=0.75pt]  [font=\Large]  {$\widetilde{G}_{\mathrm{ret}}$};
\draw (187.82,61.17) node  [font=\large,color={rgb, 255:red, 82; green, 198; blue, 22 }  ,opacity=1 ]  {$\mathbf{=0}$};
\draw (477.22,154.77) node  [font=\large,color={rgb, 255:red, 82; green, 198; blue, 22 }  ,opacity=1 ]  {$\mathbf{=0}$};
\draw    (72.71,205.55) -- (140.71,205.55) -- (140.71,236.55) -- (72.71,236.55) -- cycle  ;
\draw (106.71,221.05) node  [font=\normalsize]  {$x_{0} < \ 0$};
\draw    (358.71,205.55) -- (426.71,205.55) -- (426.71,236.55) -- (358.71,236.55) -- cycle  ;
\draw (392.71,221.05) node  [font=\normalsize]  {$x_{0}  >\ 0$};

\end{tikzpicture}

\end{center}
\caption{\label{fig:Gret_semicircle} These large semi circles in green integrate to 0 for $\widetilde{G}_{\rm ret}$.}
\end{figure}
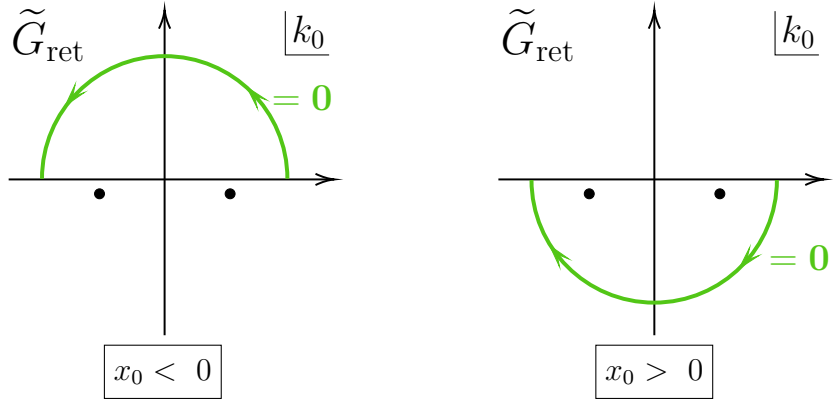

Now, our actual contour of interest is $k_0$ going from $-\infty$ to $\infty$. In the case of $x_0 < 0$ when we close the contour we don't enclose the poles, and therefore the integral is 0. However, when $x_0 > 0$ we do pick up the poles and get a non-zero result.

In order to easily find the residues of the integrand, one can use the difference of squares
\begin{equation}
    \frac{1}{(k_0 + i \eps)^2 - \bk^2} = \frac{1}{(k_0 + |\bk| + i \eps)(k_0 - |\bk| + i \eps)}.
\end{equation}
After using the residue theorem, (rembering to multiply by $-1$ due to the clockwise contour) we get
\begin{align}
    G_{\rm ret}(x) &=  \frac{\Theta(x_0)}{(2 \pi)^4} \int d^3 k \frac{2 \pi i}{2 |\bk|} e^{i \bk \bx }(e^{-i |\bk| x_0} - e^{i |\bk| x_0}) \label{Gret_eq}\\
    &= \frac{\Theta(x_0)}{(2 \pi)^3} \int d^3 k \frac{1 }{|\bk|} e^{i \bk \bx }\sin( |\bk| x_0) \\
    &= \frac{\Theta(x_0)}{(2 \pi)^2} \int \kappa^2 d \kappa \sin \theta d \theta \; \frac{1 }{\kappa} e^{i\kappa |\bx| \cos \theta}\sin( \kappa x_0)
\end{align}
where in the last line, $\kappa = |\bk|$ and $\theta$ is the angle between $\bk$ and $\bx$. Substituting $u = \cos \theta$,
\begin{align}
    G_{\rm ret}(x) &= \frac{\Theta(x_0)}{(2 \pi)^2} \int_0^\infty \kappa^2 d \kappa \int_{-1}^{+1} du \; \frac{1 }{\kappa} e^{i\kappa |\bx| u}\sin( \kappa x_0) \\
    &=  \frac{\Theta(x_0)}{(2 \pi)^2 |\bx|} \int_0^\infty d \kappa \; 2 \sin(\kappa |\bx|) \sin(\kappa x_0)\\
    &=  \frac{\Theta(x_0)}{(2 \pi)^2 |\bx|} \int_{-\infty}^{+\infty} d \kappa \; \sin(\kappa |\bx|) \sin(\kappa x_0)\\
    &= \frac{\Theta(x_0)}{(2 \pi)^2 |\bx|} \int_{-\infty}^{+\infty} d \kappa \; \frac{1}{2}( e^{ i \kappa (x_0 - |\bx|)} -  e^{i \kappa (x_0 + |\bx|)} )
\end{align}
where for the last equality you must you use the $\sin/\cos$ addition formulas and that the integral of an odd function is 0. Note that the results are just delta functions, with the second one is always being 0 because $x_0$ and $|\bx|$ are both positive. Therefore
\begin{align}
    G_{\rm ret}(x) &= \frac{\Theta(x_0)}{4 \pi |\bx|} \delta(x_0 - |\bx|) \\
    &= \frac{\Theta(x_0)}{2 \pi} \delta(x^2).
\end{align}
Note that from the definition of the $G_{\rm ret}$ and $G_{\rm adv}$ in \eqref{G_retadv_fourier}, we manifestly have
\begin{equation}\label{Gsym}
    G_{\rm adv}(x) = G_{\rm ret}(-x).
\end{equation}
Therefore
\begin{align}
    G_{\rm adv}(x) &= \frac{\Theta(-x_0)}{2 \pi} \delta(x^2).
\end{align}
We can now clearly see what the boundary conditions of $G_{\rm ret}$ and $G_{\rm adv}$ are. $G_{\rm ret}$ vanishes for $x_0 < 0$ and $G_{\rm adv}$ vanishes for $x_0 > 0$.

There is another useful way to write $G_{\rm ret}$. Using \eqref{Gret_eq}, we see
\begin{align}
    G_{\rm ret}(x) &= i \frac{\Theta(x_0)}{(2 \pi)^3} \int d^3k \frac{1}{2 |\bk|} e^{i \bk \bx} (e^{- i |\bk| x_0} - e^{i |\bk| x_0} ) \\
    &= i \frac{\Theta(x_0)}{(2 \pi)^3} \int d^4 k \; \mathrm{sign}(k_0) e^{-ikx} \delta(k^2). \label{Gret_fourier}
\end{align}

To complete the discussion, let's define the difference of the two Green's functions
\begin{equation}
    G_{(-)}(x) \equiv G_{\rm ret}(x) - G_{\rm adv}(x).
\end{equation}
Using \eqref{Gsym}, along with \eqref{Gret_fourier}, we find
\begin{align}\label{Gm_fourier}
    G_{(-)}(x) &= \frac{i}{(2 \pi)^3} \int d^4k \; \mathrm{sign}(k_0) e^{- i k x} \delta(k^2).
\end{align}
From \eqref{Gm_fourier} we can see that $G_{(-)}(x)$ manifestly solves the homogeneous wave equation $\partial^2 G_{(-)}=0$. This had to be the case because it was defined as the difference between $G_{\rm ret}$ and $G_{\rm adv}$ which each solve the inhomogeneous equation.

\subsection{EM radiation}

The following is mostly taken from section 1.3.2 of \cite{itzykson1980quantum}. If we have the gauge field $A_\mu$ and field strength tensor
\begin{equation}
    F_{\mu \nu} = \partial_\mu A_\nu - \partial_\nu A_\mu
\end{equation}
Maxwell's laws are
\begin{equation}
    \partial_\mu F^{\mu \nu} = j^\nu.
\end{equation}
If we use Lorenz gauge
\begin{equation}
    \partial_\mu A^\mu = 0
\end{equation}
then
\begin{equation}
    \partial_\mu F^{\mu \nu} = \partial_\mu ( \partial^\mu A^\nu - \partial^\nu A^\mu) = \partial_\mu \partial^\mu A^\nu
\end{equation}
and Maxwell's laws become
\begin{equation}
    \partial_\mu \partial^\mu A^\nu = j^\nu
\end{equation}
which is just four copies of the scalar wave equation \eqref{inhomo_wave}.

We can immediately write the gauge field in the presence of the source.\footnote{One might question why we exclusively use the retarded propagator, as this seems to break time reversal symmetry. The answer is simply that the advanced propagator corresponds to a non-physical situation where a perfectly arranged sphere of light (which has existed since the beginning of the time) has been converging on our charge only to get perfectly ``sucked up'' by it just as the charge accelerates. These initial conditions require an absurd degree of fine-tuning and morally violate the second law of thermodynamics \cite{wheeler1945interaction}.}
\begin{align}
    A^\mu(x) &= \int d^4 y G_{\rm ret}(x - y) j^\mu(y). \label{Amu_Gret}
\end{align}
However, going forward, we're actually going to use a different expression. Let us substitute in
\begin{equation}
    G_{\rm ret} = G_{(-)} + G_{\rm adv}.
\end{equation}
Note that $G_{\rm adv}$ only has support in the past lightcone and thus can't account for any radiated light. Therefore, assuming we are only concerned with radiation emitted towards the future, we can write
\begin{equation}\label{Arad_Gm}
    A^\mu_{\rm rad}(x) = \int d^4 y G_{(-)}(x - y) j^\mu(y).
\end{equation}
Because $G_{(-)}$ solves free $(j^\mu = 0)$ equations of motion, it is just a sum of plane waves, which means it should be thought of as describing free moving light waves. It is sometimes called $G_{\rm rad}$ for this reason.

Next, we write the current $j^\mu(x)$ in terms of its Fourier components
\begin{equation}
    j^\mu(x) = \frac{1}{(2 \pi)^4} \int d^4 k \;e^{-i k x} \tilde{j}^\mu(k).
\end{equation}
Note the reality condition
\begin{equation}
    \tilde{j}_\mu(k) = \tilde{j}_\mu^* (-k).
\end{equation}
We rewrite \eqref{Arad_Gm} using \eqref{Gm_fourier} as
\begin{equation}\label{Aradk}
    A^\mu_{\rm rad}(x) = \frac{i}{(2 \pi)^3} \int d^4 k \; \mathrm{sign}(k_0) e^{-ikx} \delta(k^2) \tilde{j}^\mu(k).
\end{equation}

We now have an expression for the total radiated field, but one might wonder how much of the field is radiated in each mode. Recall that in Lorenz gauge, light waves take the form
\begin{equation}
    A_\mu(x) \propto \vep_\mu e^{- i k x}
\end{equation}
where $k^2 = 0$. Here, $\vep_\mu$ is called a `polarization vector.' Lorenz gauge implies the following condition:
\begin{equation}
    \partial_\mu A^\mu = 0 \hspace{0.75cm} \implies \hspace{0.75cm} k_\mu \vep^\mu = 0.
\end{equation}
Furthermore, the residual gauge symmetry of
\begin{equation}
    A_\mu \sim A_\mu + \partial_\mu \lambda \hspace{0.5 cm} \text{where} \hspace{0.5 cm} \partial_\mu \partial^\mu \lambda = 0
\end{equation}
implies the following redundancy of
\begin{equation}
    \vep_\mu \sim \vep_\mu + \widetilde{\lambda} k_\mu
\end{equation}
for any number $\widetilde{\lambda}$. This leaves two linearly independent spacelike polarization vectors, which we call $\vep_\mu^a$ for $a = 1,2$. In a Fourier decomposition, these depend on $k$ as $\vep^a_\mu = \vep^a_\mu(k)$ but we will suppress this dependence in our notation. We can choose them to satisfy
\begin{equation}
    (\vep^a)^2=-1 \hspace{1cm} \vep^1_\mu \vep^{2 \mu} = 0  \hspace{1cm} (\vep^a_\mu)^* = \vep^a_\mu
\end{equation}
for each $k$. This means we have the following ``identity'' operator (up to gauge transformation) which we can act on $A^\mu_{\rm rad}$ in \eqref{Aradk}. It can be written as
\begin{equation}\label{deltaeps}
    \delta^\mu_\nu = -\sum_{a = 1,2} \vep^{a \mu} \vep^a_\nu + \mathrm{gauge}.
\end{equation}
This also means that we have the two projection operators
\begin{equation}
    (P^a)_{\; \;\nu}^{\mu} = - \vep^{a \mu} \vep^a_{\nu} \hspace{1cm} \text{(no $a$ sum)}
\end{equation}
which can use to project a wave onto its polarization components.

Next we write the field strength tensor as
\begin{align}
    F^{\mu \nu}_{\rm rad}(x) &= \frac{1}{(2 \pi )^3 } \int d^4 k\; e^{-ikx} \mathrm{sign}(k_0) \delta(k^2) \big( k^\mu \tilde{j}^\nu -k^\nu \tilde{j}^\mu \big) \\
    &= \frac{1}{(2 \pi)^3} \int \frac{d^3 k}{2 k_0} e^{i \vec{k} \cdot \vec{x}} (k^\mu \tilde{j}^\nu - k^\nu \tilde{j}^\mu) \label{Frad}
\end{align}
where we used the formula
\begin{equation}
    \int d^4 k \delta(k^2) = \int \frac{d^3 k}{2 |\vec{k}|} \sum_{k_0 = \pm |\vec{k}|}
\end{equation}
and simply chose not to write the sum over $k_0 = \pm |\vec{k}|$ because it's notationally cumbersome and commonly suppressed.

The energy density of the field is given by the stress energy tensor as
\begin{equation}
    T^{0}_{\; \; 0} = -F^{\mu 0} F_{\mu 0} + \tfrac{1}{4} \delta^0_0 F^{\mu \nu} F_{\mu \nu}.
\end{equation}
Let's now find the total energy of the field $F^{\mu \nu}_{\rm rad}$ in \eqref{Frad}. First we evaluate $\int d^3 x F^{\mu\nu} F_{\mu \nu}$.
\begin{align}
    \int d^3 x F^{\mu \nu}F_{\mu \nu} &= \frac{1}{(2 \pi)^6} \int d^3 x \frac{d^3 k}{2k_0} \frac{d^3 k'}{2k_0'} e^{i \vec{x} \cdot (\vec{k} + \vec{k}')} 2(k^\mu k'_\mu \tilde{j}^\nu \tilde{j}'_\nu - k^\mu \tilde{j}_\mu {k'}^\nu \tilde{j}_\nu') \\
    &= \frac{1}{(2 \pi)^3} \int \frac{d^3 k}{-4 k_0^2} (-2) (k^2 \tilde{j}^\nu \tilde{j}_\nu^* - k^\mu \tilde{j}_\mu k^\mu \tilde{j}_\mu^*) \\
    &= 0.
\end{align}
The whole expression is zero because (i) $k^2 = 0$ in the integral and (ii) $\partial_\mu j^\mu = 0$ implies $k^\mu \tilde{j}_\mu(k) = 0$.
Next, let's evaluate $\int d^3 x F^{\mu 0} F_{\mu 0}$.
\begin{align}
    \int d^3 x F^{\mu 0} F_{\mu 0} &= \frac{1}{(2 \pi)^6} \int d^3 x \frac{d^3 k}{2k_0} \frac{d^3 k'}{2k_0'} e^{i \vec{x} \cdot (\vec{k} + \vec{k}')} (k^\mu \tilde{j}^0 - k^0 \tilde{j}^\mu) (k_\mu' \tilde{j}_0' - k_0' \tilde{j}_\mu') \\
    &= \frac{1}{(2 \pi)^3} \int \frac{d^3 k}{- 4 k_0^2} (-1) (k_0^2 \tilde{j}^\mu \tilde j_\mu^*) \\
    &= \frac{-1}{(2 \pi)^3} \int_{k_0 > 0} \frac{d^3 k}{2 k_0} k_0 \sum_{a=1,2}\left| \vep^a_\mu \tilde{j}^\mu \right|^2.
\end{align}
To get the last line, we acted with the decomposition of the identity operator \eqref{deltaeps} on $A_{\mathrm{rad}}^\mu$ in \eqref{Aradk}, which essentially allows us to replace $\tilde{j}^\mu \mapsto -\sum_a \vep^{a \mu} \vep^a_{\; \nu} \tilde{j}^\nu$.

Therefore, the total energy of radiated photons is
\begin{align}
    \mathcal{E} &= \int d^3 T^0_{\;0} \\
    &= \frac{1}{(2 \pi)^3} \int_{k_0 > 0} \frac{d^3 k}{2 k_0} k_0 \sum_{a=1,2}\left| \vep^a_\mu \tilde{j}^\mu \right|^2.
\end{align}
The amount of energy emitted per momentum vector $k^\mu$ and polarization $\vep^a_\mu$ is
\begin{equation}
    d \mathcal{E} = \frac{1}{(2 \pi)^3} \frac{d^3 k}{2 k_0} k_0 \left| \vep^a_\mu \tilde{j}^\mu \right|^2.
\end{equation}
We can write the polarization summed version as
\begin{equation}\label{dEsummed}
    d \mathcal{E} = \frac{-1}{(2 \pi)^3} \frac{d^3 k}{2 k_0} k_0 \tilde{j}^\mu \tilde{j}^*_\mu.
\end{equation}

\subsection{The Liénard–Wiechert potential}\label{app_LW}

If we have a single point particle of charge $e$ tracing out a path in spacetime $x(\tau)$,
\begin{equation}
    j^\mu(x) = e \int d \tau \; \frac{d x^\mu(\tau)}{d \tau} \delta^4[ x - x(\tau) ].
\end{equation}
Let us quickly check that $\partial_\mu j^\mu = 0$. Note that the 1D delta function satisfies
\begin{equation}
    \partial_\tau \delta\left[x - f(\tau) \right] = - \frac{df}{d \tau} \partial_x \delta \left[x - f(\tau) \right]
\end{equation}
so
\begin{align}
    \partial_\mu j^\mu(x) &= e \int d \tau \; \frac{d x^\mu(\tau)}{d \tau} \frac{-1}{d x^\mu / d \tau} \partial_\tau \delta^4\left[ x - x(\tau) \right] \\
    &= e \int d \tau (-4) \partial_\tau \delta^4\left[ x - x(\tau) \right] \\
    &= 0
\end{align}
where in the last line we integrated by parts.

In Lorenz gauge, $A^\mu(x)$ is given by \eqref{Amu_Gret}
\begin{align}
    A^\mu(x) &= \int d^4 y \, G_{\rm ret}(x - y) j^\mu(y) \\
    &= e \int d \tau \frac{d x^\mu}{d \tau} G_{\rm ret}(x - x(\tau)) \\
    &= \frac{e}{2 \pi} \int d \tau  \frac{d x^\mu}{d \tau} \Theta(x^0 - x^0(\tau)) \delta[ (x - x(\tau))^2].
\end{align}
At this point we must use the delta function identity
\begin{equation}
    \delta( f(\tau)) = \frac{\delta(\tau - \tau_*)}{|f'(\tau_*)|}
\end{equation}
where $\tau_*$ is defined by $f(\tau_*) = 0$. The derivative of $(x - x(\tau))^2$ is
\begin{equation}
    \frac{d}{d \tau}(x - x(\tau))^2 = - 2 \dot x \cdot (x - x(\tau)).
\end{equation}
The above quantity is guaranteed to be negative because $\dot x^\mu$ is forward-pointing and timelike while $(x-x(\tau))^\mu$ is forward-pointing and null. Therefore, the absolute value introduces a minus sign and
\begin{equation}
    A^\mu(x) = \frac{e}{4 \pi} \frac{\dot{x}_{\rt}^\mu}{\dot{x}_{\rt} \cdot(x - x_{\rt})}
\end{equation}
where $x^\mu_{\rt}$ is the unique point on the path $x^\mu(\tau)$ that satisfies
\begin{align}
    x^0_{\rt} <& x^0 \\
    (x - x_{\rt})^2 &= 0.
\end{align}
Let's now find the expression for the potential created by a particle which moves at a constant velocity $\vec{\beta}$ and passes through the origin. This implies $x_{\rt} = (t_{\rt}, \vec{\beta} t_{\rt})$. To evaluate $A^\mu(t, \vec{r})$, we must first solve for $t_{\rt}$ using the equation
\begin{align}
    t - t_{\rt} &= |\vec{r} - \vec{\beta} t_{\rt} | \\
    t^2 - 2 t t_{\rt} + t_{\rt}^2 &= r^2 - 2 \vec{r} \cdot \vec{\beta} t_{\rt} + \beta^2 t_{\rt}^2
\end{align}

\begin{figure}[H]
\tikzset{every picture/.style={line width=0.75pt}} 

\begin{center}

\tikzset{every picture/.style={line width=0.75pt}} 

\begin{tikzpicture}[x=0.75pt,y=0.75pt,yscale=-1,xscale=1]

\draw    (78.49,35.33) -- (148.77,105.61) ;
\draw  [draw opacity=0] (148.92,105.66) .. controls (148.97,106.07) and (149,106.48) .. (149,106.89) .. controls (149,120.84) and (117.48,132.15) .. (78.6,132.15) .. controls (39.73,132.15) and (8.21,120.84) .. (8.21,106.89) .. controls (8.21,106.12) and (8.31,105.36) .. (8.49,104.6) -- (78.6,106.89) -- cycle ; \draw   (148.92,105.66) .. controls (148.97,106.07) and (149,106.48) .. (149,106.89) .. controls (149,120.84) and (117.48,132.15) .. (78.6,132.15) .. controls (39.73,132.15) and (8.21,120.84) .. (8.21,106.89) .. controls (8.21,106.12) and (8.31,105.36) .. (8.49,104.6) ;
\draw    (8.21,105.61) -- (78.49,35.33) ;
\draw    (127.33,12.33) -- (94.07,114.73) ;
\draw  [fill={rgb, 255:red, 0; green, 0; blue, 0 }  ,fill opacity=1 ] (75.42,35.33) .. controls (75.42,33.64) and (76.79,32.27) .. (78.49,32.27) .. controls (80.18,32.27) and (81.56,33.64) .. (81.56,35.33) .. controls (81.56,37.03) and (80.18,38.4) .. (78.49,38.4) .. controls (76.79,38.4) and (75.42,37.03) .. (75.42,35.33) -- cycle ;
\draw  [fill={rgb, 255:red, 0; green, 0; blue, 0 }  ,fill opacity=1 ] (106.75,66.67) .. controls (106.75,64.97) and (108.13,63.6) .. (109.82,63.6) .. controls (111.52,63.6) and (112.89,64.97) .. (112.89,66.67) .. controls (112.89,68.36) and (111.52,69.73) .. (109.82,69.73) .. controls (108.13,69.73) and (106.75,68.36) .. (106.75,66.67) -- cycle ;
\draw  [draw opacity=0][dash pattern={on 4.5pt off 4.5pt}] (9.06,104.71) .. controls (15.2,98.36) and (43.91,93.55) .. (78.4,93.55) .. controls (112.92,93.55) and (141.64,98.36) .. (147.75,104.72) -- (78.4,107.25) -- cycle ; \draw  [dash pattern={on 4.5pt off 4.5pt}] (9.06,104.71) .. controls (15.2,98.36) and (43.91,93.55) .. (78.4,93.55) .. controls (112.92,93.55) and (141.64,98.36) .. (147.75,104.72) ;

\draw (117.67,59.39) node [anchor=west] [inner sep=0.75pt]    {$( t_{\rt} ,\vec{\beta } t_{\rt})$};
\draw (78.2,20.83) node    {$( t,\vec{r})$};

\end{tikzpicture}

\end{center}
\caption{\label{fig:lightcone} Say a charged particle moves along the line parameterized by $(t, \vec{\beta} t)$. If an observer is located at a point $(t, \vec{r})$, then $(t_{\rt}, \vec{\beta} t_{\rt})$ is defined to be the point on its past lightcone that intersects the particle's path.}
\end{figure}
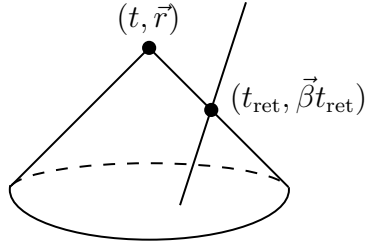

We can solve this using the quadratic equation and taking the smaller  root because we want the point in the past light cone.
\begin{equation}
    (1 - \beta^2) t_{\rt}^2 - 2(t - \vec{r} \cdot \vec{\beta} ) t_{\rt} + (t^2 - r^2) = 0
\end{equation}
\begin{equation}
    t_{\rt} = \frac{t - \vec{r} \cdot \vec{\beta} - \sqrt{(t - \vec{r} \cdot \vec{\beta} ) - (1 - \beta^2)(t^2 - r^2 ) }}{1 - \beta^2}
\end{equation}
Next, note that
\begin{align}
    &(1, \vec{\beta}) \cdot (t - t_{\rt}, \vec{r} - \vec{\beta} t_{\rt} ) \\
    &= t - \vec{r} \cdot \vec{\beta} - (1 - \beta^2) t_{\rt} \\
    &= \sqrt{(t - \vec{r} \cdot \vec{\beta} ) - (1 - \beta^2)(t^2 - r^2 ) } \\
    &= \sqrt{ (\vec{r} - \vec{\beta} t)^2 + (\vec{r} \cdot \vec{\beta} )^2 - \beta^2 r^2 }
\end{align}
so
\begin{equation} \label{LW_tr}
    A^\mu(t, \vec{r}) = \frac{e}{4 \pi} \frac{(1, \vec{\beta})}{\sqrt{ (\vec{r} - \vec{\beta} t)^2 + (\vec{r} \cdot \vec{\beta} )^2 - \beta^2 r^2 }}.
\end{equation}

\bibliography{mybib.bib}
\bibliographystyle{utphys}

\end{document}